\documentclass[aps,preprint,nofootinbib]{revtex4}%
\usepackage{amsfonts}
\usepackage{amsmath}
\usepackage{amssymb}
\usepackage{graphicx}%
\providecommand{\U}[1]{\protect\rule{.1in}{.1in}}
\begin{document}
\preprint{EFI 20-1, USC/20-1, arXiv:2001.nnnnn}
\title[H-atom and Oscillator]{Duality Between Hydrogen Atom and Oscillator Systems \\via Hidden SO(d,2) Symmetry and 2T-physics}
\author{Itzhak Bars$^{\dagger}$ and Jonathan L. Rosner$^{\ddagger}$}
\affiliation{$^{\dagger}$Department of Physics and Astronomy, University of Southern
California, Los Angeles, CA 90089-0484, USA, }
\affiliation{$^{\ddagger}$Enrico Fermi Institute and Department of Physics, University of
Chicago, Chicago, IL 60637, USA}
\author{\textit{{In memory of Peter George Oliver Freund}}}
\pacs{PACS number}

\begin{abstract}
The relation between motion in $-1/r$ and $r^{2}$ potentials, known since
Newton, can be demonstrated by the substitution $r\rightarrow r^{2}$ in the
classical/quantum radial equations of the Kepler/Hydrogen problems versus the
harmonic oscillator. This suggests a duality-type relationship between these
systems. However, when both radial and angular components of these systems are
included the possibility of a true duality seems to be remote. Indeed,
investigations that explored and generalized Newton's radial relation,
including algebraic approaches based on noncompact groups such as SO(4,2),
have never exhibited a full duality consistent with Newton's. On the other
hand, 2T-physics predicts a host of dualities between pairs of a huge set of
systems that includes Newton's two systems. These dualities take the form of
rather complicated canonical transformations that relate the full phase spaces
of these respective systems in all directions. In this paper we focus on
Newton's case by imposing his radial relation to find an appropriate basis for
2T-physics dualities, and then construct the full duality. Using the
techniques of 2T-physics, we discuss the hidden symmetry of the actions
(beyond the symmetry of Hamiltonians) for the Hydrogen atom in $D$-dimensions
and the harmonic oscillator in $\bar{D}$ dimensions. The symmetries lead us to
find the one-to-one relation between the quantum states, including angular
degrees of freedom, for specific values of $\left(  D,\bar{D}\right)  $, and
construct the explicit quantum canonical transformation in those special
cases. We find that the canonical transformation has itself a hidden gauge
symmetry that is crucial for the respective phase spaces to be dual even when
$D\neq\bar{D}$. In this way we display the surprising beautiful symmetry of
the full duality that generalizes Newton's radial duality.

\end{abstract}
\maketitle
\tableofcontents

%\volumeyear{year}
%\volumenumber{number}
%\issuenumber{number}
%\eid{identifier}
%\received[Received text]{date}
%\revised[Revised text]{date}
%\accepted[Accepted text]{date}
%\published[Published text]{date}
%\startpage{101}
%\endpage{102}
%%EFI 20-1, USC/20-1, arXiv:2001.nnnnn
\newpage

\section{Introduction and brief statement of results \label{eqn:intro}}

A relation between power-law potentials in the radial Schr\"{o}dinger equation
was encountered when considering the interaction between heavy quarks and
antiquarks (\textquotedblleft quarkonium\textquotedblright)
\cite{Quigg:1979vr}. For every potential $V(r)\sim r^{\alpha}$ in the radial
Schr\"{o}dinger equation there exists a related radial Schr\"{o}dinger
equation with a \textquotedblleft partner potential\textquotedblright%
\ $\bar{V}(\bar{r})\sim\bar{r}^{\bar{\alpha}}$ obtained from the first by the
substitution (see Section \ref{sec:dual})
\begin{equation}
r=\bar{r}^{-\bar{\alpha}/\alpha}~,~~\left(  \alpha+2\right)  \left(
\bar{\alpha}+2\right)  =4. \label{eqn:sub}%
\end{equation}
The Kepler or Hydrogen atom (Hatom) problem, with $\alpha=-1$, is thus related
to the harmonic oscillator problem (HOsc), with $\bar{\alpha}=2$. See, e.g.,
Refs.\ \cite{Bohlin:1911,Jauch:1940,Louck:1960,Bergmann:1965,Talman:1968,Moshinsky:1972wb,Rockmore:1975}%
. Moreover, every potential with $-\infty<\alpha<-2$ or $-2<\alpha<\infty$ has
a partner potential with $-\infty<\bar{\alpha}<-2$ or $-2<\bar{\alpha}<\infty
$, respectively, as noted for classical \cite{Faure,Vasilev,Arnold,Grant} and
quantum \cite{Feldman:1978si,Gazeau:1980,Johnson:1980} systems. However
\cite{Chandra}, this relation can be traced as far back as Newton
\cite{Newton} and Hooke \cite{Hooke}. Not only did Newton transform the radial
equation for the Hatom to the HOsc in order to solve it, but he noted that
pairs of potentials related by Eq.\ (\ref{eqn:sub}) gave congruent orbits with
small deviations from circular shape.

The close relationship between the Hatom-HOsc problems in $D=3$ space
dimensions described above is limited to the radial equation. Whether this
relationship could be elevated to include the angular degrees of freedom in
addition to the radial degrees of freedom remained as an unsolved problem. A
pessimism on this issue developed because when the complete set of states of
the Hatom, including the orbital angular momentum quantum numbers, are
compared to the corresponding complete set of states of the HOsc, one finds
that they are different, so the radial duality is not a true full duality
between the complete systems.

We mention parallel developments related to the Hatom and HOsc that use
spectrum-generating algebrae involving noncompact groups such as SO(2,1),
SO(4,1), and SO(4,2) \cite{Goshen:1959}-\cite{Barut:1986}. Unitary
representations of these groups contain an infinity of energy levels of the
Hamiltonian, related to each other by group transformations within the same
representation of the noncompact group. By suitable identification of
generators, one can pick out stepping operators relating eigenstates with
different energies. This is an indication that these systems may have some
hidden symmetry structure that goes beyond the well known symmetries of the
respective Hamiltonians in three spatial dimensions (SO(4) for Hatom, SU(3)
for HOsc). However, beyond being dynamical groups, the possible existence of
non-compact \textit{symmetries} remained undetermined within those
developments. Furthermore, these efforts did not establish a duality-type
relationship between the Hatom and HOsc on the basis of a \textit{common}
non-compact group and its \textit{common} representations that apply
simultaneously to \textit{both} of these systems.

A full duality between the Hatom-HOsc (and many other systems as well) in
every spatial dimension $D$ and $1$ time dimension was discovered as a simple
prediction of Two-Time Physics (2T-physics) in 1998
\cite{Bars:1998ph,Bars:1998pc,BarsReview1998}. A summary of the concepts of
2T-physics appears in Appendix A. As briefly explained in the paragraphs
containing Eqs.\ (\ref{shadows}-\ref{dualitiesF}) in the Appendix, the general
duality transformation predicted by 2T-physics between any two 1T-physics
\textit{shadows} (explained in the Appendix), that include the Hatom and HOsc
shadows, is a non-linear canonical transformation between their phase spaces
involving time and Hamiltonian, $\left(  \mathbf{r,p},t,H\right)  $
$\leftrightarrow$ $\left(  \mathbf{\bar{r},\bar{p},}\bar{t},\bar{H}\right)  $.
This is just a gauge transformation, of the underlying Sp$\left(  2,R\right)
$ local phase space gauge symmetry, that takes one fixed gauge to another
fixed gauge. We emphasize the \textit{change of the Hamiltonian and the
simultaneous change of the concept of time} as part of the canonical
transformation in which $t$ is canonically conjugate to $H$. Moreover,
2T-physics predicts that these systems (and many other shadows) have a
\textit{common} hidden symmetry SO$\left(  D+1,2\right)  $ \textit{in their
actions, beyond the symmetry of Hamiltonians}, and despite having different 1T
Hamiltonians and different 1T actions, the spectra of the respective
Hamiltonians fit into the \textit{same unitary representations} of the hidden
SO$\left(  D+1,2\right)  ,$ with the same fixed Casimir eigenvalues $C_{n}$
given in Eq.\ (\ref{CasimirsSOd2}) in the Appendix. Therefore, there is indeed
a true duality between the Hatom, HOsc and many others in every dimension $D$,
as explained in the Appendix.

In a general number of spatial dimensions, $D,$ the canonical transformation
derived from 2T-physics, including time and Hamiltonian, is rather complicated
and we can provide it at this time only at the classical level for most
systems \cite{Araya:2013bca} because of the complexities of quantum ordering
for non-linear functions of phase space. The predicted canonical
transformation that follows from Eq.\ (\ref{dualitiesF}) in the Appendix
includes angular directions beyond Eq.\ (\ref{eqn:sub}), but for general $D$
it yields a radial direction different than Eq.\ (\ref{eqn:sub}). However, in
the special case of $D=2$, as well as a few other special cases discussed in
Section \ref{dualitySec}, the canonical transformation Hatom$\leftrightarrow
$HOsc can be brought to a special phase space basis in which time does not
transform, $t=\bar{t}.$ In those cases the radial direction is identical to
Eq.\ (\ref{eqn:sub}) up to an overall constant, and it includes angular
directions beyond Eq.\ (\ref{eqn:sub}), thus becoming a full duality rather
than only a partial radial duality. Moreover, for $D=2$ there are some
beautiful SO$\left(  3,2\right)  =$ Sp$\left(  4,R\right)  $ group-theoretical
properties of the Hatom$_{2}\leftrightarrow$ HOsc$_{2}$ spectra\footnote{From
here on, subscripts in Hatom$_{D}$ or HOsc$_{\bar{D}}$ imply the corresponding
system in the indicated number of dimensions, $D$ or $\bar{D}$.} that clarify
the duality at the quantum level. These nice properties are consistent with
the expected hidden SO$\left(  3,2\right)  $ symmetry of the action predicted
by 2T-physics, as will be displayed in Section \ref{sec:Sp4R}.

Inspired by the form of the $D=2$ full canonical transformation at the quantum
level in Section \ref{sec:Sp4R}, we are able to generalize it in Section
\ref{specialDbarD} to a full canonical transformation that embeds the phase
space $\left(  \mathbf{r,p}\right)  _{D}$ of the Hatom$_{D}$ into the phase
space $\left(  \mathbf{\bar{r},\bar{p}}\right)  _{\bar{D}}$ of HOsc$_{\bar{D}%
}$ for some special values of $3\leq D<\bar{D},$ with $D\neq\bar{D},$ such
that we obtain a Hatom$_{D}\leftrightarrow$ HOsc$_{\bar{D}}$ full duality
(i.e., including angles, beyond (\ref{eqn:sub})) that is consistent with
2T-physics and the expected hidden symmetry SO$\left(  D+1,2\right)  $ of the
Hatom$_{D}$ action. In this paper we display the cases for the pairs $\left(
D,\bar{D}\right)  =\left(  2,2\right)  $ and $\left(  3,4\right)  $ and
comment on a few larger values of the $\left(  D,\bar{D}\right)  $ pairs.

The rest of this paper is organized as follows. In Section \ref{sec:dual} we
display generalized radial duality in all dimensions $D$ through radial
substitution of the form (\ref{eqn:sub}) by relating the radial equations for
two different potentials, $V\left(  r\right)  =\lambda r^{\alpha}$ and
$\bar{V}\left(  \bar{r}\right)  =\bar{\lambda}\bar{r}^{\bar{\alpha}}$,
including the cases of the Hatom$_{D}$ and HOsc$_{\bar{D}}.$ In Section
\ref{sec:noncompact} we introduce details of separate non-compact groups:
SO$\left(  D+1,2\right)  $ for the Hatom$_{D}$'s hidden symmetry of its action
\cite{Bars:1998pc},\footnote{When the action has a larger symmetry than the
Hamiltonian it is imperative that both the spectrum of the Hamiltonian as well
as the dynamics due to interactions are controlled by the symmetry of the
action. As an example, consider the Lorentz symmetry in special relativity,
which is a symmetry of the action, but not a symmetry of the Hamiltonian.
Recall that the Hamiltonian is the time component of the total momentum that
is a Lorentz vector, not a Lorentz scalar. A familiar setting that fits the
bill is relativistic field theory. The same is true also in much simpler
particle systems, such as the Lorentz-invariant worldline formalism with a
gauge symmetry under reparametrizations of proper time $\tau$. After gauge
fixing, such as $x^{0}\left(  \tau\right)  =\tau,$ the canonical conjugate
$p^{0}$ becomes the Hamiltonian that controls the evolution of the remaining
spatial degrees of freedom. The action still has Lorentz symmetry as a hidden
non-linear symmetry, but the Hamiltonian $p^{0}$ is clearly not invariant
under the boosts. The hidden symmetry SO$\left(  D+1,2\right)  $ of the
Hatom's action, as well as of all the dual shadow's actions, is easily
understood in the worldline formalism as a generalization of the statements
above. Then at the quantum level the spectrum of every shadow ends up in the
same unitary representation of the non-compact group, thus obeying a full
duality. See the Appendix to better understand this point.
\label{hiddenSymmetry}} and Sp$\left(  2\bar{D},R\right)  $ for HOsc$_{\bar
{D}}$'s dynamical symmetry. We discuss the classification of the respective
spectra under these non-compact groups. In Section \ref{sec:canonical} we
derive our main result, namely, the duality in terms of canonical
transformations. This is done by equating $L^{MN}\left(  \mathbf{r,p}\right)
=L^{MN}\left(  \mathbf{\bar{r},\bar{p}}\right)  ,$ where $L^{MN}\left(
\mathbf{r,p}\right)  $ are the SO($D+1,2$) generators expressed in terms of
the phase space for the Hatom$_{D},$ while $L^{MN}\left(  \mathbf{\bar{r}%
,\bar{p}}\right)  $ are the subgroup generators for SO($D+1,2$) $\subset$
Sp$\left(  2\bar{D},R\right)  $ expressed in terms of the phase space for the
HOsc$_{\bar{D}},$ for dimensions $D\leq\bar{D}.$ The logic behind this method
was introduced in 2T-physics as discussed in the Appendix. The method is
explicitly applied for the cases $\left(  D,\bar{D}\right)  =(2,2)$ and
$\left(  3,4\right)  .$ In Section \ref{generalizations} we generalize what
was learned in the previous sections and present further examples $\left(
D,\bar{D}\right)  =\left(  1,4\right)  ,$ $\left(  4,6\right)  $ and $\left(
5,8\right)  .$ The conclusions are in Section \ref{sec:con}, where we
summarize the information on all the cases we successfully constructed the
full duality$.$ For all these cases we established conclusively a one-to-one
correspondence between a subset of quantum states of the HOsc$_{\bar{D}}$ and
all the quantum states of the Hatom$_{D}.$ Based on this experience we
conjecture the full duality satisfies $\bar{D}=2\left(  D-1\right)  $ for all
$D\geq2$, with the same form of canonical transformation and dual quantum
states. However, there is room for the formula for$\bar{D}$ to be more general
as we indicate for $D\geq6$ that we have not analyzed in detail, so this
remains open for further investigation. The Appendix summarizes the concepts
of 2T-physics on which we have based our methods and shows the deeper
spacetime structure hidden in the systems we have discussed in this paper.

\section{Radial duality through substitution \label{sec:dual}}

The Schr\"{o}dinger equation with a spherically symmetric potential in $D$
spatial dimensions, $\left(  -\frac{1}{2\mu}\mathbf{\nabla}^{2}+V(\left\vert
\mathbf{r}\right\vert )\right)  \psi\left(  \mathbf{r}\right)  =E\psi\left(
\mathbf{r}\right)  ,$ is solved in spherical coordinates in a complete angular
momentum and energy basis as follows: \cite{Bars:1998pc,BarsQMbook}%
\footnote{For alternative approaches in $D$ dimensions with equivalent
conclusions, see also \cite{Kostelecky:1985fx,Kostelecky:1995fh,Frye:2012jj}%
.)}
\begin{equation}%
\begin{array}
[c]{l}%
\psi\left(  \mathbf{r}\right)  =r^{-\frac{D-1}{2}}u\left(  r\right)
T_{i_{1}i_{2}\cdots,i_{l}}\left(  \mathbf{\hat{r}}\right)  ,~~\text{with
}\;r\equiv\left\vert \mathbf{r}\right\vert ,\;\mathbf{\hat{r}\equiv
r/}\left\vert \mathbf{r}\right\vert ,\\
T_{i_{1}i_{2}\cdots,i_{l}}\left(  \mathbf{\hat{r}}\right)  =\left[  \left(
\mathbf{\hat{r}}_{i_{1}}\mathbf{\hat{r}}_{i_{2}}\mathbf{\cdots\hat{r}}_{i_{l}%
}-\text{trace}\right)  +\text{permutations}\right]  ,\\
-\frac{\hbar^{2}}{2\mu}u^{\prime\prime}(r)+\left[  V\left(  r\right)
+\frac{l_{D}(l_{D}+1)\hbar^{2}}{2\mu r^{2}}-E\right]  u(r)=0,\\
l_{D}\equiv l+\frac{D-3}{2},\text{ with }l=0,1,2,\cdots.
\end{array}
\label{eqn:se}%
\end{equation}
Here, the integer $l$ parametrizes the eigenvalues of angular momentum in $D$
dimensions, $\frac{1}{2}L^{ij}L_{ij}\rightarrow l\left(  l+D-2\right)  ,$
while the symbol $T_{i_{1}i_{2}\cdots,i_{l}}\left(  \mathbf{\hat{r}}\right)  $
is the angular momentum wavefunction in $D$ dimensions.\footnote{Here are
examples for $l=1,2,3,4,$ taken from \cite{BarsQMbook}:
\begin{equation}%
\begin{array}
[c]{l}%
T_{i}=\mathbf{\hat{r}}_{i},\;T_{ij}=\mathbf{\hat{r}}_{i}\mathbf{\hat{r}}%
_{j}-\frac{1}{D}\delta_{ij}\,\,\\
T_{ijk}=\mathbf{\hat{r}}_{i}\mathbf{\hat{r}}_{j}\mathbf{\hat{r}}_{k}-\frac
{1}{D+2}\left(  \delta_{ij}\mathbf{\hat{r}}_{k}+\delta_{ki}\mathbf{\hat{r}%
}_{j}+\delta_{jk}\mathbf{\hat{r}}_{i}\right) \\
T_{ijkl}=\left[
\begin{array}
[c]{l}%
\mathbf{\hat{r}}_{i}\mathbf{\hat{r}}_{j}\mathbf{\hat{r}}_{k}\mathbf{\hat{r}%
}_{l}-\frac{1}{D+4}\left(
\begin{array}
[c]{c}%
\delta_{ij}\mathbf{\hat{r}}_{k}\mathbf{\hat{r}}_{l}+\delta_{ik}\mathbf{\hat
{r}}_{l}\mathbf{\hat{r}}_{j}+\delta_{il}\mathbf{\hat{r}}_{j}\mathbf{\hat{r}%
}_{k}\\
+\delta_{jk}\mathbf{\hat{r}}_{l}\mathbf{\hat{r}}_{i}+\delta_{jl}%
\mathbf{\hat{r}}_{i}\mathbf{\hat{r}}_{k}+\delta_{kl}\mathbf{\hat{r}}%
_{i}\mathbf{\hat{r}}_{J}%
\end{array}
\right) \\
+\frac{D}{\left(  D+2\right)  \left(  D+4\right)  }\left(  \delta_{ij}%
\delta_{kl}+\delta_{ik}\delta_{lj}+\delta_{il}\delta_{jk}\right)
\end{array}
\right]
\end{array}
~~. \label{tensors}%
\end{equation}
} It is constructed from direct products of the unit vector $\mathbf{\hat
{r}\equiv r/}\left\vert \mathbf{r}\right\vert $ and the SO$\left(  D\right)  $
metric $\delta_{ij},$ as a traceless completely symmetric tensor of rank $l,$
such that its indices $\left(  i_{1}i_{2}\cdots,i_{l}\right)  $ take values in
$D$ dimensions, $i_{k}=1,2,\cdots,D.$ This is an irreducible representation of
SO$\left(  D\right)  ,$ and up to an overall normalization, plays the same
role as the complete set of spherical harmonics in $D=3$ dimensions,
$Y_{lm}\left(  \theta,\phi\right)  .$ The degeneracy of the angular momentum
eigenstate with fixed angular momentum $l$ in $D$ dimensions is the dimension
of this SO$\left(  D\right)  $ irreducible representation,
\begin{equation}
N_{l}\left(  D\right)  =\frac{\left(  l+D-3\right)  !}{\left(  D-2\right)
!~l!}\left(  2l+D-2\right)  . \label{NlD}%
\end{equation}
For $D=3$ this reduces to the familiar, $N_{l}\left(  3\right)  =\left(
2l+1\right)  ,$ while for $D=2$ it reduces to $N_{0}\left(  2\right)  =1$ or
$N_{l\neq0}\left(  2\right)  =2,$ as expected (i.e., angular momentum spin up
or down, $\pm l,$ in two spatial dimensions). Having taken into account the
overall factor $r^{\frac{D-1}{2}}$ in the radial wavefunction, $R\left(
r\right)  =$ $r^{-\frac{D-1}{2}}u\left(  r\right)  ,$ the normalization of the
wavefunction in $D$ dimensions, $\int d^{D}r\left\vert \psi\left(
\mathbf{r}\right)  \right\vert ^{2}=1,$ reduces to an integral on the half
line in one dimension, $\int_{0}^{\infty}dr\left\vert u\left(  r\right)
\right\vert ^{2}=1.$ Note the effective potential in (\ref{eqn:se}),
$V_{eff}=V\left(  r\right)  +\frac{\hbar^{2}}{2\mu}\frac{l_{D}(l_{D}+1)}%
{r^{2}},$ that includes the angular momentum barrier in $D$ dimensions
parametrized by $l_{D}\equiv l+\frac{D-3}{2}.$

The relation between the radial Schr\"{o}dinger equation solutions in a
power-law potential $V(r)=\lambda r^{\alpha}$ and a potential $\bar{V}(\bar
{r})=\bar{\lambda}\bar{r}^{\bar{\alpha}}$ may be derived by substituting
$r=\bar{r}^{-\bar{\alpha}/\alpha}$ in the radial Schr\"{o}dinger equation in
$D$ spatial dimensions given above, taking $u(r)=\bar{r}^{\beta}\bar{u}%
(\bar{r}),$ demanding that $(\alpha+2)(\bar{\alpha}+2)=4,$ and with
$\beta=\frac{1}{2}\left(  1+\frac{\bar{\alpha}}{\alpha}\right)  =-\frac
{\bar{\alpha}}{4},$ chosen so that no terms with $\bar{u}^{\prime}(\bar{r})$
occur. Then one recovers a radial Schr\"{o}dinger equation of the same form
for $\bar{u}(\bar{r})$ but with a change of parameters, $\left(  \lambda
,l_{D},E\right)  \rightarrow\left(  \bar{\lambda},\bar{l}_{\bar{D}},\bar
{E}\right)  ,$ that are related to each other as follows:\footnote{The version
of radial duality in Eq.\ (\ref{eqn:sebar}), involving $l_{D}$ (see
(\ref{eqn:se})) rather than $l,$ is a generalization of the same equation in
\cite{Quigg:1979vr} from $D=3$ to general $D$ dimensions. Moreover, if the
substitution in Eq.\ (\ref{eqn:sub}) is slightly generalized to $r=\left(
\bar{r}/b\right)  ^{-\bar{\alpha}/a},$ with $b>0$ instead of $b=1,$ then
Eq.\ (\ref{eqn:sebar}) is further modified to an uglier form, $\bar
{E}=-\lambda\frac{\bar{\alpha}^{2}}{\alpha^{2}}\frac{1}{b^{2}},~~\bar{\lambda
}=-E\frac{\bar{\alpha}^{2}}{\alpha^{2}}\frac{1}{b^{2+\bar{\alpha}}},$ but this
generalization, with $b=\sqrt{2}$ for all $D,$ will be needed to fit the
canonical transformation derived in Eq.\ (\ref{RandAngle}) when $\bar{\alpha
}=2$ and $\alpha=-1$. \label{bfactor}}
\begin{equation}%
\begin{array}
[c]{l}%
-\frac{\hbar^{2}}{2\mu}\bar{u}^{\prime\prime}(\bar{r})+\left[  \bar{\lambda
}\bar{r}^{\bar{\alpha}}+\frac{\hbar^{2}\bar{l}_{\bar{D}}(\bar{l}_{\bar{D}}%
+1)}{2\mu~\bar{r}^{2}}-\bar{E}\right]  \bar{u}(\bar{r})=0~,\\
\bar{E}=-\lambda\frac{\bar{\alpha}^{2}}{\alpha^{2}},~~\bar{\lambda}%
=-E\frac{\bar{\alpha}^{2}}{\alpha^{2}},\;\left\vert \bar{\ell}_{\bar{D}}%
+\frac{1}{2}\right\vert =\left\vert \frac{\bar{\alpha}}{\alpha}\right\vert
\left\vert l_{D}+\frac{1}{2}\right\vert .
\end{array}
\label{eqn:sebar}%
\end{equation}
The last relation in (\ref{eqn:sebar}) is the solution of the quadratic
equation, $\frac{\bar{\alpha}^{2}}{^{\alpha^{2}}}l_{D}(l_{D}+1)+\frac{1}%
{4}\left(  \frac{\bar{\alpha}^{2}}{^{\alpha^{2}}}-1\right)  =\bar{l}_{\bar{D}%
}(\bar{l}_{\bar{D}}+1),$ that imposes the same form of angular momentum
barrier in the effective potential.

This defines the radial duality, under which, coupling constant $\lambda$ and
energy eigenvalue $E$ trade places up to the factor $-(\bar{\alpha}%
/\alpha)^{2};$ furthermore the positive integer angular momenta $l,\bar{l}$
are related by
\begin{equation}
\left\vert \bar{l}+\frac{\bar{D}-2}{2}\right\vert =\left\vert \frac
{\bar{\alpha}}{\alpha}\right\vert \left\vert l+\frac{D-2}{2}\right\vert ,
\label{eqn:lrelation}%
\end{equation}
It is important to emphasize that the dimension $D$ need not be equal to
$\bar{D}$ but both must be positive integers; hence $\bar{l}$ need not be
equal to $l$ while satisfying Eq.\ (\ref{eqn:lrelation}), but both must be
positive integers since they determine the ranks of the angular tensors,
$T_{i_{1}i_{2}\cdots,i_{l}}\left(  \mathbf{\hat{r}}\right)  $ and $\bar
{T}_{i_{1}i_{2}\cdots,i_{\bar{l}}}(\widehat{\mathbf{\bar{r}}})$ respectively.

We have not yet given an explicit transformation rule that relates the angular
variables $\mathbf{\hat{r}}$ and $\widehat{\mathbf{\bar{r}}}.$ We would like
to keep the possibility of $\bar{D}\neq D$ open if the claim for duality is
only for the radial equation rather than for the complete system. Complete
duality requires that the degeneracy of the states should match when
$l,\bar{l}$ are related as in (\ref{eqn:lrelation}), but given that
$N_{l}\left(  D\right)  \neq\bar{N}_{\bar{l}}\left(  \bar{D}\right)  $ when
$\bar{D}\neq D,$ this non-linear requirement is clearly much too strong. So
complete duality guided by the radial equations (\ref{eqn:se}%
-\ref{eqn:lrelation}) seems impossible to satisfy except for special cases of
$D,\bar{D}$ and $\alpha,\bar{\alpha}$. Later, when we provide an explicit
nonlinear relation between the angles $\mathbf{\hat{r}}$ and $\widehat
{\mathbf{\bar{r}}},$ we will show how for certain dimensions, $D\leq\bar{D},$
there is a duality between an \textit{appropriate subset of the degenerate
quantum states} by embedding SO$\left(  D\right)  \subset$ SO$\left(  \bar
{D}\right)  .$

\subsection{Spectra in D,~\={D} dimensions}

Let's consider the case of the Hatom$_{D}$ with $\alpha=-1$ and HOsc$_{\bar
{D}}$ with $\bar{\alpha}=2.$ The spectra of the respective Hamiltonians in $D$
dimensions are well known (using units $c=1,~\hbar=1,~\mu=1$)
\begin{equation}%
\begin{tabular}
[c]{|l|c|c|}\hline
& Hatom$_{D}$ & HOsc$_{\bar{D}}$\\\hline
\multicolumn{1}{|c|}{$V\left(  r\right)  ,\bar{V}\left(  \bar{r}\right)  $} &
$-\frac{Z}{r}$ & $\frac{1}{2}\omega^{2}\bar{r}^{2}$\\\hline
\multicolumn{1}{|c|}{$E_{n},\bar{E}_{\bar{n}}$} & $-\frac{Z^{2}}{2\left(
n+\frac{D-3}{2}\right)  ^{2}},\;n=1,2,3,\cdots~$ & $\;\omega\left(  \bar
{n}+\frac{\bar{D}}{2}\right)  ,\;\left\{
\begin{array}
[c]{c}%
\bar{n}_{\text{even}}=0,2,4,\cdots\\
\bar{n}_{\text{odd}}=1,3,5,\cdots
\end{array}
\right.  $\\\hline
\multicolumn{1}{|c|}{$l,\bar{l}$} & $l=0,1,2,\cdots,\left(  n-1\right)  $ &
$\bar{l}=\left\{
\begin{array}
[c]{c}%
\bar{l}_{\text{even}}=0,2,4,\cdots,\bar{n}_{\text{even}}\\
\bar{l}_{\text{odd}}=1,3,5,\cdots,\bar{n}_{\text{odd}}%
\end{array}
\right.  $\\\hline
\multicolumn{1}{|c|}{radial q.n.} & $n=\left(  1+l+n_{r}\right)
,\;n_{r}=0,1,2,\cdots$ & $\bar{n}=\bar{l}+2n_{r},\;n_{r}=0,1,2,\cdots$\\\hline
\end{tabular}
\ \ \ \ \ \ \label{spectra}%
\end{equation}
The spectra for the Hatom$_{2}$ and HOsc$_{2}$ are graphically displayed in
Eq.\ (\ref{H+HO}), where $\left(  n,l\right)  ,$ respectively $\left(  \bar
{n},\bar{l}\right)  ,$ label rows and columns. For the HOsc$_{2}$ case the
$\bar{n}_{\text{even}},\bar{l}_{\text{even}}$ labels are shown in large bold
numbers, 0,2,4,$\cdots,$ while the $\bar{n}_{\text{odd}},\bar{l}_{\text{odd}}$
labels are shown in smaller numbers, 1,3,5,$\cdots$. The entry at each
$\left(  n,l\right)  $ or $\left(  \bar{n},\bar{l}\right)  $ pigeon holes is
the SO$\left(  2\right)  $ angular momentum degeneracy of the state which is
in accordance with the dimensions of SO(D) representations in (\ref{NlD})$.$
The leftmost column of each table lists the total degeneracy for each energy
level labelled by $n$ or $\bar{n}.$ Note that the total degeneracy at level
$n$ for the Hatom$_{2}$ is $\left(  2\left(  n-1\right)  +1\right)  $ while
for the HOsc$_{2}$ it is $\left(  2\frac{\bar{n}}{2}+1\right)  .$ These match
the dimensions of representations for SO$\left(  3\right)  $ or SU$\left(
2\right)  ,$ namely $\left(  2J+1\right)  ,$ where we identify $J=\left(
n-1\right)  $ for Hatom$_{2}$ and $J=\frac{\bar{n}}{2}$ for HOsc$_{2}$.
\begin{equation}%
\begin{tabular}
[c]{||c||lllllllll||}\hline\hline
$\overset{\text{Hatom}_{2}}{\text{{\scriptsize SO(3)}}{\scriptsize \supset
}\text{{\scriptsize SO(2)}}}$ & $\underset{\downarrow}{~{\large n}%
~~~}^{{\Large l\rightarrow}}$ & ${\small 0}$ & ${\small 1}$ & ${\small 2}$ &
${\small 3}$ & ${\small 4}$ & ${\small 5}$ & ${\small 6}$ & ${\small 7}\cdots
$\\\cline{1-1}\cline{3-10}%
$\vdots$ & \multicolumn{1}{||c}{$\vdots$} & \multicolumn{1}{|c}{$\vdots$} &
\multicolumn{1}{c}{$\vdots$} & \multicolumn{1}{c}{$\vdots$} &
\multicolumn{1}{c}{$\vdots$} & \multicolumn{1}{c}{$\vdots$} &
\multicolumn{1}{c}{$\vdots$} & \multicolumn{1}{c}{$\vdots$} &
\multicolumn{1}{c||}{$\vdots$}\\
$13$ & \multicolumn{1}{||c}{$7$} & \multicolumn{1}{|c}{$1$} &
\multicolumn{1}{c}{$2$} & \multicolumn{1}{c}{$2$} & \multicolumn{1}{c}{$2$} &
\multicolumn{1}{c}{$2$} & \multicolumn{1}{c}{$2$} & \multicolumn{1}{c}{$2$} &
\multicolumn{1}{c||}{}\\
$11$ & \multicolumn{1}{||c}{$6$} & \multicolumn{1}{|c}{$1$} &
\multicolumn{1}{c}{$2$} & \multicolumn{1}{c}{$2$} & \multicolumn{1}{c}{$2$} &
\multicolumn{1}{c}{$2$} & \multicolumn{1}{c}{$2$} & \multicolumn{1}{c}{} &
\multicolumn{1}{c||}{}\\
$9$ & \multicolumn{1}{||c}{$5$} & \multicolumn{1}{|c}{$1$} &
\multicolumn{1}{c}{$2$} & \multicolumn{1}{c}{$2$} & \multicolumn{1}{c}{$2$} &
\multicolumn{1}{c}{$2$} & \multicolumn{1}{c}{} & \multicolumn{1}{c}{} &
\multicolumn{1}{c||}{}\\
$7$ & \multicolumn{1}{||c}{$4$} & \multicolumn{1}{|c}{$1$} &
\multicolumn{1}{c}{$2$} & \multicolumn{1}{c}{$2$} & \multicolumn{1}{c}{$2$} &
\multicolumn{1}{c}{} & \multicolumn{1}{c}{} & \multicolumn{1}{c}{} &
\multicolumn{1}{c||}{}\\
$5$ & \multicolumn{1}{||c}{$3$} & \multicolumn{1}{|c}{$1$} &
\multicolumn{1}{c}{$2$} & \multicolumn{1}{c}{$2$} & \multicolumn{1}{c}{} &
\multicolumn{1}{c}{} & \multicolumn{1}{c}{} & \multicolumn{1}{c}{} &
\multicolumn{1}{c||}{}\\
$3$ & \multicolumn{1}{||c}{$2$} & \multicolumn{1}{|c}{$1$} &
\multicolumn{1}{c}{$2$} & \multicolumn{1}{c}{} & \multicolumn{1}{c}{} &
\multicolumn{1}{c}{} & \multicolumn{1}{c}{} & \multicolumn{1}{c}{} &
\multicolumn{1}{c||}{}\\
$1$ & \multicolumn{1}{||c}{$1$} & \multicolumn{1}{|c}{$1$} &
\multicolumn{1}{c}{} & \multicolumn{1}{c}{} & \multicolumn{1}{c}{} &
\multicolumn{1}{c}{} & \multicolumn{1}{c}{} & \multicolumn{1}{c}{} &
\multicolumn{1}{c||}{}\\\hline\hline
\end{tabular}
\ \ \ \ \ ~~%
\begin{tabular}
[c]{||c||lcccccccc||}\hline\hline
$\overset{\text{HOsc}_{2}}{\text{{\scriptsize SU(2)}}{\scriptsize \supset
}\text{{\scriptsize SO(2)}}}$ & $~\underset{\downarrow}{{\large \bar{n}}%
}^{~~~~{\Large \bar{l}\rightarrow}}$ & \textbf{0} & {\scriptsize 1} &
\textbf{2} & {\scriptsize 3} & \textbf{4} & {\scriptsize 5} & \textbf{6} &
{\scriptsize 7}$\cdots$\\\cline{1-1}\cline{1-1}\cline{3-10}%
$\vdots$ & \multicolumn{1}{||c}{$\vdots$} & \multicolumn{1}{|c}{$\vdots$} &
$\vdots$ & $\vdots$ & $\vdots$ & $\vdots$ & $\vdots$ & $\vdots$ & $\vdots$\\
\textbf{7} & \multicolumn{1}{||c}{\textbf{6}} & \multicolumn{1}{|c}{\textbf{1}%
} &  & \textbf{2} &  & \textbf{2} &  & \textbf{2} & \\
{\scriptsize 6} & \multicolumn{1}{||c}{{\scriptsize 5}} &
\multicolumn{1}{|c}{} & {\scriptsize 2} &  & {\scriptsize 2} &  &
{\scriptsize 2} &  & \\
\textbf{5} & \multicolumn{1}{||c}{\textbf{4}} & \multicolumn{1}{|c}{\textbf{1}%
} &  & \textbf{2} &  & \textbf{2} &  &  & \\
{\scriptsize 4} & \multicolumn{1}{||c}{{\scriptsize 3}} &
\multicolumn{1}{|c}{} & {\scriptsize 2} &  & {\scriptsize 2} &  &  &  & \\
\textbf{3} & \multicolumn{1}{||c}{\textbf{2}} & \multicolumn{1}{|c}{\textbf{1}%
} &  & \textbf{2} &  &  &  &  & \\
{\scriptsize 2} & \multicolumn{1}{||c}{{\scriptsize 1}} &
\multicolumn{1}{|c}{} & {\scriptsize 2} &  &  &  &  &  & \\
\textbf{1} & \multicolumn{1}{||c}{\textbf{0}} & \multicolumn{1}{|c}{\textbf{1}%
} &  &  &  &  &  &  & \\\hline\hline
\end{tabular}
\ \ \ \label{H+HO}%
\end{equation}

As shown in \cite{Bars:1998pc}, for general $D$ or $\bar{D}$ the pigeon holes
would be filled with the numbers $N_{l}\left(  D\right)  $ or $\bar{N}%
_{\bar{l}}\left(  \bar{D}\right)  $ respectively as given in (\ref{NlD}). The
total degeneracies at the leftmost column for each fixed $n$ or $\bar{n}$ are
then%
\begin{equation}%
\begin{array}
[c]{l}%
\text{Hatom}_{D}:\sum_{l=0}^{n-1}N_{l}(D)=\frac{\left(  n+D-3\right)
!}{\left(  D-1\right)  !~\left(  n-1\right)  !}\left(  2n+D-3\right)
=N_{n-1}\left(  D+1\right)  ,\\
\text{HOsc}_{\bar{D}}\;:\sum_{\bar{l}=\text{even or odd}}^{\bar{n}}\bar
{N}_{\bar{l}}(\bar{D})=\frac{\left(  \bar{n}+\bar{D}-1\right)  !}{\bar
{n}!\left(  \bar{D}-1\right)  !}\text{.}%
\end{array}
\label{degeneracies}%
\end{equation}
For $D=\bar{D}=2$ these total degeneracies reproduce the results of the
previous paragraph and tables in (\ref{H+HO}), while for $D=\bar{D}=3$ they
match the well known degeneracies, $n^{2}$ for Hatom$_{3},$ and $\frac{1}%
{2}\left(  \bar{n}+2\right)  \left(  \bar{n}+1\right)  $ for HOsc$_{3}.$ For
the Hatom$_{D},$the total degeneracy at each $n$ matches the dimension of the
SO$\left(  D+1\right)  $ representation for the completely symmetric traceless
tensor, $T_{I_{1}I_{2}\cdots I_{n-1}}$, of rank $\left(  n-1\right)  $ in
$\left(  D+1\right)  $ dimensions (single-row Young tableau, $\left(
n-1\right)  $ boxes, with trace removed). Similarly, for the HOsc$_{\bar{D}},$
the total degeneracy matches the dimension of the completely symmetric
SU$\left(  \bar{D}\right)  $ tensor with $\bar{n}$ indices (single-row Young
tableau, $\bar{n}$ boxes). The underlying reason for these degeneracies is the
well known hidden symmetries of the \textit{Hamiltonians}: SO$\left(
D+1\right)  $ for the Hatom$_{D}$ Hamiltonian and SU$\left(  \bar{D}\right)  $
for the HOsc$_{\bar{D}}$ Hamiltonian, as discussed in Section
\ref{sec:noncompact}.

This result is only a small part of the group-theoretical properties of the
respective spectra for the Hatom$_{D}$ or HOsc$_{\bar{D}}.$ As seen in the
tables above in (\ref{H+HO}), for each state in an SO$\left(  D\right)  $
multiplet labelled by a fixed value of $l$ or $\bar{l},$ there exists an
infinite tower of states of increasing values of $n$ or $\bar{n}$. It was
shown in \cite{Bars:1998pc} that these towers form infinite-dimensional
irreducible representations of the non-compact groups SO$\left(  1,2\right)  $
or Sp$\left(  2,R\right)  $ corresponding to the positive discrete series
\cite{Bargmann:1947,Biedenharn:1965} labelled by
%TCIMACRO{\TEXTsymbol{\vert}}%
%BeginExpansion
$\vert$%
%EndExpansion
$j,m\rangle$ (similar to SU$\left(  2\right)  $ quantum numbers), with
\begin{equation}
m\left(  j\right)  =j+1+n_{r},\;\;n_{r}=0,1,2,3,\cdots\label{mj}%
\end{equation}
where\ the integer $n_{r}$\ coincides with the usual radial quantum number in
Eq.\ (\ref{spectra}) that emerges when solving the radial equations in
(\ref{eqn:se}) or (\ref{eqn:sebar}). For the towers associated with the
SO$\left(  D\right)  $ or SO$\left(  \bar{D}\right)  $ multiplets $l$ or
$\bar{l},$ the value of $j$ depends on $l$ or $\bar{l}$ as follows
\cite{Bars:1998pc} (see (\ref{jlH}) and (\ref{jl}) for the derivations of
$j\left(  l\right)  $ and $\bar{j}\left(  \bar{l}\right)  $ respectively):%
\begin{equation}
\text{Hatom}_{D}\text{:}\left\{
\begin{array}
[c]{l}%
j=0\text{ if }D=1\\
j\left(  l\right)  =l+\frac{D-3}{2},\text{ if }D\geq2
\end{array}
\right.  ,\;\;\text{HOsc}_{\bar{D}}:\bar{j}\left(  \bar{l}\right)  =\frac
{1}{2}\left(  \bar{l}+\frac{\bar{D}-4}{2}\right)  . \label{j(l)}%
\end{equation}
Hence the overall spectra are direct sums of irreducible representations of
direct product groups as follows:%
\begin{equation}%
\begin{array}
[c]{l}%
\text{Hatom}_{D}\text{:~}\sum_{l=0}^{\infty}\oplus\text{%
%TCIMACRO{\TEXTsymbol{\vert}}%
%BeginExpansion
$\vert$%
%EndExpansion
SO}\left(  1,2\right)  _{j\left(  l\right)  },\text{SO}\left(  D\right)
_{l}\rangle,\;\text{with SO}\left(  1,2\right)  \otimes\text{SO}\left(
D\right)  \subset\text{ SO}\left(  D+1,2\right)  ,\\
\text{HOsc}_{\bar{D}}\text{:~}\left\{
\begin{array}
[c]{l}%
\sum_{l_{\text{even}}}^{\infty}\oplus\text{%
%TCIMACRO{\TEXTsymbol{\vert}}%
%BeginExpansion
$\vert$%
%EndExpansion
Sp}\left(  2,R\right)  _{\bar{j}\left(  \bar{l}_{\text{even}}\right)
},\text{SO}\left(  \bar{D}\right)  _{\bar{l}_{\text{even}}}\rangle\\
\sum_{l_{\text{odd}}}^{\infty}\oplus\text{%
%TCIMACRO{\TEXTsymbol{\vert}}%
%BeginExpansion
$\vert$%
%EndExpansion
Sp}\left(  2,R\right)  _{\bar{j}\left(  \bar{l}_{\text{odd}}\right)
},\text{SO}\left(  \bar{D}\right)  _{\bar{l}_{\text{odd}}}\rangle
\end{array}
\right.  ,\;\text{with Sp}\left(  2,R\right)  \otimes\text{SO}\left(  \bar
{D}\right)  \subset\text{ Sp}\left(  2\bar{D},R\right)  .
\end{array}
\label{SpectraGroups}%
\end{equation}
The direct product groups that classify the spectra are themselves subgroups
of larger non-compact groups, SO$\left(  D+1,2\right)  $ and Sp$\left(
2\bar{D},R\right)  $ respectively as indicated in (\ref{SpectraGroups}). These
non-compact groups will be discussed in Section \ref{sec:noncompact} in more
detail. In fact, the full spectrum of the Hatom$_{D}$ corresponds to a single
irreducible representation of SO$\left(  D+1,2\right)  ,$ while the even/odd
states of the HOsc$_{\bar{D}}$ correspond to two distinct irreducible
representations of Sp$\left(  2\bar{D},R\right)  $ as will be explained in
Section \ref{sec:noncompact}$.$ In both cases these are called singleton
representations that have the following Casimir eigenvalues (see
Eqs.\ (\ref{CasimirsSp2D},\ref{CasimirsSOd2})):
\begin{equation}%
\begin{array}
[c]{l}%
\text{SO}\left(  D+1,2\right)  :\;C_{2}=-\left(  \frac{\left(  D+1\right)
^{2}}{4}-1\right)  ,\;C_{3},C_{4}=\cdots\\
\text{Sp}\left(  2\bar{D}\right)  :\;\bar{C}_{2}^{\text{even}}=\bar{C}%
_{2}^{\text{odd}}=-\frac{\bar{D}}{2}\left(  \frac{\bar{D}}{2}+\frac{1}%
{4}\right)  ,\;\bar{C}_{3},\bar{C}_{4}=\cdots
\end{array}
\label{casmirsSOSp}%
\end{equation}
These facts about the spectra of Hatom$_{D}$ and HOsc$_{\bar{D}}$ will be
relevant for the full duality we are seeking in this paper, namely a full
duality that would be consistent with Newton's radial duality $r\sim\bar
{r}^{2}$ discussed in the Introduction and details produced in Sections
\ref{hints}, \ref{sec:Sp4R}, \ref{specialDbarD}, \ref{sec:con}.

\subsection{Hints of full duality \label{hints}}

Armed with the full spectrum, including angles and angular momentum, we now
return to the radial duality displayed in Eqs.\ (\ref{eqn:se}%
-\ref{eqn:lrelation}). Specializing to $D=\bar{D}=2,$ the angular momentum
relations (\ref{eqn:lrelation}) become%
\begin{equation}
D=\bar{D}=2:\;\bar{l}=2l\;. \label{lOsc}%
\end{equation}
It is seen graphically in Eq.\ (\ref{H+HO}) that only the $\bar{l}=$ even
(equivalently the $\bar{n}$ = even) HOsc$_{2}$ states are in one to one
correspondence with all the states of the Hatom$_{2},$ including matching
representations of the hidden symmetries SO$\left(  3\right)  =$ SU$\left(
2\right)  $ level by level, at each $\left(  n-1\right)  =\frac{\bar
{n}_{\text{even}}}{2}=J,$ with degeneracy $\left(  2J+1\right)  $. In
particular, the infinite vertical towers for SO$\left(  1,2\right)  =$
Sp$\left(  2,R\right)  $ also match at each $l_{\text{even}}=2l.$

Moreover, the $D=2$ noncompact group SO$\left(  3,2\right)  $ is the same as
the $\bar{D}=2$ non-compact group Sp$\left(  4,R\right)  ,$ and according to
Eq.\ (\ref{casmirsSOSp}) the quadratic Casimir is the same, $C_{2}=-\frac
{5}{4},$ and so is the only other cubic Casimir, $C_{3}=-\frac{5}{8},$
according to (\ref{CasimirsSOd2},\ref{CasimirsSp2D}). Hence the full
Hatom$_{2}$ and even-HOsc$_{2}$ spectra are in the same irreducible
representation of SO$\left(  3,2\right)  =$ Sp$\left(  4,R\right)  ,$ just as
expected on the basis of 2T-physics dualities as explained in the paragraphs
containing Eqs.\ (\ref{L2fixed}-\ref{dualitiesF}) in the Appendix. These are
very encouraging indications of a full duality between the Hatom$_{2}$ and the
even half of the HOsc$_{2}.$

This result raises the question: what is the dual of the odd half of the
HOsc$_{2}$? Amazingly, the answer is provided in 2T-physics with spin, that
yields a generalization of the Hatom$_{D}$ in a particular gauge (see Section
V in \cite{Bars2Tspinning}). In the current paper we will call this case the
dyonic-Hatom$_{D}$. The hidden symmetry in this case is again SO$\left(
D+1,2\right)  ,$ but the representation is different than the zero spin case,
and has a quadratic Casimir given by (see Eq.\ (80) in \cite{Bars2Tspinning}
and substitute $d=D+1$) as outlined in the Appendix around
Eq.\ (\ref{casimirs-S}):
\begin{equation}
\text{Dyonic-Hatom}_{D},~\text{SO}\left(  D+1,2\right)  ,\;C_{2}^{\text{spin
}s=1/2}=-\frac{1}{8}D\left(  D+3\right)  .
\end{equation}
Although the physical interpretation of this model was not fully grasped in
\cite{Bars2Tspinning}, it was later understood that it corresponds to a
hypothetical Hatom whose nucleus is a dyon that has both electric and magnetic
charges instead of the usual proton. For $D=3$ this matches the model
discussed in \cite{Zwanziger}. The spectrum of the dyonic-Hatom$_{2}$
resembles that of the Hatom$_{2}$ but instead of $l$ there appears $\left(
l+\frac{1}{2}\right)  ,$ where the additional 1/2 is generated by the dyon.
Furthermore, we find that the dyonic-Hatom$_{2}$ has Casimir $C_{2}=-\frac
{5}{4},$ which is the same as the $C_{2}$ for Hatom$_{2}$ or HOsc$_{2},$ and
the spectrum matches the spectrum of odd-HOsc$_{2}$ since now the
generalization of Eq.\ (\ref{lOsc}) is, $\bar{l}=2\left(  l+\frac{1}%
{2}\right)  ,$ where $\bar{l}$ is odd.

We have found very strong hints that the duals for the even and odd parts of
the HOsc$_{2}$ are given by Hatom$_{2,s}$ with $s=0,\frac{1}{2}$ respectively.
In the next Section we will display a quantum canonical transformation that
establishes the full duality transformation, Hatom$_{2}\leftrightarrow
$ even-HOsc$_{2},$ and show that it is compatible with the simple radial
substitution, $r\sim\bar{r}^{2}$ in Eq.\ (\ref{eqn:sub}) with $\alpha=-1,$
$\bar{\alpha}=2$, that started the current investigation.

After displaying the canonical transformation, we will generalize the method
to a few other special values of the pair $D<\bar{D}$ that are compatible with
the simple radial substitution, $r \sim \bar{r}^{2}$. There are also other
dualities as non-linear canonical transformations that connect Hatom$_{D},$
HOsc$_{D}$ (i.e., $\bar{D}=D$) and many other systems in $D$ spatial and one
time dimensions as predicted by 2T-physics \cite{Araya:2013bca}, but those
predicted more general cases, that apply in every dimension $D,$ are at first
sight not compatible with Newton's simple radial substitution, $r\sim\bar
{r}^{2}$. However, we suspect a further canonical transformation partly
related to the one discussed in Section \ref{dualitySec} must make the
general 2T dualities and Newton's case compatible as well.

\section{Non-compact symmetries of the Hatom$_{D}$ and HOsc$_{\bar{D}}$
\label{sec:noncompact}}

In this section we discuss the SO$\left(  D+1,2\right)  $ and Sp$\left(
2\bar{D},R\right)  $ generators constructed from the quantum phase space
degrees of freedom of the Hatom$_{D}$ and HOsc$_{\bar{D}}.$ Following the
method in \cite{Araya:2013bca} (as stated in the paragraphs that contain
Eqs.\ (\ref{L3Fixed}-\ref{dualitiesF}) in the Appendix) we compare these
gauge-invariant generators for different shadows (see Appendix) to one another
when $D=\bar{D}=2$, and from this we obtain the sought-after canonical
transformation that relates the two phase spaces $\left(  \mathbf{r,p}\right)
\leftrightarrow~\left(  \mathbf{\bar{r},\bar{p}}\right)  $, as shown in
Section \ref{sec:Sp4R}.

\subsection{SO$\left(  D+1,2\right)  $ and the Hatom$_{D}$ \label{so(D+1,2)}}

The subgroup SO$\left(  D+1\right)  $ is the well known hidden symmetry for
the Hatom$_{D},$ which is best explained in the context of 2T-physics because
of its extra space dimension (in addition to the extra time dimension)
\cite{Bars:1998ph}. The hidden symmetry SO$\left(  4\right)  $ in Hatom$_{3}$,
associated with a conserved Runge-Lenz vector, was recognized already in the
19$^{th}$ century in the study of the Kepler problem in celestial mechanics,
and used by Pauli to understand the \textquotedblleft accidental
degeneracy\textquotedblright\ in the spectrum of the Hatom$_{3}$
\cite{Pauli:1926qp,Fock:1935vv,Bargmann:1936}. Later, in the context of
spectrum-generating algebraic techniques, the SO$\left(  4\right)  =$
SU$\left(  2\right)  \otimes$ SU$\left(  2\right)  $ hidden symmetry was
embedded in the non-compact group SO$\left(  4,2\right)  $
\cite{Mukunda:1965wt,Bander:1965rz,Bander:1965im,Bacry:1966,Musto:1966bw,Pratt:1966,Barut:1967zza,Barut:1972,Barut:1986,Wybourne:1974,Barut:1971pf,Kibler:2004}%
.

Eventually, the underlying reason for the existence of the spectrum-generating
algebra was finally understood with the advent of 2T-physics. Namely,
SO$\left(  D+1,2\right)  $ is far more than an algebraic tool; it is actually
a hidden symmetry of the action (not Hamiltonian) for the Hatom$_{D}$ for any
dimension $D$ (see Eq.\ (20) in \cite{Bars:1998pc}) and for this reason the
spectrum of Hatom$_{D}$ must be described in terms of irreducible
representations of SO$\left(  D+1,2\right)  $ (see remarks in footnote
\ref{hiddenSymmetry}). Part of this symmetry, namely SO$\left(  D+1\right)
\times$ U$\left(  1\right)  $, is also a symmetry of the Hatom$_{D}$
Hamiltonian, where SO$\left(  D+1\right)  $ rotates all spatial dimensions in
2T-physics on an equal footing, and U$\left(  1\right)  =$ SO$\left(
2\right)  $ rotates the two temporal dimensions. In fact, 2T-physics shows
that the Hamiltonian is proportional to $\left(  -1\right)  /\left(
L^{00^{\prime}}\right)  ^{2}$ where $L^{00^{\prime}}$ is the SO$\left(
2\right)  =$U$\left(  1\right)  $ generator. In this way the Hatom$_{D}$
system is a very transparent window to all spatial and temporal dimensions of
2T-physics, and its spectrum displays the action of the remaining $L^{0I}$ or
$L^{0^{\prime}I}$ generators that mix spatial and temporal dimensions with
each other.

The generators of SO$\left(  D+1,2\right)  $ are computed in 2T-physics as
$L^{MN}=\left(  X^{M}P^{N}-X^{N}P^{M}\right)  .$ The phase space $\left(
X^{M},P^{M}\right)  $ is not gauge-invariant under the Sp$\left(  2,R\right)
$ gauge transformations of the phase space (for each $M$ they transform as a
doublet of the \textit{local} gauge group Sp$\left(  2,R\right)  $), but the
combination $L^{MN}$ for the \textit{global }symmetry SO$\left(  D+1,2\right)
$ are gauge-invariant, so the $L^{MN}$ can be evaluated in any gauge (see
Appendix). It is customary in 2T-physics to label the $\left(  D+1,2\right)  $
indices $M$ by $M=\left(  0^{\prime},0,1^{\prime},1,2,\cdots D\right)  ,$ or
by $M=\left(  +^{\prime},-^{\prime},0,1,2,\cdots D\right)  ,$ where
\begin{equation}
X^{+^{\prime}}=\frac{X^{0^{\prime}}+X^{1^{\prime}}}{2},\;X^{-^{\prime}%
}=X^{0^{\prime}}-X^{1^{\prime}},\;\Leftrightarrow~X^{0^{\prime}}=X^{+^{\prime
}}+\frac{X^{-^{\prime}}}{2},\;X^{1^{\prime}}=X^{+^{\prime}}-\frac
{X^{-^{\prime}}}{2}. \label{lightcone}%
\end{equation}
and similarly for $P^{M}.$ Then SO$\left(  D+1,2\right)  $-invariant dot
products in flat spacetime take the form
\begin{equation}
X\cdot P=\left\{
\begin{array}
[c]{l}%
\left[  -X^{0^{\prime}}P^{0^{\prime}}-X^{0}P^{0}+X^{1^{\prime}}P^{1^{\prime}%
}+\mathbf{X\cdot P}\right] \\
=\left[  -X^{+^{\prime}}P^{-^{\prime}}-X^{-^{\prime}}P^{+^{\prime}}-X^{0}%
P^{0}+\mathbf{X\cdot P}\right]
\end{array}
\right.  , \label{dotLightcone}%
\end{equation}
and similarly for $X\cdot X$ and $P\cdot P,$ where the dot product in bold
letters is the Euclidean dot product in D spatial dimensions.

In the Hatom shadow (see Eqs.\ (12-21) in \cite{Bars:1998pc}), the gauge-fixed
version of the D-dimensional Euclidean phase space $\left(  \mathbf{X,P}%
\right)  $ is relabeled as $\left(  \widetilde{\mathbf{r}}\mathbf{,}%
\widetilde{\mathbf{p}}\right)  ,$ while the remaining 6 functions of the
worldline proper time $\tau$ introduced in the Appendix$,$ $\left(
X^{0,0^{\prime},1^{\prime}},P^{0,0^{\prime},1^{\prime}}\right)  ,$ are
gauge-fixed as functions of $\left(  \widetilde{\mathbf{r}}\mathbf{,}%
\widetilde{\mathbf{p}}\mathbf{,}\widetilde{t}\right)  .$ To get there, three
gauge parameters of Sp$\left(  2,R\right)  $ are used to gauge-fix three
functions$,$ and the three constraints $X\cdot X=P\cdot P=X\cdot P=0$ are
explicitly solved to fix 3 more functions. Therefore $\left(  X^{0,0^{\prime
},1^{\prime}},P^{0,0^{\prime},1^{\prime}}\right)  $ are all dependent on
$\left(  \widetilde{\mathbf{r}}\left(  \widetilde{t}\right)  \mathbf{,}%
\widetilde{\mathbf{p}}\left(  \widetilde{t}\right)  \right)  $ and
$\widetilde{t}\left(  \tau\right)  =\tau$. The gauge-invariant 2T-physics
action is then evaluated in this gauge, and it is shown in \cite{Bars:1998pc}
that it reduces to the 1T-physics action for the Hatom$_{D}$, $\int
d\widetilde{t}\left(  \partial_{\widetilde{t}}\widetilde{\mathbf{r}%
}\mathbf{\cdot}\widetilde{\mathbf{p}}\mathbf{-}\left(  \frac{1}{2}%
\widetilde{\mathbf{p}}^{2}-\frac{Z}{\left\vert \widetilde{\mathbf{r}%
}\right\vert }\right)  \right)  .$ Note that the original 2T action has no
parameters, so the mass and coupling constants in the Hatom Hamiltonian (and
similarly in actions for other shadows) emerge from the gauge-fixing of the
phase space $\left(  X^{0,0^{\prime},1^{\prime}},P^{0,0^{\prime},1^{\prime}%
}\right)  $ in a way similar to the emergence of parameters from
\textquotedblleft moduli\textquotedblright\ in M-theory. The gauge-invariant
action that one starts with in Eq.\ (\ref{Lagr}) in the Appendix is explicitly
invariant under the global symmetry SO$\left(  D+1,2\right)  $ that acts
linearly on the original phase space $\left(  X^{M},P^{M}\right)  .$ Since the
global symmetry commutes with the local symmetry, the gauge-fixed action,
namely the Hatom$_{D}$ action in this paragraph, must have the same
non-linearly realized hidden SO$\left(  D+1,2\right)  $ symmetry. The
generators of this symmetry have to be the gauge-fixed form of the
gauge-invariant $L^{MN}$ which is now a non-linear function of the gauge-fixed
$\left(  X^{M}\left(  \widetilde{\mathbf{r}}\mathbf{,}\widetilde{\mathbf{p}%
},\widetilde{t}\right)  ,P^{M}\left(  \widetilde{\mathbf{r}}\mathbf{,}%
\widetilde{\mathbf{p}}\mathbf{,}\widetilde{t}\right)  \right)  $ in terms of
the $D$-dimensional Euclidean phase space, $L^{MN}\left(  \mathbf{\tilde{r}%
},\mathbf{\tilde{p}}\right)  .$ Indeed, it was shown \cite{Bars:1998pc} that
the Hatom$_{D}$ action is invariant under the non-linear transformations
obtained by applying Poisson brackets between $\frac{1}{2}\omega_{MN}L^{MN}$
and the phase space $\left(  \mathbf{\widetilde{r}},\mathbf{\widetilde{p}}
\right)  ,$ namely
\begin{equation}
\delta_{\omega}\widetilde{\mathbf{r}}\mathbf{=}\frac{1}{2}\omega_{MN}%
\frac{\partial L^{MN}\left(  \widetilde{\mathbf{r}}\mathbf{,}\widetilde
{\mathbf{p}}\mathbf{,}\widetilde{t}\right)  }{\partial\widetilde{\mathbf{p}}%
},\;\delta_{\omega}\mathbf{\widetilde{p}=-}\frac{1}{2}\omega_{MN}%
\frac{\partial L^{MN}\left(  \widetilde{\mathbf{r}}\mathbf{,}\widetilde
{\mathbf{p}}\mathbf{,}\widetilde{t}\right)  }{\partial\widetilde{\mathbf{r}}}.
\end{equation}
where the constant $\omega_{MN}$ are the global SO$\left(  D+1,2\right)  $
parameters. One may reverse this approach by starting from the invariance of
the action and use Noether's theorem to build the $L^{MN}\left(
\widetilde{\mathbf{r}}\mathbf{,}\widetilde{\mathbf{p}}\mathbf{,}\widetilde
{t}\right)  .$ Either way, one finds that $L^{00^{\prime}}=$ $\left(
X^{0}P^{0^{\prime}}-X^{0^{\prime}}P^{0}\right)  ,$ evaluated in the Hatom
shadow, yields classically (i.e., ignoring quantum ordering)
\begin{equation}
L^{00^{\prime}}=\frac{Z}{\sqrt{-2H}},\text{ with }H=\left(  \frac{1}%
{2}\widetilde{\mathbf{p}}^{2}-\frac{Z}{\left\vert \widetilde{\mathbf{r}%
}\right\vert }\right)  =-\frac{Z^{2}/2}{\left(  L^{00^{\prime}}\right)  ^{2}}.
\label{HL00}%
\end{equation}
Hence the Hatom$_{D}$ Hamiltonian can be written very simply in terms of the
gauge-invariant generator $L^{00^{\prime}}.$ Now, this is the generator of a
compact SO$\left(  2\right)  $ that rotates the two times into each other, so
its eigenvalues must be parametrized by an integer, just like orbital angular
momentum, but due to quantum ordering issues the integer may be shifted by a
constant that depends on $D.$

Resolving the quantum ordering is too complicated in the Hatom shadow using
$\left(  \widetilde{\mathbf{r}}\mathbf{,}\widetilde{\mathbf{p}}\mathbf{,}%
\widetilde{t}\right)  $. However, since $L^{00^{\prime}}$ is gauge-invariant,
and its commutation rules with all other $L^{MN}$ are also gauge-invariant,
one may choose any convenient gauge to evaluate the $L^{MN},$ resolve all
quantum ordering ambiguities, and then diagonalize $L^{00^{\prime}}$ to find
its gauge-invariant eigenvalues algebraically by using only the commutation
rules of the hidden symmetry SO$\left(  D+1,2\right)  $ in any shadow. This
was done in \cite{Bars:1998pc} by choosing the following gauge (evaluated at
zero time for that gauge\footnote{Normally, in choosing the gauges there is a
non-trivial explicit dependence on time, $t\left(  \tau\right)  =\tau,$ where
$\tau$ is the proper time in the original Lagrangian (\ref{Lagr}) in the
Appendix. Since at this stage we are interested in the equal-time commutation
rules of observables, we have chosen $t\left(  \tau\right)  =\tau=0$ to
simplify as much as possible the gauge-fixed versions of $\left(  X^{M}%
,P^{M}\right)  $ as shown in (\ref{gaugeH}).})
\begin{equation}
\left(
\begin{array}
[c]{l}%
X^{M}\\
P^{M}%
\end{array}
\right)  =\left(  \overset{+^{\prime}}{%
\begin{array}
[c]{l}%
0\\
1
\end{array}
~~~}\overset{-^{\prime}}{%
\begin{array}
[c]{l}%
\mathbf{r\cdot p}\\
\mathbf{p}^{2}/2
\end{array}
~~~}\overset{0}{%
\begin{array}
[c]{l}%
\left\vert \mathbf{r}\right\vert \\
0
\end{array}
}~\overset{i=1,\cdots,D}{%
\begin{array}
[c]{l}%
\mathbf{r}\\
\mathbf{p}%
\end{array}
}\right)  \label{gaugeH}%
\end{equation}
The three numerical entries, $X^{+^{\prime}}=0,$ $P^{+^{\prime}}=1,$
$P^{0}=0,$ are gauge choices, while $\left(  \mathbf{X,P}\right)  =\left(
\mathbf{r},\mathbf{p}\right)  $ is just renaming symbols to indicate that we
are in another gauge, and the remaining three entries $\left(  \mathbf{r\cdot
p,}\left\vert \mathbf{r}\right\vert ,\mathbf{p}^{2}/2\right)  $ are computed
by solving the tree constraints $X^{2}=P^{2}=X\cdot P=0$ by using the
lightcone version (\ref{dotLightcone}) of the SO$\left(  D+1,2\right)  $
invariant dot product.

Now we evaluate the gauge-invariant $L^{MN}$ in this gauge and perform the
necessary quantum ordering to insure that $\left(  i\right)  $ the $L^{MN}$
obey the correct SO$\left(  D+1,2\right)  $ commutation rules by using only
the quantum rules $\left[  \mathbf{r}_{i},\mathbf{p}_{j}\right]  =i\delta
_{ij},$ and $\left(  ii\right)  $ the quadratic Casimir eigenvalue for
$C_{2}=\frac{1}{2}L^{MN}L_{MN},$ as computed in this gauge, gives the same
gauge-invariant result in the Appendix, namely $C_{2}=1-\left(  D+1\right)
^{2}/4,$ that was obtained in covariant quantization for physical states,
without choosing any gauge. The result that satisfies these physical
conditions, with all quantum ordering issues resolved, and insuring
hermiticity of the $\frac{1}{2}\left(  D+3\right)  \left(  D+2\right)  $
generators, is as follows \cite{Bars:1998pc}:%
\begin{equation}%
\begin{array}
[c]{ll}%
\text{SO}\left(  1,2\right)  : & L^{mn}=\left\{
\begin{array}
[c]{l}%
L^{0+^{\prime}}=\left\vert \mathbf{r}\right\vert \\
L^{-^{\prime}+^{\prime}}=\frac{\mathbf{r\cdot p+p\cdot r}}{2}\\
L^{0-^{\prime}}=\sqrt{\left\vert \mathbf{r}\right\vert }\frac{\mathbf{p}^{2}%
}{2}\sqrt{\left\vert \mathbf{r}\right\vert }%
\end{array}
\right.  ,~m,n=\left(  +^{\prime},-^{\prime},0\right) \\
\text{SO}\left(  D\right)  : & L^{ij}=\left(  \mathbf{r}^{i}\mathbf{p}%
^{j}\mathbf{-r}^{j}\mathbf{p}^{i}\right)  ,\;\;i,j=1,2,\cdots,D.\\
\text{Coset}: & L^{im}=\left\{
\begin{array}
[c]{l}%
L^{i+^{\prime}}=\mathbf{r}^{i}\\
L^{i0}=-\frac{\left\vert \mathbf{r}\right\vert \mathbf{p}^{i}\mathbf{+p}%
^{i}\left\vert \mathbf{r}\right\vert }{2},\;\\
L^{i-^{\prime}}=-\frac{\left(  \mathbf{p\cdot r}\right)  \mathbf{p}%
^{i}+\mathbf{p}^{i}\left(  \mathbf{r\cdot p}\right)  }{2}+\frac{\mathbf{p}%
^{2}\mathbf{r}^{i}+\mathbf{r}^{i}\mathbf{p}^{2}}{4}-\frac{\mathbf{r}^{i}%
}{8\left\vert \mathbf{r}\right\vert ^{2}}%
\end{array}
\right.
\end{array}
\label{HquantumL}%
\end{equation}
The subgroup structure, SO$\left(  1,2\right)  \otimes$ SO$\left(  D\right)
\subset$ SO$\left(  D+1,2\right)  ,$ that is used to classify the Hatom$_{D}$
spectrum as in Eqs.\ (\ref{H+HO}-\ref{SpectraGroups}) is indicated above. The
coset is in the $\left(  \text{vector}\otimes\text{vector}\right)  $
representation of the subgroup. It is also possible to reorganize the
generators according to the subgroup, SO$\left(  2\right)  \otimes$ SO$\left(
D+1\right)  \subset$ SO$\left(  D+1,2\right)  ,$ where the SO$\left(
2\right)  $ generator is $L^{0^{\prime}0}$ that rotates the two temporal
coordinates $\left(  X^{0^{\prime}},X^{0}\right)  $ into each other and
SO$\left(  D+1\right)  $ rotates the spatial coordinates $\left(
X^{1^{\prime}},\mathbf{X}\right)  $ into each other. These subalgebra
operators are related to the above according to the lightcone map in
(\ref{lightcone}) as follows:%
\begin{equation}%
\begin{array}
[c]{l}%
\text{SO}\left(  2\right)  :\;~L^{0^{\prime}0}=\left(  L^{0-^{\prime}}%
+\frac{1}{2}L^{0+^{\prime}}\right)  =\sqrt{\left\vert \mathbf{r}\right\vert
}\frac{\mathbf{p}^{2}+1}{2}\sqrt{\left\vert \mathbf{r}\right\vert },\\
\text{SO}\left(  D+1\right)  :\;L^{ij},\;L^{i1^{\prime}}=\left(
L^{i-^{\prime}}-\frac{1}{2}L^{i+^{\prime}}\right)  ,
\end{array}
\end{equation}
where $L^{i1^{\prime}}$ is related (by gauge invariance) to the famous
Runge-Lenz vector when $\left(  \mathbf{r,p}\right)  $ are rewritten in terms
of $\left(  \widetilde{\mathbf{r}}\mathbf{,}\widetilde{\mathbf{p}}\right)  $
(see Eq.\ (24) in \cite{Bars:1998pc}). Recall that the aim is to find the
eigenvalues of $L^{0^{\prime}0}$. This is easily done algebraically
\cite{Bars:1998pc} by noting that $L^{0^{\prime}0}$ is the compact generator
of the SO$\left(  1,2\right)  $ Lie algebra, whose quadratic Casimir is
related to SO$\left(  D\right)  $ angular momentum as follows:\footnote{To
raise/lower SO$\left(  1,2\right)  $ indices $L_{mn}$ use the metric
$\eta_{mn}$ with nonzero entries: $\eta_{+^{\prime}-^{\prime}}=\eta
_{-^{\prime}+^{\prime}}=\eta_{00}=-1.$}
\begin{equation}%
\begin{array}
[c]{l}%
C_{2}^{\text{SO}\left(  1,2\right)  }=\frac{1}{2}L^{mn}L_{mn}=L^{0+^{\prime}%
}L^{0-^{\prime}}+L^{0-^{\prime}}L^{0+^{\prime}}-\left(  L^{+^{\prime}%
-^{\prime}}\right)  ^{2}\\
\;\;\;=\left[  \frac{\mathbf{r}^{2}\mathbf{p}^{2}+\mathbf{p}^{2}\mathbf{r}%
^{2}}{2}-\left(  \frac{\mathbf{r\cdot p+p\cdot r}}{2}\right)  ^{2}+\frac{3}%
{4}\right] \\
\;\;\;=\left[  \frac{1}{2}L^{ij}L_{ij}+\frac{\left(  D-1\right)  \left(
D-3\right)  }{4}\right]  .
\end{array}
\end{equation}
The computation above that yields the numerical contribution $\frac{3}{4}$
(proportional to $\hbar^{2}\rightarrow1$) is performed by watching the orders
of operators and using the commutators, $\left[  \mathbf{r}^{i},\mathbf{p}%
^{j}\right]  =i\delta^{ij},$ to change their orders. Since the right hand side
shows that $C_{2}^{\text{SO}\left(  1,2\right)  }\geq-\frac{1}{4}$ for all
$D=1,2,3,\cdots,$ only the unitary positive discrete series representation of
SO$\left(  1,2\right)  =$ SL$\left(  2,R\right)  =$ SU$\left(  1,1\right)  $
can occur. Then according to known representation theory of SL$\left(
2,R\right)  $ \cite{Bargmann:1936}, the eigenvalues of $C_{2}^{\text{SO}%
\left(  1,2\right)  }$ and $L^{0^{\prime}0}$ are respectively, $j\left(
j+1\right)  $ and $m\left(  j\right)  =j+1+n_{r},$ with $n_{r}=0,1,2,\cdots.$
Simultaneously, SO$\left(  D\right)  $ angular momentum is also diagonal,
$\frac{1}{2}L^{ij}L_{ij}\rightarrow l\left(  l+D-2\right)  .$ Therefore, we
have the following relations among $\left(  j,l,n_{r},D\right)  $
\cite{Bars:1998pc}%
\begin{equation}%
\begin{array}
[c]{l}%
j\left(  j+1\right)  =\left[  l\left(  l+D-2\right)  +\frac{\left(
D-1\right)  \left(  D-3\right)  }{4}\right]  ,\;\Rightarrow j\left(  l\right)
=-\frac{1}{2}\pm\left\vert l+\frac{D-2}{2}\right\vert ,\\
L^{0^{\prime}0}\rightarrow m\left(  j\right)  =j\left(  l\right)  +1+n_{r}.
\end{array}
\label{jlH1}%
\end{equation}
Since this is the positive discrete series we must have $m\left(  j\right)
>0,$ which requires $\left(  j\left(  l\right)  +1\right)  >0.$ Accordingly,
we can choose the $\pm\rightarrow+,$ resolve the absolute value sign, and
write:
\begin{equation}%
\begin{array}
[c]{l}%
j\left(  l\right)  =\left\{
\begin{array}
[c]{l}%
0,\;\text{if }D=1\text{ (and }l=0\text{ necessarily)}\\
l+\frac{D-3}{2},\;\text{if }D\geq2
\end{array}
\right.  ,\;\\
m\left(  j\right)  =j\left(  l\right)  +1+n_{r}=\left\{
\begin{array}
[c]{l}%
\left(  0+1+n_{r}\right)  =n,~\text{for }D=1\\
\left(  l+1+n_{r}\right)  +\frac{D-3}{2}=n+\frac{D-3}{2},~\text{for }D\geq2
\end{array}
\right.  ,
\end{array}
\label{jlH}%
\end{equation}
where $n=\left(  l+1+n_{r}\right)  \geq1$ since both $l,n_{r}=0,1,2,\cdots.$
This explains the Hatom part in Eq.\ (\ref{j(l)}).

Thus, we have computed algebraically the desired eigenvalue of $L^{0^{\prime
}0},$ which then, according to 2T-physics,\footnote{Historically, the
algebraic computation of the Hatom spectrum followed a different path. It
relied on the Runge-Lenz vector that together with orbital angular momentum,
complete an SO$\left(  D+1\right)  $ algebra. The quadratic Casimir of this
algebra can be shown to be related to the Hatom Hamiltonian, so the spectrum
of the Hamiltonian was computed by computing $C_{2}^{\text{SO(D+1)}}.$ The
2T-physics approach showed another way to get there, namely by computing the
eigenvalues of $L^{0^{\prime}0}.$} determines the quantum eigenvalue of the
Hatom Hamiltonian given in terms of the gauge-invariant $L^{0^{\prime}0}$ in
Eq.\ (\ref{HL00}),
\begin{equation}
H=-\frac{Z^{2}/2}{\left(  L^{00^{\prime}}\right)  ^{2}}\rightarrow\left\{
\begin{array}
[c]{l}%
-\frac{Z^{2}/2}{n^{2}},\text{ for }D=1\\
-\frac{Z^{2}/2}{\left(  n+\frac{D-3}{2}\right)  ^{2}},\text{ for }D\geq2
\end{array}
\right.  ,\;n=1,2,3,\cdots.
\end{equation}
As anticipated, quantum ordering did produce a quantum shift of the integer
$n$ by the amount $\left(  D-3\right)  /2.$ This result agrees with solving
the Hatom$_{D}$ radial differential equation, hence the quantum-ordered
generators given above are correct since they produce not only the correct
spectrum, but also the correct SO$\left(  D+1,2\right)  $ Casimir eigenvalue,
as well as the correct Lie algebra for SO$\left(  D+1,2\right)  $.

A useful final observation that follows from (\ref{HquantumL}) is to realize
that the phase space $\left(  \mathbf{r,p}\right)  $ can be written in terms
of the gauge-invariants $L^{0+^{\prime}}=\left\vert \mathbf{r}\right\vert $
and $L^{i0}=-\frac{1}{2}\left(  \left\vert \mathbf{r}\right\vert
\mathbf{p}^{i}+\mathbf{p}^{i}\left\vert \mathbf{r}\right\vert \right)  :$
\begin{equation}
\mathbf{r}^{i}=L^{i+^{\prime}},\;\mathbf{p}^{i}=-\left(  L^{0+^{\prime}%
}\right)  ^{-1/2}L^{i0}\left(  L^{0+^{\prime}}\right)  ^{-1/2}. \label{rpL}%
\end{equation}
The second relation is verified at the quantum level as follows:
\begin{equation}
-\left(  L^{0+^{\prime}}\right)  ^{-1/2}L^{i0}\left(  L^{0+^{\prime}}\right)
^{-1/2}=\frac{1}{2}\left(  \sqrt{\left\vert \mathbf{r}\right\vert }%
\mathbf{p}^{i}\frac{1}{\sqrt{\left\vert \mathbf{r}\right\vert }}+\frac
{1}{\sqrt{\left\vert \mathbf{r}\right\vert }}\mathbf{p}^{i}\sqrt{\left\vert
\mathbf{r}\right\vert }\right)  =\mathbf{p}^{i}.
\end{equation}
Eq.\ (\ref{rpL}) will be very important to extract the desired canonical transformation.

\subsection{Sp$\left(  2\bar{D},R\right)  $ and the Harmonic Oscillator
\label{sp2DR}}

The HOsc$_{\bar D}$ has a dynamical symmetry Sp$\left(  2\bar{D},R\right)  $
that controls its spectrum as described in this section. The harmonic
oscillator in $\bar{D}$ space dimensions, with phase space $\left(
\mathbf{\bar{r}}_{\alpha}\mathbf{,\bar{p}}_{\alpha}\right)  ,$ $\alpha
=1,2,\cdots,\bar D,$ has dynamics described by the Hamiltonian
\begin{equation}
H=\left(  \frac{\mathbf{\bar{p}^{2}}}{2\mu}+\frac{\mu\omega^{2}}%
{2}\mathbf{\bar{r}}^{2}\right)  =\left(  \mathbf{a}^{\dag}\cdot\mathbf{a}%
+\frac{\bar{D}}{2}\right)  ,\;\mathbf{a}\equiv\left(  \sqrt{\frac{\hbar\omega
}{2\mu}}\mathbf{\bar{r}}+i\sqrt{\frac{\mu}{2\hbar\omega}}\mathbf{\bar{p}%
}\right)  \mathbf{.} \label{HOhamiltonian}%
\end{equation}
For convenience (or by a rescaling of the coordinates and momenta) we will
take $\mu=1,$ $\omega=1$ and $\hbar=1.$ The Hamiltonian is invariant under
unitary transformations $a_{\alpha}\rightarrow U_{\alpha\beta}a_{\beta}$ where
$U^{\dag}U=1$. These transformations belong to the group SU$\left(  \bar
{D}\right)  \otimes$ U$\left(  1\right)  \subset$ Sp$\left(  2\bar
{D},R\right)  $. Excited levels are of the form $a_{\alpha_{1}}^{\dag
}a_{\alpha_{2}}^{\dag}\ldots a_{\alpha_{\bar{n}}}^{\dag}|0\rangle$, where
$|0\rangle$ denotes the ground state. The totally symmetric tensor form of the
quantum states implies that only totally symmetric representations of
SU($\bar{D}$) will occur (single-row Young tableaux with $\bar{n}$ boxes).
Thus the spectrum will have degeneracy
\begin{equation}
\frac{(\bar{D}+\bar{n}-1)!}{\bar{n}!(\bar{D}-1)!},\text{or }\left\{  1,\bar
{D},\frac{\bar{D}(\bar{D}+1)}{2},\frac{\bar{D}(\bar{D}+1)(\bar{D}+2)}%
{6},\ldots\right\}  ,
\end{equation}
corresponding to total excitation number, $\bar{n}=0,1,2,3,\ldots,$ and
energies, $E_{\bar{n}}=\hbar\omega(\bar{n}+\frac{\bar{D}}{2})$.

Sp$\left(  2\bar{D},R\right)  $ acts linearly on a $2\bar{D}$ dimensional
column consisting of $\bar{D}$ position and $\bar{D}$ momentum real degrees of
freedom in a real basis $\left(
%TCIMACRO{\QATOP{\QTR{bf}{\bar{r}}}{\QTR{bf}{\bar{p}}}}%
%BeginExpansion
\genfrac{}{}{0pt}{}{\mathbf{\bar{r}}}{\mathbf{\bar{p}}}%
%EndExpansion
\right)  ,$ or equivalently in a pseudo-complex basis, $\left(
%TCIMACRO{\QATOP{\QTR{bf}{a}}{\QTR{bf}{a}^{\dag}}}%
%BeginExpansion
\genfrac{}{}{0pt}{}{\mathbf{a}}{\mathbf{a}^{\dag}}%
%EndExpansion
\right)  $. The generators of Sp$\left(  2\bar{D},R\right)  $ are formed by
the symmetric product of two such columns, so there are $\bar{D}(2\bar{D}+1$)
generators that take the form
\begin{equation}
\left\{  \frac{\bar{r}_{\alpha}\bar{r}_{\beta}}{2},\frac{\bar{r}_{\alpha}%
\bar{p}_{\beta}+\bar{p}_{\beta}\bar{r}_{\alpha}}{4},\frac{\bar{p}_{\alpha}%
\bar{p}_{\beta}}{2}\right\}  \;\text{or\ }\left\{  \frac{a_{\alpha}^{\dagger
}a_{\beta}^{\dagger}}{2},\frac{a_{\alpha}^{\dagger}a_{\beta}+a_{\beta
}a_{\alpha}^{\dagger}}{4},\frac{a_{\alpha}a_{\beta}}{2}\right\}  .
\label{sp2Dgens}%
\end{equation}
These form the Sp$\left(  2\bar{D},R\right)  $ Lie algebra under classical
Poisson brackets or quantum commutators. The subset of operators $\frac{1}%
{2}\left(  a_{\alpha}^{\dagger}a_{\beta}+a_{\beta}a_{\alpha}^{\dagger}\right)
$ or $\frac{1}{2}\left(  \bar{r}_{\alpha}\bar{p}_{\beta}+\bar{p}_{\beta}%
\bar{r}_{\alpha}\right)  $ that can be decomposed into one symmetric and one
antisymmetric tensor form the SU$\left(  \bar{D}\right)  \otimes$ U$\left(
1\right)  $ subalgebra, where the U$\left(  1\right)  $ operator, which is the
trace of the symmetric tensor, is the Hamiltonian $H=\frac{1}{2}\left(
\mathbf{\bar{p}^{2}}+\mathbf{\bar{r}}^{2}\right)  =\left(  \mathbf{a}%
^{\dagger}\cdot\mathbf{a}+\bar{D}/2\right)  .$ Acting on the harmonic
oscillator quantum states in Fock space, the step-up or step-down generators
$\left(  a_{\alpha}^{\dagger}a_{\beta}^{\dagger},a_{\alpha}a_{\beta}\right)  $
doubly excite or doubly de-excite any given state with total excitation number
$\bar{n}$. For this reason the Sp$\left(  2\bar{D},R\right)  $ action cannot
mix $\bar{n}_{\text{odd}}$ states with $\bar{n}_{\text{even}}$ states.
Moreover, it is evident that all even (odd) states mix among the even (odd)
sets under repeated action of the step up/down operators, so even/odd states
form disjoint irreducible representation bases. The Casimir operators $\bar
{C}_{k}$ that commute with all the generators, in this construction, are all
pure numbers (i.e., not operators),
\begin{equation}
\bar{C}_{k}=\frac{2}{4^{k}k!}\left(  2\bar{D}+1\right)  \left(  1-\left(
2\bar{D}+1\right)  ^{k-1}\right)  ,\;k=2,3,\cdots,\bar{D}.
\label{CasimirsSp2D}%
\end{equation}
For example, the quadratic Casimir is computed explicitly as follows by taking
into account the order of harmonic oscillator operators,
\begin{equation}
\bar{C}_{2}=\left(  2\frac{a_{\alpha}^{\dagger}a_{\beta}+a_{\beta}a_{\alpha
}^{\dagger}}{4}\frac{a_{\beta}^{\dagger}a_{\alpha}+a_{\alpha}a_{\beta
}^{\dagger}}{4}-\frac{a_{\alpha}a_{\beta}}{2}\frac{a_{\beta}^{\dagger
}a_{\alpha}^{\dagger}}{2}-\frac{a_{\alpha}^{\dagger}a_{\beta}^{\dagger}}%
{2}\frac{a_{\beta}a_{\alpha}}{2}\right)  =-\frac{\bar{D}}{2}\left(  \frac
{\bar{D}}{2}+\frac{1}{4}\right)  .
\end{equation}
So, the Fock space of HOsc$_{\bar{D}}$ corresponds to two specific fixed
representations of Sp$\left(  2\bar{D},R\right)  $ such that both even and odd
states have the same Casimir eigenvalues, $\bar{C}_{k},$ even though they are
distinct irreducible representations.

Another commuting set of subalgebras of Sp$\left(  2\bar{D},R\right)  $ that
is relevant for our analysis is Sp$\left(  2,R\right)  \otimes$ SO$\left(
\bar{D}\right)  $. The three Sp$\left(  2,R\right)  $ generators
$G_{\mu=0,1,2}$ are obtained from the traces of the tensors listed in
(\ref{sp2Dgens}), while the SO$\left(  \bar{D}\right)  $ generators
$L_{\alpha\beta}$ correspond to the only antisymmetric tensor constructed from
those listed in (\ref{sp2Dgens}), namely $L_{\alpha\beta}=\left(  \bar
{r}_{\alpha}\bar{p}_{\beta}-\bar{r}_{\beta}\bar{p}_{\alpha}\right)  =\left(
a_{\alpha}^{\dagger}a_{\beta}-a_{\beta}^{\dagger}a_{\alpha}\right)  .$ The
remaining coset generators are labelled by representations of Sp$\left(
2,R\right)  \otimes$ SO$\left(  \bar{D}\right)  $, as $S_{\mu\left(
\alpha\beta\right)  },$ where the pair $\left(  \alpha\beta\right)  $
corresponds to the irreducible symmetric traceless tensor of SO$\left(
\bar{D}\right)  .$ So, the Sp$\left(  2\bar{D},R\right)  $ generators are
given as follows (these may be rewritten in terms of $\left(  a_{\alpha
}^{\dagger},a_{\alpha}\right)  $), for a total of $\bar{D}\left(  2\bar
{D}+1\right)  $ Hermitian operators:
\begin{equation}%
\begin{array}
[c]{l}%
\text{Sp}\left(  2,R\right)  :G_{\mu}=\left\{
\begin{array}
[c]{l}%
G_{+}=\left(  G_{0}+G_{1}\right)  =\frac{\mathbf{\bar{r}\cdot\bar{r}}}{2}\\
G_{2}=-\frac{\mathbf{\bar{r}\cdot\bar{p}+\bar{p}\cdot\bar{r}}}{4}\\
G_{-}=\frac{1}{2}\left(  G_{0}-G_{1}\right)  =\frac{\mathbf{\bar{p}\cdot
\bar{p}}}{4}%
\end{array}
\right.  ,\\
\text{SO}\left(  \bar{D}\right)  :L_{\alpha\beta}=\left(  \bar{r}_{\alpha}%
\bar{p}_{\beta}-\bar{r}_{\beta}\bar{p}_{\alpha}\right)  ,\\
\text{Coset}:\text{\ }S_{\mu\left(  \alpha\beta\right)  }=\left\{
\begin{array}
[c]{l}%
S_{+\left(  \alpha\beta\right)  }=\frac{\bar{r}_{\alpha}\bar{r}_{\beta}}%
{2}-\frac{\delta_{\alpha\beta}}{\bar{D}}\frac{\mathbf{\bar{r}\cdot\bar{r}}}%
{2}\\
S_{2\left(  \alpha\beta\right)  }=\frac{\bar{r}_{\alpha}\bar{p}_{\beta}%
+\bar{r}_{\beta}\bar{p}_{\alpha}}{4}-\frac{\delta_{\alpha\beta}}{\bar{D}}%
\frac{\mathbf{\bar{r}\cdot\bar{p}+\bar{p}\cdot\bar{r}}}{4}\\
S_{-\left(  \alpha\beta\right)  }=\frac{\bar{p}_{\alpha}\bar{p}_{\beta}}%
{2}-\frac{\delta_{\alpha\beta}}{\bar{D}}\frac{\mathbf{\bar{p}\cdot\bar{p}}}{2}%
\end{array}
\right.  .
\end{array}
\label{GLS-HOsc}%
\end{equation}
Note that the $G_{\pm,0}$ are Hermitian combinations of the Hermitian $G_{\mu
}.$ The Sp$\left(  2,R\right)  \otimes$ SO$\left(  \bar{D}\right)  $
subalgebras commute with each other, $\left[  G_{\mu},L_{\alpha\beta}\right]
=0,$ because the $G_{\mu}$ are constructed from SO$\left(  \bar{D}\right)
$-invariant dot products. It can be checked that the compact generator $G_{0}$
of Sp$\left(  2,R\right)  =$ SL$\left(  2,R\right)  =$ SU$\left(  1,1\right)
,$ given by $G_{0}=\frac{1}{2}G_{+}+G_{-},$ is related to half the Hamiltonian
$H$ in (\ref{HOhamiltonian}), and the quadratic Casimir operator, $\bar{C}%
_{2}\left(  \text{Sp}\left(  2,R\right)  \right)  =\left(  G_{0}^{2}-G_{1}%
^{2}-G_{2}^{2}\right)  =\left(  G_{+}G_{-}+G_{-}G_{+}-G_{2}^{2}\right)  ,$ is
related to angular momentum $\frac{1}{2}L^{\alpha\beta}L_{\alpha\beta}$:
\begin{equation}%
\begin{array}
[c]{l}%
G_{0}=\frac{\mathbf{\bar{p}}^{2}\mathbf{+\bar{r}}^{2}}{4}=\frac{1}{2}H,\\
\bar{C}_{2}\left(  \text{Sp}\left(  2,R\right)  \right)  =\frac{1}{4}\left[
\frac{\mathbf{\bar{r}}^{2}\mathbf{\bar{p}}^{2}+\mathbf{\bar{p}}^{2}%
\mathbf{\bar{r}}^{2}}{2}-\left(  \frac{\mathbf{\bar{r}}\cdot\mathbf{\bar{p}%
}+\mathbf{\bar{p}}\cdot\mathbf{\bar{r}}}{2}\right)  ^{2}\right]  =\frac{1}%
{4}\left(  \frac{1}{2}L^{\alpha\beta}L_{\alpha\beta}+\frac{\bar{D}\left(
\bar{D}-4\right)  }{4}\right)  .
\end{array}
\label{c2Spc2L}%
\end{equation}
The eigenstates of Sp$\left(  2,R\right)  =$ SL$\left(  2,R\right)  =$
SU$\left(  1,1\right)  $ simultaneously diagonalize $\bar{C}_{2}\left(
\text{Sp}\left(  2,R\right)  \right)  \rightarrow\bar{j}\left(  \bar
{j}+1\right)  ,$ and the compact generator $G_{0}\rightarrow m\left(
j\right)  =\left(  j+1+n_{r}\right)  $. Angular momentum is also
simultaneously diagonalized, $\frac{1}{2}L^{\alpha\beta}L_{\alpha\beta
}\rightarrow\bar{l}\left(  \bar{l}+\bar{D}-2\right)  .$ Hence
Eq.\ (\ref{c2Spc2L}) relates the quantum numbers $\left(  \bar{j},\bar{l}%
,\bar{D},\bar{n},n_{r}\right)  $ as follows:%
\begin{equation}%
\begin{array}
[c]{l}%
\bar{j}\left(  \bar{j}+1\right)  =\frac{1}{4}\left(  \bar{l}\left(  \bar
{l}+\bar{D}-2\right)  +\frac{\bar{D}\left(  \bar{D}-4\right)  }{4}\right)  ,\\
G_{0}\rightarrow\left(  \bar{j}\left(  \bar{l}\right)  +1+n_{r}\right)
=\frac{1}{2}\left(  \bar{n}+\frac{\bar{D}}{2}\right)  ,
\end{array}
\Rightarrow%
\begin{array}
[c]{l}%
\bar{j}\left(  \bar{l}\right)  =\frac{1}{2}\left(  \bar{l}+\frac{\bar{D}-4}%
{2}\right)  ,\\
\bar{n}=\left(  \bar{l}+2n_{r}\right)  \geq0.
\end{array}
\label{jl}%
\end{equation}
In this case we see that the smallest eigenvalue of $G_{0}$ (which corresponds
to half the Hamiltonian of HOsc$_{\bar{D}}$) occurs for the vacuum of Fock
space, and is givem by, $\bar{j}\left(  0\right)  +1=\frac{\bar{D}}{4}.$ This
is positive for all values of $\bar{D}=1,2,3,\cdots,$ so only one solution of
the quadratic equation for $j$ is given in (\ref{jl}). This explains the HOsc
part in Eq.\ (\ref{j(l)}).

\section{Duality as canonical transformations and SO$\left(  D+1,2\right)  $
\label{dualitySec}}

\label{sec:canonical}

We are now ready to compare the SO$\left(  D+1,2\right)  $ generators
$L^{MN}\left(  \mathbf{r,p}\right)  $ in (\ref{HquantumL}) to the Sp$\left(
2\bar{D},R\right)  $ generators in (\ref{GLS-HOsc}). These clearly are
different Lie algebras; however, they both share the crucial subgroups
\begin{equation}
\text{Hatom}_{D}:\text{SO}\left(  1,2\right)  \otimes\text{SO}\left(
D\right)  ,\text{ and}\;\text{HOsc}_{\bar{D}}:\text{Sp}\left(  2,R\right)
\otimes\text{SO}\left(  \bar{D}\right)  , \label{crucial}%
\end{equation}
that classify the spectra in the generalized version of Eq.\ (\ref{H+HO}) as
described following that equation. The fact that SO$\left(  1,2\right)  =$
Sp$\left(  2,R\right)  $ suggests that, for a duality to exist, we should
identify \textit{some subsets} of the infinite vertical towers of the
HOsc$_{\bar{D}}$ to all the towers of the Hatom$_{D}.$ We already saw in
Section \ref{hints} that when $D=\bar{D}=2$ this idea actually works. For more
general $3\leq D\leq\bar{D},$ identifying the towers requires at the very
least that we require they are in the same representation of SO$\left(
1,2\right)  =$ Sp$\left(  2,R\right)  ,$ which means $j\left(  l\right)
=\bar{j}\left(  \bar{l}\right)  .$ Using Eqs.\ (\ref{jlH},\ref{jl}), this
gives
\begin{equation}%
\begin{array}
[c]{l}%
\text{for }D=1:\;j\left(  l\right)  =0=\frac{1}{2}\left(  0+\frac{\bar{D}%
-4}{2}\right)  \text{ and }l,\bar{l}=0.\\
\text{for }D\geq2:\;j\left(  l\right)  =l+\frac{D-3}{2}=\frac{1}{2}\left(
\bar{l}+\frac{\bar{D}-4}{2}\right)  .
\end{array}
\label{identify}%
\end{equation}
This condition reproduces Eq.\ (\ref{eqn:lrelation}) for $\frac{\bar{\alpha}%
}{\alpha}=-2,$ that was based on the radial duality. This is a very
encouraging observation, so we pursue it in this section.

Thus, for sufficiently large $\bar{D}\geq D$ we can find an appropriate
subgroup SO$\left(  D+1,2\right)  \subset$ Sp$\left(  2\bar{D},R\right)  $
with generators $L^{MN}\left(  \mathbf{\bar{r},\bar{p}}\right)  $ that are
some linear combinations of the Sp$\left(  2\bar{D},R\right)  $ generators in
(\ref{GLS-HOsc}). These will look different than the Hatom's $L^{MN}\left(
\mathbf{r,p}\right)  $ in (\ref{HquantumL}) as functions of different phase
spaces $\left(  \mathbf{r,p}\right)  _{D}$ versus $\left(  \mathbf{\bar
{r},\bar{p}}\right)  _{\bar{D}},$ but will have the crucial subalgebras in
(\ref{crucial}), with the imposed condition (\ref{identify}). By the
2T-physics gauge-invariance argument in the Appendix, we consider the pair
$\left(  L^{MN}\left(  \mathbf{r,p}\right)  ,L^{MN}\left(  \mathbf{\bar
{r},\bar{p}}\right)  \right)  $ to be the algebraic description of the hidden
SO$\left(  D+1,2\right)  $ symmetry of two shadows, and on this basis we
equate them as described in the Appendix:
\begin{equation}
L^{MN}\left(  \mathbf{r,p}\right)  =L^{MN}\left(  \mathbf{\bar{r},\bar{p}%
}\right)  . \label{equate}%
\end{equation}
From this equation we will derive the sought-after canonical transformation
that relates the two phase spaces $\left(  \mathbf{r,p}\right)
\leftrightarrow~\left(  \mathbf{\bar{r},\bar{p}}\right)  $, as shown below. We
will see that this scheme works for certain dimensions $\bar{D}\geq D$.

To say that Hatom$_{D}$ is dual to HOsc$_{\bar{D}\geq D}$ via a canonical
transformation may seem incomplete since the phase space $\left(
\mathbf{\bar{r},\bar{p}}\right)  _{\bar{D}>D}$ has certainly more degrees of
freedom as compared to $\left(  \mathbf{r,p}\right)  _{D}.$ Hence we must
expect some constraints on the phase space $\left(  \mathbf{\bar{r},\bar{p}%
}\right)  _{\bar{D}>D}$ that reduce the degrees of freedom such that the
spectrum of HOsc$_{\bar{D}\geq D}$ becomes compatible with Hatom$_{D}.$ This
means that the canonical transformation that we will display must have a gauge
symmetry that gives rise to the constraints in the form of vanishing gauge
generators (thus distinguishing the gauge-invariant subspace). Hence the
duality Hatom$_{D}\leftrightarrow$ HOsc$_{\bar{D}>D}$ can be true only on the
gauge invariants for which the gauge generators vanish. We will display the
gauge symmetry in precisely the form stated in this paragraph. Because of the
gauge symmetry, the gauge-invariant sector of HOsc$_{\bar{D}\geq D}$ is
effectively equivalent to a phase space in $D$ dimensions.

\subsection{Case of $D=\bar{D}=2$ and SO$\left(  3,2\right)  =$ Sp$\left(
4,R\right)  $ \label{sec:Sp4R}}

For $D=\bar{D}=2$ we already have SO$\left(  3,2\right)  =$ Sp$\left(
4,R\right)  $ so there is no need to search for a subgroup. However the
Sp$\left(  4,R\right)  $ generators are expressed in the 4-dimensional spinor
basis as the product of two 4-dimensional columns $\psi_{A}=\left(
%TCIMACRO{\QATOP{\QTR{bf}{\bar{r}}}{\QTR{bf}{\bar{p}}}}%
%BeginExpansion
\genfrac{}{}{0pt}{}{\mathbf{\bar{r}}}{\mathbf{\bar{p}}}%
%EndExpansion
\right)  _{A},$ whereas the SO$\left(  3,2\right)  $ generators $L^{MN}$ are
expressed in the antisymmetric product of the 5-dimensional basis of
SO$\left(  3,2\right)  .$ All we need to do is convert the symmetric product
of the spinor basis $\frac{1}{2}\left(  \psi_{A}\psi_{B}+\psi_{B}\psi
_{A}\right)  $ discussed in Section \ref{sp2DR} to the antisymmetric product
in the vector basis $L^{MN}$ discussed in Section \ref{so(D+1,2)}. This is
done by using the antisymmetric product of SO$\left(  3,2\right)  $ gamma
matrices, $\left(  \gamma^{MN}\right)  _{AB}=\frac{1}{2}\left(  \gamma
^{M}\gamma^{N}-\gamma^{N}\gamma^{M}\right)  _{AB},$ where the 4$\times4$ gamma
matrices $\gamma_{AB}^{M}$ satisfy the Clifford algebra $\left\{  \gamma
^{M},\gamma^{N}\right\}  =2\eta^{MN}$, with the SO$\left(  3,2\right)  $
Minkowski metric $\eta^{MN}.$ Fortunately, for SO$\left(  3,2\right)  ,$
$\left(  C\gamma^{MN}\right)  _{AB}$ where $C$ is the \textquotedblleft charge
conjugation matrix\textquotedblright, is antisymmetric in the pair $\left[
MN\right]  $ and symmetric in the pair $\left(  AB\right)  ,$ so we can write%
\begin{equation}
L^{MN}\sim\left(  C\gamma^{MN}\right)  _{AB}\left(  \frac{\psi_{A}\psi
_{B}+\psi_{B}\psi_{A}}{2}\right)  , \label{vectorSpinor}%
\end{equation}
up to a normalization to preserve the correct commutation rules. We can now
focus on just a subset of the generators that lead to the desired canonical
transformation. These are $\left(  L^{i+^{\prime}},L^{i0},L^{0+^{\prime}%
}\right)  $ for the Hatom$_{2}$ as given in (\ref{HquantumL}), and $\left(
S_{+\left(  \alpha\beta\right)  },S_{2\left(  \alpha\beta\right)  }%
,G_{+}\right)  $ for the HOsc$_{2}$ as given in (\ref{GLS-HOsc}). Relating
them as in (\ref{vectorSpinor}) we obtain a simplified version of
(\ref{vectorSpinor}) for this subset:
\begin{equation}
L^{i+^{\prime}}=\left(  \gamma^{i}\right)  _{\alpha\beta}S_{+\left(
\alpha\beta\right)  }\text{ },\text{\ }L^{i0}=\left(  \gamma^{i}\right)
_{\alpha\beta}S_{2\left(  \alpha\beta\right)  },\;L^{0+^{\prime}}=G_{+},
\label{canonical1}%
\end{equation}
where now $\gamma_{\alpha\beta}^{i}$ are simply the SO$\left(  2\right)  $
gamma matrices given by 2$\times2$ Pauli matrices, $\gamma^{1}=\sigma_{3}$ and
$\gamma^{2}=\sigma_{1}$ that are correctly normalized. Recalling (\ref{rpL})
that says $\mathbf{r}^{i}=L^{i+^{\prime}}$ and $\mathbf{p}^{i}=-\left(
L^{0+^{\prime}}\right)  ^{-1/2}L^{i0}\left(  L^{0+^{\prime}}\right)  ^{-1/2},$
and using Eqs.\ (\ref{HquantumL},\ref{GLS-HOsc},\ref{canonical1}), we now can
write%
\begin{equation}%
\begin{array}
[c]{l}%
\mathbf{r}^{i}=L^{i+^{\prime}}=\left(  \gamma^{i}\right)  _{\alpha\beta
}S_{+\left(  \alpha\beta\right)  }=\gamma_{\alpha\beta}^{i}\left(
\frac{\mathbf{\bar{r}}^{\alpha}\mathbf{\bar{r}}^{\beta}}{2}-\frac
{\delta^{\alpha\beta}}{\bar{D}}\frac{\mathbf{\bar{r}\cdot\bar{r}}}{2}\right)
,\\
\mathbf{p}^{i}=-\left(  L^{0+^{\prime}}\right)  ^{-1/2}L^{i0}\left(
L^{0+^{\prime}}\right)  ^{-1/2}=-\left(  G_{+}\right)  ^{-1/2}\left(
\gamma^{i}\right)  _{\alpha\beta}S_{2\left(  \alpha\beta\right)  }%
L^{i0}\left(  G_{+}\right)  ^{-1/2}\\
\;\;\;=-\gamma_{\alpha\beta}^{i}\frac{\sqrt{2}}{\left\vert \mathbf{\bar{r}%
}\right\vert }\left(  \frac{\mathbf{\bar{r}}^{\alpha}\mathbf{\bar{p}}^{\beta
}\mathbf{+\bar{r}}^{\beta}\mathbf{\bar{p}}^{\alpha}}{4}-\frac{\delta
_{\alpha\beta}}{\bar{D}}\frac{\mathbf{\bar{r}\cdot\bar{p}+\bar{p}\cdot\bar{r}%
}}{4}\right)  \frac{\sqrt{2}}{\left\vert \mathbf{\bar{r}}\right\vert }.
\end{array}
\end{equation}
For traceless $\gamma_{\alpha\beta}^{i},$ this simplifies to
\begin{equation}
\mathbf{r}^{i}=\frac{1}{2}\mathbf{\bar{r}}\gamma^{i}\mathbf{\bar{r}%
},\;\mathbf{p}^{i}=\frac{1}{2}\frac{1}{\left\vert \mathbf{\bar{r}}\right\vert
}\left(  \mathbf{\bar{r}}\gamma^{i}\mathbf{\bar{p}}+\mathbf{\bar{p}}\gamma
^{i}\mathbf{\bar{r}}\right)  \frac{1}{\left\vert \mathbf{\bar{r}}\right\vert
}. \label{canon2}%
\end{equation}
Explicitly, using $\gamma^{i}=\left(  \sigma_{3},\sigma_{1}\right)  $ we
obtain%
\begin{equation}%
\begin{array}
[c]{l}%
\left(
\begin{array}
[c]{c}%
r_{1}\\
r_{2}%
\end{array}
\right)  =\left(
\begin{array}
[c]{c}%
\frac{1}{2}\left(  \bar{r}_{1}^{2}-\bar{r}_{2}^{2}\right) \\
\bar{r}_{1}\bar{r}_{2}%
\end{array}
\right)  \;,\\
\left(
\begin{array}
[c]{c}%
p_{1}\\
p_{2}%
\end{array}
\right)  =\frac{1}{\left\vert \mathbf{\bar{r}}\right\vert }\left(
\begin{array}
[c]{c}%
\frac{\bar{r}_{1}\bar{p}_{1}+\bar{p}_{1}\bar{r}_{1}}{2}-\frac{\bar{r}_{2}%
\bar{p}_{2}+\bar{p}_{2}\bar{r}_{2}}{2}\\
\bar{r}_{1}\bar{p}_{2}+\bar{r}_{2}\bar{p}_{1}%
\end{array}
\right)  \frac{1}{\left\vert \mathbf{\bar{r}}\right\vert },
\end{array}
\label{canon3}%
\end{equation}
where $\left\vert \mathbf{\bar{r}}\right\vert \equiv\sqrt{\bar{r}_{1}^{2}%
+\bar{r}_{2}^{2}}.$ We can verify that this is indeed a canonical
transformation by computing the Poisson brackets or quantum commutators as
follows:%
\begin{equation}%
\begin{array}
[c]{l}%
\left[  \mathbf{r}^{i},\mathbf{p}^{j}\right]  =\left[  \frac{1}{2}%
\mathbf{\bar{r}}^{\alpha}\gamma_{\alpha\beta}^{i}\mathbf{\bar{r}}^{\beta
},\frac{1}{2}\frac{1}{\left\vert \mathbf{\bar{r}}\right\vert }\left(
\mathbf{\bar{r}}^{\kappa}\gamma_{\kappa\lambda}^{j}\mathbf{\bar{p}}^{\lambda
}+\mathbf{\bar{p}}^{\lambda}\gamma_{\lambda\kappa}^{j}\mathbf{\bar{r}}%
^{\kappa}\right)  \frac{1}{\left\vert \mathbf{\bar{r}}\right\vert }\right] \\
\;\;\;=i\frac{1}{2}\frac{1}{\left\vert \mathbf{\bar{r}}\right\vert
}\mathbf{\bar{r}}\left(  \gamma^{i}\gamma^{j}+\gamma^{j}\gamma^{i}\right)
\mathbf{\bar{r}}\frac{1}{\left\vert \mathbf{\bar{r}}\right\vert }=\frac{i}%
{2}2\delta^{ij}\frac{1}{\left\vert \mathbf{\bar{r}}\right\vert }%
\mathbf{\bar{r}}\cdot\mathbf{\bar{r}}\frac{1}{\left\vert \mathbf{\bar{r}%
}\right\vert }=i\delta^{ij}.
\end{array}
\end{equation}
This shows that we obtain the right result, $\left[  \mathbf{r}^{i}%
,\mathbf{p}^{j}\right]  =i\delta^{ij},$ that a canonical transformation should
obey. Moreover, one can verify that all components of the generators in Eqs.
(\ref{HquantumL},\ref{GLS-HOsc}) do satisfy Eq.\ (\ref{vectorSpinor}) at the
quantum level (i.e., with correct ordering of operators) when the canonical
transformation (\ref{canon2} or \ref{canon3}) is inserted in
(\ref{vectorSpinor}).

We can rewrite the canonical transformation (\ref{canon3}) in cylindrical
coordinates by defining%
\begin{equation}
r_{1}=r\cos\theta,\;r_{2}=r\sin\theta,\text{ and }\;\bar{r}_{1}=\bar{r}%
\cos\bar{\theta},\;\bar{r}_{2}=\bar{r}\sin\bar{\theta}.
\end{equation}
Then (\ref{canon3}) for the transformation of the coordinates reduces to%
\begin{equation}
r=\frac{1}{2}\bar{r}^{2},\;\theta=2\bar{\theta}. \label{RandAngle}%
\end{equation}
This reproduces Newton's radial substitution (\ref{eqn:sub}) for the case
$\alpha=-1$ and $\bar{\alpha}=2,$ and adds the transformation of the angles as
well, thus completing the radial duality to a full duality in the case of
$D=\bar{D}=2.$ Moreover, we point out the fact that $\theta=2\bar{\theta}$ is
consistent with the shapes of classical orbits in the Kepler and harmonic
oscillator potentials: The perigee is reached once per Kepler orbit, but twice
per harmonic oscillator orbit.

Relative to (\ref{eqn:sub}) we ended up with an extra factor of $1/2$ in the
relation $r=\frac{1}{2}\bar{r}^{2}$ in Eq.\ (\ref{RandAngle}). This is of no
concern: the extra factor of 1/2 slightly alters only the radial duality rules
in (\ref{eqn:sebar}) as explained in footnote \ref{bfactor}.

As a final remark we identify a gauge symmetry that commutes with the
canonical transformation. Recall that according to Eq.\ (\ref{H+HO}) only the
even $\left(  \bar{n}_{\text{even}},\bar{l}_{\text{even}}\right)  $ quantum
states of HOsc$_{2}$ are dual to all the quantum states of the Hatom$_{2}$. As
asserted in the paragraph following Eq.\ (\ref{equate}), there must be a gauge
symmetry in the canonical transformation that can be used to project out the
$\left(  \bar{n}_{\text{odd}},\bar{l}_{\text{odd}}\right)  $ quantum states of
the HOsc$_{2}.$ Indeed there is such a gauge symmetry in Eqs. (\ref{canon2})
or (\ref{canon3}) that amounts to the discrete rotation of the HOsc$_{2}$
phase space by 180 degrees, $\left(  \mathbf{\bar{r},\bar{p}}\right)
\rightarrow\left(  -\mathbf{\bar{r},-\bar{p}}\right)  .$ Only the even quantum
states of the HOsc$_{2}$ are gauge-invariant under this discrete
transformation as can be seen from the wavefunction in Eq.\ (\ref{eqn:se})
that transforms under the gauge symmetry according to
\begin{equation}
T_{\alpha_{1}\cdots\alpha_{\bar{l}}}\left(  -\widehat{\mathbf{\bar{r}}%
}\right)  =\left(  -1\right)  ^{\bar{l}}T_{\alpha_{1}\cdots\alpha_{\bar{l}}%
}\left(  -\widehat{\mathbf{\bar{r}}}\right)  .
\end{equation}
Imposing this gauge symmetry on the quantum states of the HOsc$_{2}$
eliminates the non-invariant odd states, and gives precisely the duality that
is observed in the spectrum for the gauge-invariant even states, as displayed
in Eq.\ (\ref{H+HO}).

For the odd states of the HOsc$_{2}$, we anticipated in Section \ref{hints}
that the dyonic-Hatom is the correct dual, based on 2T-physics properties. We
will not discuss the details of this structure here in the same manner as
above, because we have no space to introduce more technical tools in this
paper to deal with the spin degrees of freedom required in the 2T formalism
\cite{Bars2Tspinning}.

\subsection{Case of D = 3, \={D} = 4, and SO$\left(  4,2\right)  \otimes$
U$\left(  1\right)  \subset$ Sp$\left(  8,R\right)  $~\label{specialDbarD}}

The form of the canonical transformation (\ref{canon2}) suggests applying it
to higher dimensions $D\geq2:$
\begin{equation}
\mathbf{r}^{i}=\frac{1}{2}\mathbf{\bar{r}}\gamma^{i}\mathbf{\bar{r}%
},\;\mathbf{p}^{i}=\frac{1}{2}\frac{1}{\left\vert \mathbf{\bar{r}}\right\vert
}\left(  \mathbf{\bar{r}}\gamma^{i}\mathbf{\bar{p}}+\mathbf{\bar{p}}\gamma
^{i}\mathbf{\bar{r}}\right)  \frac{1}{\left\vert \mathbf{\bar{r}}\right\vert
}. \label{canon34}%
\end{equation}
So, starting with Hatom$_{3}$ let's try it for $D=3$ and some $\bar{D}\geq3$.
A series of questions are: what is $\bar{D}$, what are the three $\bar
{D}\times\bar{D}$-symmetric matrices, $\gamma_{\alpha\beta}^{i},$ and in what
SO$\left(  3\right)  $ representation embedded in SO$\left(  \bar{D}\right)
,$ compatible with $\gamma_{\alpha\beta}^{i},$ should the components of the
SO$\left(  \bar{D}\right)  $ vector $\mathbf{\bar{r}}^{\alpha}$ be classified?
As a first try, consider $\bar{D}=3.$ The relation $\mathbf{r}^{i}=\frac{1}%
{2}\mathbf{\bar{r}}\gamma^{i}\mathbf{\bar{r}}$ indicates that the product of
two $\mathbf{\bar{r}}$ vectors is desired to be a vector of SO$\left(
3\right)  $, but $\bar{r}\gamma^{i}\bar{r}$ is necessarily a symmetric product
of two $\bar{r}$ vectors which yields SO(3) spin either 0 or 2 but not 1
(vector). Therefore, if (\ref{canon34}) which is the only form consistent with
Eq.\ (\ref{eqn:sub}), is adopted as part of the canonical transformation,
$\bar{D}$ cannot be 3.\footnote{There are other forms of canonical
transformations between these two systems predicted by 2T-physics that are
more general than (\ref{eqn:sub}) (see Section \ref{general}). However, in
this section we are focusing on dualities consistent with the radial duality
in (\ref{eqn:sub}).}

In search for $\bar{D},$ we observe that the phase space $\left(
\mathbf{r}^{i}\mathbf{,p}^{i}\right)  $ for the Hatom$_{3}$ is in the vector
basis of SO$\left(  3\right)  ,$ while the spinor basis (which has the
smallest dimension possible) is the doublet $Z=\left(
%TCIMACRO{\QATOP{z_{1}}{z_{2}}}%
%BeginExpansion
\genfrac{}{}{0pt}{}{z_{1}}{z_{2}}%
%EndExpansion
\right)  $ of SU$\left(  2\right)  =$ SO$\left(  3\right)  .$ The spinor basis
is complex so it contains 4 real numbers. Therefore, for (\ref{canon34}) to
work for the smallest possible $\bar{D}$, we must take $\bar{D}=4$ for the
phase space vectors $\left(  \mathbf{\bar{r},\bar{p}}\right)  $ of the
HOsc$_{4}$. This suggests to rearrange the 4 real components of $\mathbf{\bar
{r},}$ which is a vector of SO$\left(  4\right)  ,$ into two complex numbers
of an SO$\left(  3\right)  $ spinor $Z=\left(
%TCIMACRO{\QATOP{z_{1}}{z_{2}}}%
%BeginExpansion
\genfrac{}{}{0pt}{}{z_{1}}{z_{2}}%
%EndExpansion
\right)  ,$ and similarly for $\mathbf{\bar{p}.}$ Then, using the 3 Pauli
matrices $\sigma^{i},$ we can write an SU$\left(  2\right)  =$ SO$\left(
3\right)  $ covariant relation
\begin{equation}%
\begin{array}
[c]{l}%
\mathbf{r}^{i}\mathbf{=}Z^{\dagger}\frac{\sigma^{i}}{2}Z,\;\text{with }%
z_{1}=\bar{r}_{4}+i\bar{r}_{3},\;z_{2}=-\bar{r}_{2}+i\bar{r}_{1},\\
\mathbf{r}\cdot\mathbf{r=}\left(  Z^{\dagger}\frac{\sigma^{i}}{2}Z\right)
\left(  Z^{\dagger}\frac{\sigma^{i}}{2}Z\right)  =\left(  \frac{Z^{\dagger}%
Z}{2}\right)  ^{2}=\left(  \frac{\mathbf{\bar{r}}\cdot\mathbf{\bar{r}}}%
{2}\right)  ^{2},
\end{array}
\label{rZZ}%
\end{equation}
The second line yields $\left\vert \mathbf{r}\right\vert =\frac{\left\vert
\mathbf{\bar{r}}\right\vert ^{2}}{2},~$consistent with radial duality
(\ref{eqn:sub}), including the extra factor of $1/2$ that emerged in
(\ref{RandAngle}). The first line of (\ref{rZZ}) can be rewritten in the form
(\ref{canon34}) to derive three 4$\times4$ real symmetric $\gamma_{\alpha
\beta}^{i}$ matrices with $\alpha,\beta=1,2,3,4$. Furthermore, from
(\ref{rZZ}) we obtain the relation between the angular coordinates
$\mathbf{\hat{r}}$\textbf{ }for the Hatom$_{3}$ and the angular coordinates
$\widehat{\mathbf{\bar{r}}}$\textbf{ }for the HOsc$_{4}$%
\begin{equation}
\mathbf{\hat{r}}^{i}=\widehat{\mathbf{\bar{r}}}\gamma^{i}\widehat
{\mathbf{\bar{r}}}=\hat{Z}^{\dagger}\sigma^{i}\hat{Z},\text{ with }\hat
{Z}\equiv\sqrt{2}\frac{Z\left(  \mathbf{\bar{r}}\right)  }{\left\vert
\mathbf{\bar{r}}\right\vert }=\sqrt{2}Z\left(  \widehat{\mathbf{\bar{r}}%
}\right)  , \label{angular34}%
\end{equation}
where the doublet $\hat{Z}$ contains only angular coordinates, thus
generalizing the radial duality relation (\ref{eqn:sub}) by including the
angular relation. This embeds the angular spatial rotations SO$\left(
3\right)  $ of the Hatom$_{3}$ into the spatial rotations SO$\left(  4\right)
$ of the HOsc$_{4}.$We will give more clarifying details about this embedding
below, but first let's complete and verify the canonical transformation.

In addition to the $Z=\left(
%TCIMACRO{\QATOP{z_{1}}{z_{2}}}%
%BeginExpansion
\genfrac{}{}{0pt}{}{z_{1}}{z_{2}}%
%EndExpansion
\right)  $ doublet we introduce a doublet of canonical conjugates $\Pi=\left(
%
%TCIMACRO{\QATOP{\pi_{1}}{\pi_{2}}}%
%BeginExpansion
\genfrac{}{}{0pt}{}{\pi_{1}}{\pi_{2}}%
%EndExpansion
\right)  ,$ to write the second half of the canonical transformation
\begin{equation}
\mathbf{p}_{i}=\frac{Z^{\dagger}\sigma_{i}\Pi+\Pi^{\dagger}\sigma_{i}%
Z}{2Z^{\dagger}Z},\;\text{with},\pi_{1}=\bar{p}_{4}+i\bar{p}_{3},\;\pi
_{2}=-\bar{p}_{2}+i\bar{p}_{1}. \label{pZPi}%
\end{equation}
Using the HOsc$_{4}$ commutation rules $\left[  \mathbf{\bar{r}}^{\alpha
}\mathbf{,\bar{p}}^{\beta}\right]  =i\delta^{\alpha\beta},$ we see that%
\begin{equation}
\left[  Z_{a},\Pi_{b}^{\dagger}\right]  =\left[  Z_{a}^{\dagger},\Pi
_{b}\right]  =i\delta_{ab},\;\text{while ~}\left[  Z_{a},\Pi_{b}\right]
=\left[  Z_{a}^{\dagger},\Pi_{b}^{\dagger}\right]  =0\text{, for }a,b=1,2.
\label{canonZPi}%
\end{equation}
Now we verify that Eqs. (\ref{rZZ},\ref{pZPi}), that are equivalent to
(\ref{canon34}), amount to a canonical transformation, by computing $\left[
\mathbf{r}^{i}\mathbf{,p}^{j}\right]  $ by using only the commutators
(\ref{canonZPi}) and obtain $\left[  \mathbf{r}^{i}\mathbf{,p}^{j}\right]
=i\delta^{ij}$ as follows:
\begin{equation}
\left[  \mathbf{r}^{i}\mathbf{,p}^{j}\right]  =\left[  \left(  Z^{\dagger
}\frac{\sigma^{i}}{2}Z\right)  ,\frac{Z^{\dagger}\sigma_{i}\Pi+\Pi^{\dagger
}\sigma_{i}Z}{2Z^{\dagger}Z}\right]  =i\delta^{ij}.
\end{equation}

Now we give a group-theoretical analysis of the duality Hatom$_{3}%
\leftrightarrow$ HOsc$_{4}$ via the canonical transformation above. Given the
reasoning at the beginning of Section \ref{dualitySec}, and using $\left(
D=3,\bar{D}=4\right)  $, we infer that this duality involves the non-compact
groups SO$\left(  4,2\right)  $ and Sp$\left(  8,R\right)  $ that classify the
corresponding spectra as discussed in Sections \ref{sec:dual} and
\ref{sec:noncompact}$,$ provided these obey a subgroup relationship. Indeed in
the spinor basis SO$\left(  4,2\right)  =$ SU$\left(  2,2\right)  $ is easily
recognized as a subgroup of Sp$\left(  8,R\right)  \supset$ SO$\left(
4,2\right)  \otimes U\left(  1\right)  .$ Then, in the sense of
Eq.\ (\ref{H+HO}), the entire spectrum of Hatom$_{3},$ in the form of
SO$\left(  1,2\right)  $ towers with angular momentum $l$ for SO$\left(
3\right)  ,$ should correspond to part of the HOsc$_{4}$ spectrum in the form
of Sp$\left(  2,R\right)  $ towers with angular momentum $\bar{l}$ for
SO$\left(  4\right)  .$ Furthermore, the condition for identical SO$\left(
1,2\right)  =$ Sp$\left(  2,R\right)  $ representations for the towers, namely
$j\left(  l\right)  =\bar{j}\left(  \bar{l}\right)  $ given in (\ref{identify}%
), must also be satisfied. The last requirement boils down to simply
\begin{equation}
j\left(  l\right)  =l,\;\bar{j}\left(  \bar{l}\right)  =\frac{1}{2}\bar
{l};\text{ }\Rightarrow\text{ }\bar{l}=2l. \label{JlL}%
\end{equation}
This means the entire Hatom$_{3}$ spectrum, in the SO$\left(  4,2\right)  $
singleton representation with Casimir eigenvalues $C_{k}\left(  D=3\right)  $
in Eq.\ (\ref{CasimirsSOd2}), must correspond to a \textit{subset} of the
states of the HOsc$_{4}$ in the even sector with $\left(  \bar{n}%
_{\text{even}},\bar{l}_{\text{even}}\right)  ,$ i.e., within the \textit{even}
singleton representation of Sp$\left(  8,R\right)  $ with Casimir eigenvalues
$\bar{C}_{k}\left(  \bar{D}=4\right)  $ in Eq.\ (\ref{CasimirsSp2D}). This is
then consistent with SO$\left(  4,2\right)  =$ SU$\left(  2,2\right)  \subset$
Sp$\left(  8,R\right)  .$
\begin{equation}%
\begin{tabular}
[c]{||c||llllllll||}\hline\hline
$\overset{\text{Hatom}_{3}}{\text{{\scriptsize SO(4)}}{\scriptsize \supset
}\text{{\scriptsize SO(3)}}}$ & $\underset{\downarrow}{~{\large n}%
~~~}^{{\Large l\rightarrow}}$ & $0$ & $1$ & $2$ & $3$ & $4$ & $5$ &
$~6~\cdots$\\\cline{1-1}\cline{3-9}%
$\vdots$ & \multicolumn{1}{||c}{$\vdots$} & \multicolumn{1}{|c}{$\vdots$} &
\multicolumn{1}{c}{$\vdots$} & \multicolumn{1}{c}{$\vdots$} &
\multicolumn{1}{c}{$\vdots$} & \multicolumn{1}{c}{$\vdots$} &
\multicolumn{1}{c}{$\vdots$} & \multicolumn{1}{c||}{$\vdots$}\\
$49$ & \multicolumn{1}{||c}{$7$} & \multicolumn{1}{|c}{$1$} &
\multicolumn{1}{c}{$3$} & \multicolumn{1}{c}{$5$} & \multicolumn{1}{c}{$7$} &
\multicolumn{1}{c}{$9$} & \multicolumn{1}{c}{$11$} & \multicolumn{1}{c||}{$13$%
}\\
$36$ & \multicolumn{1}{||c}{$6$} & \multicolumn{1}{|c}{$1$} &
\multicolumn{1}{c}{$3$} & \multicolumn{1}{c}{$5$} & \multicolumn{1}{c}{$7$} &
\multicolumn{1}{c}{$9$} & \multicolumn{1}{c}{$11$} & \multicolumn{1}{c||}{}\\
$25$ & \multicolumn{1}{||c}{$5$} & \multicolumn{1}{|c}{$1$} &
\multicolumn{1}{c}{$3$} & \multicolumn{1}{c}{$5$} & \multicolumn{1}{c}{$7$} &
\multicolumn{1}{c}{$9$} & \multicolumn{1}{c}{} & \multicolumn{1}{c||}{}\\
$16$ & \multicolumn{1}{||c}{$4$} & \multicolumn{1}{|c}{$1$} &
\multicolumn{1}{c}{$3$} & \multicolumn{1}{c}{$5$} & \multicolumn{1}{c}{$7$} &
\multicolumn{1}{c}{} & \multicolumn{1}{c}{} & \multicolumn{1}{c||}{}\\
$9$ & \multicolumn{1}{||c}{$3$} & \multicolumn{1}{|c}{$1$} &
\multicolumn{1}{c}{$3$} & \multicolumn{1}{c}{$5$} & \multicolumn{1}{c}{} &
\multicolumn{1}{c}{} & \multicolumn{1}{c}{} & \multicolumn{1}{c||}{}\\
$4$ & \multicolumn{1}{||c}{$2$} & \multicolumn{1}{|c}{$1$} &
\multicolumn{1}{c}{$3$} & \multicolumn{1}{c}{} & \multicolumn{1}{c}{} &
\multicolumn{1}{c}{} & \multicolumn{1}{c}{} & \multicolumn{1}{c||}{}\\
$1$ & \multicolumn{1}{||c}{$1$} & \multicolumn{1}{|c}{$1$} &
\multicolumn{1}{c}{} & \multicolumn{1}{c}{} & \multicolumn{1}{c}{} &
\multicolumn{1}{c}{} & \multicolumn{1}{c}{} & \multicolumn{1}{c||}{}%
\\\hline\hline
\end{tabular}
\ \ \ \ \ \ \ \ ~~%
\begin{tabular}
[c]{||c||lccccccc||}\hline\hline
$\overset{\text{HOsc}_{4}}{\text{{\scriptsize SU(4)}}{\scriptsize \supset
}\text{{\scriptsize SO(4)}}}$ & $~\underset{\downarrow}{{\large \bar{n}}%
}^{~~~~{\Large \bar{l}\rightarrow}}$ & \textbf{0} & {\scriptsize 1} &
\textbf{2} & {\scriptsize 3} & \textbf{4} & {\scriptsize 5} & ~\textbf{6~}%
$\cdots$\\\cline{1-1}\cline{1-1}\cline{3-9}%
$\vdots$ & \multicolumn{1}{||c}{$\vdots$} & \multicolumn{1}{|c}{$\vdots$} &
$\vdots$ & $\vdots$ & $\vdots$ & $\vdots$ & $\vdots$ & $\vdots$\\
\textbf{84} & \multicolumn{1}{||c}{\textbf{6}} &
\multicolumn{1}{|c}{\textbf{1}} &  & \textbf{9} &  & \textbf{25} &  &
\textbf{49}\\
{\scriptsize 56} & \multicolumn{1}{||c}{{\scriptsize 5}} &
\multicolumn{1}{|c}{} & {\scriptsize 4} &  & {\scriptsize 16} &  &
{\scriptsize 36} & \\
\textbf{35} & \multicolumn{1}{||c}{\textbf{4}} &
\multicolumn{1}{|c}{\textbf{1}} &  & \textbf{9} &  & \textbf{25} &  & \\
{\scriptsize 20} & \multicolumn{1}{||c}{{\scriptsize 3}} &
\multicolumn{1}{|c}{} & {\scriptsize 4} &  & {\scriptsize 16} &  &  & \\
\textbf{10} & \multicolumn{1}{||c}{\textbf{2}} &
\multicolumn{1}{|c}{\textbf{1}} &  & \textbf{9} &  &  &  & \\
{\scriptsize 4} & \multicolumn{1}{||c}{{\scriptsize 1}} &
\multicolumn{1}{|c}{} & {\scriptsize 4} &  &  &  &  & \\
\textbf{1} & \multicolumn{1}{||c}{\textbf{0}} & \multicolumn{1}{|c}{\textbf{1}%
} &  &  &  &  &  & \\\hline\hline
\end{tabular}
\ \ \ \ \ \ \label{HHO2}%
\end{equation}

To see this more clearly, we display in Eq.\ (\ref{HHO2}) the full spectra of
Hatom$_{3}$ and HOsc$_{4}$ similar to Eq.\ (\ref{H+HO}) with all their states
included. The entry in each pigeon hole labelled by $\left(  n,l\right)  $ for
Hatom$_{3}$ is the dimension of the SO$\left(  3\right)  $ $l$-representation,
i.e., $\left(  2l+1\right)  $, while the entry in each pigeon hole labelled by
$\left(  \bar{n},\bar{l}\right)  $ for HOsc$_{4}$ is the dimension of the
SO$\left(  4\right)  $ $\bar{l}$-representation given in (\ref{NlD}), i.e.,
$\left(  \bar{l}+1\right)  ^{2}$. Similarly, the entries at fixed values of
$n$ or $\bar{n}$ at the leftmost columns of each table are the dimensions of
the hidden symmetries of the Hamiltonians, SO$\left(  4\right)  $ for the
Hatom$_{3}$ in the $\left(  n-1\right)  $-representation, i.e., dimension
$n^{2}$ , and SU$\left(  4\right)  $ for the HOsc$_{4}$ in the totally
symmetric $\bar{n}$-representation, i.e., dimension $\frac{\left(  \bar
{n}+3\right)  !}{\bar{n}!3!}$. These correspond to the total degeneracies of
the energy levels of the respective systems, as discussed in Section
\ref{sec:noncompact} for general $D,\bar{D}$.

We still need to indicate the \textit{specific subset} of HOsc$_{4}$-even
states that correspond to all the states of Hatom$_{3}.$ We do this by
comparing each pair of towers, at $\left(  l=j,\bar{l}=2j\right)  $ for
$j=0,1,2,\cdots,$ since we have already guaranteed that the towers $\left(
l=j,\bar{l}=2j\right)  $ have the same SO$\left(  1,2\right)  _{j}=$
Sp$\left(  2,R\right)  _{j}$ representation as in (\ref{identify}), and that
this is a common subgroup of SO$\left(  4,2\right)  \subset$ Sp$\left(
8,R\right)  .$ For the pair of towers $\left(  l=j,\bar{l}=2j\right)  $ the
$\left(  \text{SO}\left(  3\right)  _{j},\text{SO}\left(  4\right)
_{2j}\right)  $ dimensions are generally different as seen in the tables in
Eq.\ (\ref{H+HO}). For example for $l=j=2$ the tables show that there are 5
Hatom$_{3}$ $l=2$ towers, versus for the same $j=2,$ there are 25 HOsc$_{4}$
$\bar{l}=4$ towers. Hence for $j=2,$ only 5 linear combinations of the 25
HOsc$_{4}$ towers correspond to the 5 Hatom$_{3}$ towers.

To figure out methodically the subset of HOsc$_{4}$ towers that correspond to
the Hatom$_{3}$ towers at each $\bar{l}=2l$ we must identify the gauge
symmetry inherent in the canonical transformation given above in Eqs.
(\ref{rZZ},\ref{pZPi},\ref{canonZPi}). This gauge symmetry must commute with
the SO$\left(  3\right)  \subset$ SO$\left(  4\right)  $ embedding provided by
Eq.\ (\ref{rZZ}), on which we now expand. To do so, note that the spinor
$Z=\left(
%TCIMACRO{\QATOP{z_{1}}{z_{2}}}%
%BeginExpansion
\genfrac{}{}{0pt}{}{z_{1}}{z_{2}}%
%EndExpansion
\right)  $ is part of a 2$\times2$ matrix $M$ that is constructed from the
4-component vector $\mathbf{\bar{r}}$%
\begin{equation}
M=\left(  i\sigma_{1}\bar{r}_{1}+i\sigma_{2}\bar{r}_{2}+i\sigma_{3}\bar{r}%
_{3}+\bar{r}_{4}\right)  =\left(
\begin{array}
[c]{cc}%
\bar{r}_{4}+i\bar{r}_{3} & ~\bar{r}_{2}+i\bar{r}_{1}\\
-\bar{r}_{2}+i\bar{r}_{1} & ~\bar{r}_{4}-i\bar{r}_{3}%
\end{array}
\right)  =\left(
\begin{array}
[c]{cc}%
z_{1} & -z_{2}^{\ast}\\
z_{2} & z_{1}^{\ast}%
\end{array}
\right)  . \label{M}%
\end{equation}
The SO$\left(  4\right)  $ rotations on $\mathbf{\bar{r}}^{\alpha}$ amount now
to SU$\left(  2\right)  _{L}\otimes$ SU$\left(  2\right)  _{R}=$ SO$\left(
4\right)  $ transformations on the left and right sides of the matrix $M.$ The
SO$\left(  3\right)  $ transformation on the doublet spinor $Z=\left(
%TCIMACRO{\QATOP{z_{1}}{z_{2}}}%
%BeginExpansion
\genfrac{}{}{0pt}{}{z_{1}}{z_{2}}%
%EndExpansion
\right)  $ is the SU$\left(  2\right)  _{L}$ applied on the left side of the
matrix $M$ as seen clearly from (\ref{M}). The second doublet $Z_{c}=\left(
%TCIMACRO{\QATOP{-z_{2}^{\ast}}{z_{1}^{\ast}}}%
%BeginExpansion
\genfrac{}{}{0pt}{}{-z_{2}^{\ast}}{z_{1}^{\ast}}%
%EndExpansion
\right)  $ is the SU$\left(  2\right)  _{L}$ \textquotedblleft charge
conjugate\textquotedblright\ of the first doublet, $Z_{c}=\left(  -i\sigma
_{2}Z^{\ast}\right)  ,$ and transforms also as a doublet under SU$\left(
2\right)  _{L}.$ The SU$\left(  2\right)  _{R}$ interchanges the two
SU$\left(  2\right)  _{L}$ doublets $\left(  Z,Z_{c}\right)  $ and this
transformation commutes with SU$\left(  2\right)  _{L}=$ SO$\left(  3\right)
.$ The matrix $M$ satisfies $MM^{\dagger}=M^{\dagger}M=\mathbf{\bar{r}%
\cdot\bar{r}}.$ The canonical transformation (\ref{rZZ}) for the position
coordinates may now be rewritten in terms of $M$ in the form:%
\begin{equation}
\mathbf{r\cdot\sigma=}M\frac{\sigma_{3}}{2}M^{\dagger}\text{ \ or
\ }\mathbf{r}^{i}=Tr\left(  \frac{\sigma^{i}}{2}M\frac{\sigma_{3}}%
{2}M^{\dagger}\right)  =\frac{1}{2}\left(  Z^{\dagger}\frac{\sigma^{i}}%
{2}Z-Z_{c}^{\dagger}\frac{\sigma^{i}}{2}Z_{c}\right)  =Z^{\dagger}\frac
{\sigma^{i}}{2}Z. \label{rM}%
\end{equation}
Hence the \textit{gauge symmetry} that commutes with SU$\left(  2\right)
_{L}=$ SO$\left(  3\right)  $ must be part of SU$\left(  2\right)  _{R}$ since
SU$\left(  2\right)  _{R}$ commutes with SU$\left(  2\right)  _{L}.$ The form
of the canonical transformation in Eq.\ (\ref{rM}) shows clearly that the
SU$\left(  2\right)  _{R}$ is broken down to U$\left(  1\right)  $ due to the
$\sigma_{3}$ insertion on the right side of $M.$ Therefore the gauge symmetry
operator is just the third generator of SU$\left(  2\right)  _{R}$ that we
will call $J_{3R}.$ We can build this operator as a Hermitian operator that
commutes with the expressions for the canonical transformation in
(\ref{rZZ},\ref{pZPi})$.$ That is, we can verify the gauge-invariance of the
canonical transformation by computing
\begin{equation}
J_{3R}=\frac{1}{2i}\left(  Z^{\dagger}\Pi-\Pi^{\dagger}Z\right)  ,\;\left[
\mathbf{r,}J_{3R}\right]  =\left[  \mathbf{p,}J_{3R}\right]  =0.
\end{equation}
where $\left(  \mathbf{r,p}\right)  $ are written in terms of $\left(
Z,\Pi,Z^{\dagger},\Pi^{\dagger}\right)  $ before performing the commutators.
It is easy to see that the finite transformation generated by $J_{3R}$ amounts
to overall phase transformations on the doublets%
\begin{equation}
\left(  Z,\Pi,Z^{\dagger},\Pi^{\dagger}\right)  \rightarrow\left(  e^{i\alpha
}Z,e^{i\alpha}\Pi,e^{-i\alpha}Z^{\dagger},e^{-i\alpha}\Pi^{\dagger}\right)  ,
\end{equation}
which is a symmetry of the the canonical transformation in (\ref{rZZ}%
,\ref{pZPi},\ref{canonZPi}). This $U\left(  1\right)  $ is precisely the same
$U\left(  1\right)  $ that appears in SO$\left(  4,2\right)  \otimes$
U$\left(  1\right)  \subset$ Sp$\left(  8,R\right)  $, so it commutes with all
properties of the hidden SO$\left(  4,2\right)  $ symmetry of the Hatom$_{3}.$
The vanishing of the operator $J_{3R}$ is the constraint that must be applied
on the quantum states of the HOsc$_{4}$ in order to project out the
gauge-dependent states and keep only the gauge-invariant \textquotedblleft
physical states\textquotedblright. Only the \textquotedblleft physical
states\textquotedblright\ of the HOsc$_{4}$ are dual to all the Hatom$_{3}$
quantum states.

To see this projection and identification of the \textquotedblleft physical
states\textquotedblright\ explicitly, one may decompose the SO$\left(
4\right)  $ traceless symmetric tensors, whose dimensions for the $\bar{l}%
$-representations appear in (\ref{HHO2}). We remark that in this case the
$\bar{l}$-representation can also be presented as the $\left(  \frac{\bar{l}%
}{2},\frac{\bar{l}}{2}\right)  $ representation of SU$\left(  2\right)
_{L}\otimes$ SU$\left(  2\right)  _{R}$ = SO$\left(  4\right)  $, labelled by
quantum numbers $|j_{L},m_{L};j_{R},m_{R}\rangle,$ with $-j_{L}\leq m_{L}\leq
j_{L}$ and $-j_{R}\leq m_{R}\leq j_{R}$ and where $j_{L}=j_{R}=\frac{\bar{l}%
}{2}=l.$ So the multiplets $|l,m_{L};l,m_{R}\rangle$ have dimension, $\left(
2j_{L}+1\right)  \left(  2j_{R}+1\right)  =\left(  2l+1\right)  ^{2}=\left(
\bar{l}+1\right)  ^{2},$ that matches the entries in the pigeon holes in the
right-side table in Eq.\ (\ref{HHO2}). The gauge-invariant states are only
those that have $J_{3R}\rightarrow m_{R}=0,$ hence the physical states are
$|l,m_{L},l,0\rangle$ while all others with $m_{L}\neq0$ are to be projected
out by demanding gauge-invariant states. Now we see that the $|l,m_{L}%
,l,0\rangle$ HOsc$_{4}$ states are in one-to-one correspondence with the
$|l,m\rangle$ states of the Hatom$_{3}.$ This proves clearly that we have
constructed the duality transformation correctly.

It is also interesting to illustrate the role of the gauge symmetry at the
classical level (i.e., not keeping track of operator ordering) as follows. We
have already established in (\ref{rZZ}) that $\left\vert \mathbf{r}\right\vert
=\frac{\left\vert \mathbf{\bar{r}}\right\vert ^{2}}{2}$; we now compute also
all the SO$\left(  3\right)  $ dot products of the Hatom$_{3}$ phase space in
terms of the HOsc$_{4}$ phase space as follows:%
\begin{equation}%
\begin{array}
[c]{l}%
\mathbf{r}\cdot\mathbf{p=}Z^{\dagger}\frac{\sigma_{i}}{2}Z\frac{Z^{\dagger
}\sigma_{i}\Pi+\Pi^{\dagger}\sigma_{i}Z}{2Z^{\dagger}Z}=\frac{Z^{\dagger}%
\Pi+\Pi^{\dagger}Z}{4}=\frac{\mathbf{\bar{r}\cdot\bar{p}}}{4},\\
\mathbf{p}^{2}=\frac{Z^{\dagger}\sigma_{i}\Pi+\Pi^{\dagger}\sigma_{i}%
Z}{2Z^{\dagger}Z}\frac{Z^{\dagger}\sigma_{i}\Pi+\Pi^{\dagger}\sigma_{i}%
Z}{2Z^{\dagger}Z}=\frac{\Pi^{\dagger}\Pi}{Z^{\dagger}Z}+\frac{\left(
Z^{\dagger}\Pi-\Pi^{\dagger}Z\right)  ^{2}}{4\left(  Z^{\dagger}Z\right)
^{2}}=\frac{\mathbf{\bar{p}}^{2}}{\mathbf{\bar{r}}^{2}}-\frac{\left(
J_{3R}\right)  ^{2}}{\left(  \mathbf{\bar{r}}^{2}\right)  ^{2}}.
\end{array}
\end{equation}
Then compute the classical Hatom$_{3}$ Hamiltonian in terms of the HOsc$_{4}$
phase space by using the relations above,
\begin{equation}
H_{\text{Hatom}}=\left(  \frac{1}{2}\mathbf{p}^{2}-\frac{Z}{r}\right)
=\frac{1}{\mathbf{\bar{r}}^{2}}\left(  \frac{1}{2}\mathbf{\bar{p}}^{2}%
-\frac{\left(  J_{3R}\right)  ^{2}}{2\mathbf{\bar{r}}^{2}}-2Z\right)  .
\end{equation}
At a fixed bound energy state (planetary-type classical orbits) take
$H_{\text{Hatom}}\rightarrow-\left\vert E\right\vert ,$ and rewrite the above
relation by multiplying both sides by $\bar{r}^{2}$ and re-arranging, to
obtain
\begin{equation}
\left(  \frac{1}{2}\mathbf{\bar{p}}^{2}+\left\vert E\right\vert \mathbf{\bar
{r}}^{2}\right)  -\frac{\left(  J_{3R}\right)  ^{2}}{2\mathbf{\bar{r}}^{2}%
}=2Z.
\end{equation}
For gauge-invariant states of the HOsc$_{4}$ we must set $J_{3R}\rightarrow0,$
resulting in a bound energy state of the HOsc$_{4},$ $\left(  \frac{1}%
{2}\mathbf{\bar{p}}^{2}+\left\vert E\right\vert \mathbf{\bar{r}}^{2}\right)
=2Z,$ that is dual to a bound energy state of the Hatom$_{3}.$ In this way we
see again the role of the gauge symmetry, now at the classical level.

Next, we can also keep track of the angular behavior in those orbits. From
(\ref{M},\ref{rM}) we can write the relation between the unit vectors
$\mathbf{\hat{r}}$ and $\widehat{\mathbf{\bar{r}}},$ in $D=3$ and $\bar{D}=4$
dimensions respectively, as a unitary transformation $U$:%
\begin{equation}
\mathbf{\sigma\cdot\hat{r}=}U\sigma_{3}U^{\dagger},\;U=\widehat{\mathbf{\bar
{r}}}_{4}+i\sigma_{1}\widehat{\mathbf{\bar{r}}}_{1}+i\sigma_{2}\widehat
{\mathbf{\bar{r}}}_{2}+i\sigma_{3}\widehat{\mathbf{\bar{r}}}_{3},\;
\label{angles34}%
\end{equation}
that satisfies $UU^{\dagger}=U^{\dagger}U=\widehat{\mathbf{\bar{r}}}%
\cdot\widehat{\mathbf{\bar{r}}}=1.$ We parametrize the $\mathbf{\hat{r}}$ unit
vector in 3 dimensions in the usual way%
\begin{equation}
\mathbf{\hat{r}}_{1}=\sin\theta\cos\phi,\;\mathbf{\hat{r}}_{2}=\sin\theta
\sin\phi,\;\mathbf{\hat{r}}_{3}=\cos\theta.
\end{equation}
With some hindsight, we parametrize the components of $\widehat{\mathbf{\bar
{r}}}_{\alpha}$ of the unit vector in 4 dimensions as follows:%
\begin{equation}%
\begin{array}
[c]{l}%
\widehat{\mathbf{\bar{r}}}_{1}=\sin\bar{\theta}\sin\left(  \bar{\phi}%
+\bar{\chi}\right)  ,\;\widehat{\mathbf{\bar{r}}}_{2}=\sin\bar{\theta}%
\cos\left(  \bar{\phi}+\bar{\chi}\right)  ,\\
\widehat{\mathbf{\bar{r}}}_{3}=\cos\bar{\theta}\sin\left(  \bar{\chi}%
-\bar{\phi}\right)  ,\;\widehat{\mathbf{\bar{r}}}_{4}=\cos\bar{\theta}%
\cos\left(  \bar{\chi}-\bar{\phi}\right)  ,
\end{array}
\end{equation}
so that the SU$\left(  2\right)  $ unitary 2$\times2$ matrix $U$ takes the
following form:%
\begin{equation}
U=\left(
\begin{array}
[c]{ll}%
\cos\bar{\theta}~e^{i\left(  \bar{\chi}-\bar{\phi}\right)  } & -\sin
\bar{\theta}~e^{-i\left(  \bar{\chi}+\bar{\phi}\right)  }\\
\sin\bar{\theta}~e^{i\left(  \bar{\chi}+\bar{\phi}\right)  } & ~~\cos
\bar{\theta}~e^{-i\left(  \bar{\chi}-\bar{\phi}\right)  }%
\end{array}
\right)  =e^{-i\bar{\phi}\sigma_{3}}e^{i\bar{\theta}\sigma_{2}}e^{i\bar{\chi
}\sigma_{3}}%
\end{equation}
Now computing explicitly, $U\sigma_{3}U^{\dagger}=\left(
%TCIMACRO{\QATOP{\cos2\bar{\theta}}{e^{i2\bar{\phi}}\sin2\bar{\theta}}}%
%BeginExpansion
\genfrac{}{}{0pt}{}{\cos2\bar{\theta}}{e^{i2\bar{\phi}}\sin2\bar{\theta}}%
%EndExpansion%
%TCIMACRO{\QATOP{e^{-i2\bar{\phi}}\sin2\bar{\theta}}{\cos2\bar{\theta}}}%
%BeginExpansion
\genfrac{}{}{0pt}{}{e^{-i2\bar{\phi}}\sin2\bar{\theta}}{\cos2\bar{\theta}}%
%EndExpansion
\right)  ,$ and comparing to $\left(  \mathbf{\sigma\cdot\hat{r}}\right)
=\left(
%TCIMACRO{\QATOP{\cos\theta}{e^{i\phi}\sin\theta}}%
%BeginExpansion
\genfrac{}{}{0pt}{}{\cos\theta}{e^{i\phi}\sin\theta}%
%EndExpansion%
%TCIMACRO{\QATOP{e^{-i\phi}\sin\theta}{\cos\theta}}%
%BeginExpansion
\genfrac{}{}{0pt}{}{e^{-i\phi}\sin\theta}{\cos\theta}%
%EndExpansion
\right)  $ in Eq.\ (\ref{angles34}), we find the following simple relations
between the angles $\left(  \theta,\phi\right)  $ and $\left(  \bar{\theta
},\bar{\phi},\bar{\chi}\right)  :$%
\begin{equation}
\theta=2\bar{\theta},\;\phi=2\bar{\phi}. \label{double}%
\end{equation}
Note that $\mathbf{\hat{r}}\left(  \theta,\phi\right)  $ is independent of the
angle $\bar{\chi}$ that drops out in the expression $U\sigma_{3}U^{\dagger}$
because of the gauge symmetry generated by $J_{3R}$ (i.e., phase symmetry
$U\left(  1\right)  \subset$ SU$\left(  2\right)  _{R}$). So, $\bar{\chi}$ is
the pure gauge freedom in the expression of the canonical transformation
(\ref{rZZ}). This means an infinite number of HOsc$_{4}$ closed trajectories
map to the same Hatom$_{3}$ trajectories but with correlations that involve
the double circling in the HOsc$_{4}$ versus the single circling in the
Hatom$_{3}$ as indicated by Eq.\ (\ref{double}).

\subsection{2T-physics and more general dualities in any D \label{general}}

The Appendix summarizes how 2T-physics predicts many shadow 1T systems in the
same number of spatial dimensions $D,$ and that these are dual to each other
for any pair of shadows. Each shadow in $D$ space and 1 time dimensions, being
merely a gauge choice of the Sp$\left(  2,R\right)  $ gauge symmetry,
holographically captures all the gauge-invariant details of the $\left(
\left(  D+1\right)  +2\right)  $ higher-dimensional system. Because of this
holography, the gauge invariants of the unique $\left(  \left(  D+1\right)
+2\right)  $ dimensional structure predict dualities among the multitude of
1T-physics systems $\left\{  \left(  \mathbf{r,p,}t,H\right)  \right\}  $. The
duality transformations are nothing but Sp$\left(  2,R\right)  $ gauge
transformations that take one fixed gauge to another fixed gauge, and in the
traditional 1T-physics language, these correspond to canonical transformations
between two shadows, $\left(  \mathbf{r,p,}t,H\right)  \leftrightarrow~\left(
\mathbf{\bar{r},\bar{p},}\bar{t},\bar{H}\right)  .$ We emphasize that the
meaning of time and Hamiltonian are different in the two shadows and the
canonical transformations involve the time and Hamiltonian, thus explaining
how two very different 1T-physics systems (different Hamiltonians), that are
traditionally presented as very different \textquotedblleft
physics\textquotedblright, are actually not independent of each other because
they are holographic shadows of the same system in $\left(  \left(
D+1\right)  +2\right)  $ dimensions. The usual 1T formalism hides such
relationships that are not evident but actually exist as predicted by
2T-physics. The predictions come in the form of hidden symmetries and
dualities that are also measurable features of 1T physics. Establishing the
existence of the predicted hidden symmetries and dualities in 1T-physics is
tantamount to establishing the existence of the $\left(  \left(  D+1\right)
+2\right)  $ higher-dimensional structure with its Sp$\left(  2,R\right)  $
\textit{gauge symmetry in phase space} (beyond well studied local gauge
symmetry in position space). Among these dualities, the duality for the pair
Hatom $\leftrightarrow$ HOsc is only one case among many dualities for a
multitude of similar pairs.

When $\left(  \left(  D+1\right)  +2\right)  $ is flat (a very special but
very broad case), the corresponding \textit{actions} (not Hamiltonians) for
the 1T systems all have an exact hidden global symmetry SO$\left(
D+1,2\right)  .$ This has its origins as the \textit{global} symmetry in
\textit{flat} $\left(  \left(  D+1\right)  +2\right)  $ dimensions for the 2T
action. The generators $L^{MN}$ of this global non-compact symmetry are
gauge-invariant because they commute with the SL$\left(  2,R\right)  $
generators as seen in (\ref{invariant}). So, each $L^{MN}$ is independent of
the shadow, even when it is evaluated in terms of a given shadow labelled by
$k:$ $\left(  X_{\left(  k\right)  }^{M},P_{\left(  k\right)  }^{M}\right)  ,$
$k=1,2,3,\cdots$. An infinite set of duality relations between gauge-invariant
observables of shadow $k_{1}$ and shadow $k_{2}$ are predicted by evaluating
any given function of the $L^{MN}$ in those two different shadows as in Eq.
\ (\ref{dualitiesF}):%
\begin{equation}%
\begin{array}
[c]{c}%
L_{\left(  k_{1}\right)  }^{MN}=L_{\left(  k_{2}\right)  }^{MN}=L^{MN},\\
\text{Dualities for every function }F\;\text{: }F\left(  L_{\left(
k_{1}\right)  }^{MN}\right)  =F\left(  L_{\left(  k_{2}\right)  }^{MN}\right)
.
\end{array}
\end{equation}
These are an infinite set of measurable predictions from 2T-physics for the
dynamics of 1T-physics. From these gauge-invariant predictions we can extract
the canonical transformations for the phase spaces $\left(  \mathbf{r}%
_{\left(  k_{1}\right)  }\mathbf{,p}_{\left(  k_{1}\right)  },t_{\left(
k_{1}\right)  },H_{\left(  k_{1}\right)  }\right)  \leftrightarrow~\left(
\mathbf{r}_{\left(  k_{2}\right)  }\mathbf{,p}_{\left(  k_{2}\right)
},t_{\left(  k_{2}\right)  },H_{\left(  k_{2}\right)  }\right)  .$ For
examples, see \cite{Araya:2013bca}.

Turning this relation around, the Hamiltonian $H_{\left(  k\right)  }$ of each
shadow can be expressed as some function $H_{k}\left(  L^{MN}\right)  $ of the
gauge-invariant $L^{MN}$ generators of SO$\left(  D+1,2\right)  .$ Examples of
such systems that have been explicitly discussed in the past 2T-physics
literature include Hatom, HOsc, free relativistic massless/massive particles,
particles moving in various curved spaces (including some black holes),
twistor equivalents for all these, etc. These systems appear to be disjoint in
the traditional 1T physics formalism. Actually, they are simply 1T
\textquotedblleft shadows\textquotedblright, with different meanings of 1
time,\ resulting from various gauge choices of the Sp($2,R$) symmetry, thus
embedding each shadow differently within a given representation of the
underlying global symmetry SO$\left(  D+1,2\right)  $. Furthermore, since the
Hamiltonian is a function of the gauge invariant $L^{MN},$ the spectrum of
energy states in each dual shadow system is captured in the same
infinite-dimensional unitary representation of SO$\left(  D+1,2\right)  ,$
whose quadratic, cubic, quartic and higher Hermitian Casimir operators are
predicted to have the fixed eigenvalues given in Eq.\ (\ref{CasimirsSOd2})
that identify the singleton representation. Thus, notably, the Hilbert space
of one shadow is mapped to the Hilbert space of another one by a unitary
transformation within the fixed singleton representation of SO$\left(
D+1,2\right)  .$ A further notable unification fact, unique to 2T-physics, is
that the shadows listed above, and many others, satisfy the same equations in
$\left(  D+1,2\right)  $ dimensions, namely simply $X\cdot X=P\cdot P=X\cdot
P=0,$ that's all. See (\ref{consFlat}) in the Appendix for clarification how
shadows in $D+1$ dimensions emerge just from these equations.

The $D=2$ case in Section \ref{sec:Sp4R} is a particularly clean example of
dualities in flat $\left(  \left(  D+1\right)  +2\right)  $ dimensions,
including the spinning case, because there is a clean choice of two
dimensional phase space to describe the two shadows Hatom$_{2}\leftrightarrow
$ HOsc$_{2}$ such that the respective time coordinates do not transform
$\bar{t}=t.$ Similarly, the $\left(  D,\bar{D}\right)  =\left(  3,4\right)  $
case is the simplest example in which there is a leftover gauge symmetry that
makes the respective phase spaces dual to each other. This set of examples, in
which $t=\bar{t}$, can be generalized to higher dimensions using similar
methods involving beautiful group theory as outlined in the next section. For
the more general cases in which $t\neq\bar{t}$ the canonical transformations
are more complicated, dramatic and surprising and were not expected to be
possible in 1T-physics, but they do exist. For examples of 5 shadows and
related canonical transformations see \cite{Araya:2013bca}. Among these there
are in particular two versions of dualities that relate to the Hatom
$\leftrightarrow$ HOsc duality discussed here but the details of the canonical
transformation are totally different. These cases include the dualities
Hatom$_{D}\leftrightarrow~($HOsc$_{\left(  D-1\right)  }+$1 dim phase space)
that was treated in \cite{Bars:1998pc} and the Hatom$_{D}\leftrightarrow$
HOsc$_{D}$ that is implicitly given in \cite{Araya:2013bca}.\footnote{The
nonrelativistic HOsc$_{D}$ is briefly discussed in \cite{Araya:2013bca} as a
specialized case of the more general \textquotedblleft
shadow-5\textquotedblright. See Eq.\ (59) and related discussion in
\cite{Araya:2013bca}. Thus HOsc$_{D}$ is dual to all 5 shadows, including the
Hatom$_{D}.$ The corresponding canonical transformation can be extracted in
the same manner as the other cases explicitly discussed in
\cite{Araya:2013bca}. This is a pretty complicated expression that we may
discuss in another paper.}

\section{Generalizations \label{generalizations} }

We set out searching for a principle underlying the radial duality
Eq.\ (\ref{eqn:sub}) between the hydrogen atom (radial coordinate $r$) and the
harmonic oscillator (radial coordinate $\bar{r}$) which follows from Newton's
substitution $r\propto\bar{r}^{2}$. The generalization of radial duality to
other power-law potentials and any pair of dimensions $\left(  D,\bar
{D}\right)  ,$ as given in (\ref{eqn:se}-\ref{eqn:lrelation}), showed an
important quantum mechanical restriction, $\left\vert \bar{l}+\frac{\bar{D}%
-2}{2}\right\vert =\left\vert \frac{\bar{\alpha}}{\alpha}\right\vert
\left\vert l+\frac{D-2}{2}\right\vert ,$ where the quantized orbital angular
momenta $\left(  l,\bar{l}\right)  $ and the dimensions $\left(  D,\bar
{D}\right)  $ had to be integers.

The analysis in Section \ref{sec:dual} of the quantum spectra for the
Hatom$_{D}$ and HOsc$_{\bar{D}},$ with $\left(  -\frac{\bar{\alpha}}{\alpha
}\right)  =2,$ showed the disparity between the two systems when angular
degrees of freedom are included, so a full duality consistent with Newton's
substitution, $r\propto\bar{r}^{2},$ could not be expected. However, hints did
emerge on how a full quantum duality may be possible between a subset of the
HOsc$_{\bar{D}}$ quantum states and the full Hatom$_{D}$ quantum states. The
non-compact group analysis SO$\left(  D+1,2\right)  $ and Sp$\left(  2\bar
{D},R\right)  $ of the respective spectra made it clear precisely how to
proceed to construct the full duality by using 2T-physics as the guiding
principle. To implement the 2T-physics perspective, it required the embedding
of SO$\left(  D+1,2\right)  $ into Sp$\left(  2\bar{D},R\right)  $ and
demanding the identification of the corresponding generators written in terms
of the different phase spaces $\left(  \mathbf{r,p}\right)  _{D}$ and $\left(
\mathbf{\bar{r},\bar{p}}\right)  _{\bar{D}}$ as in Eq.\ (\ref{identify}). From
this we could extract the canonical transformation that relates the two phase
spaces, thus constructing a full duality that includes all directions rather
than only the radial direction, while at the same time obtaining a remarkable
beautiful symmetry perspective of the full duality.

The explicit canonical transformation between phase spaces (\ref{canon34}) in
different dimensions, $D<\bar{D},$ was clarified by identifying a gauge
symmetry group $G_{D}$ in the canonical transformation, $\left(
\mathbf{r,p}\right)  _{D}\overset{G_{D}}{\leftrightarrow}~\left(
\mathbf{\bar{r},\bar{p}}\right)  _{\bar{D}},$ such that only the
\textit{gauge-invariant subsector} of the HOsc$_{\bar{D}}$ could be dual to
the full Hatom$_{D}.$ This gauge symmetry, that we now call $G_{D},$ showed
precisely which subset of states of HOsc$_{\bar{D}},$ that are invariant under
the gauge symmetry $G_{D},$ are dual to the full Hatom$_{D}$ spectrum.

This program was carried out explicitly in the previous sections for the
pairs, $\left(  D,\bar{D}\right)  _{G_{D}}=\left[  \left(  2,2\right)
_{\text{discrete}};\left(  3,4\right)  _{\text{U}\left(  1\right)  }\right]
.$ Here we sketch how to generalize to the case $D=5$ and then to $D=1,4.$ 

The case $D=5$ works exactly the same way as the case $\left(  3,4\right)  $.
The spinor of SO$\left(  5\right)  =$ USp$\left(  4\right)  $ is a quartet of 4
complex numbers, so this suggests to consider $\bar{D}=8$. Accordingly, we
introduce a quartet $Z$ and let the four complex numbers be constructed from
the 8 components of the real vector $\mathbf{r}^{\alpha}$ for the HOsc$_{8}.$
Introduce the charge conjugate spinor $Z_{c}\equiv CZ^{\ast},$ where $C$ is
the antisymmetric charge-conjugation matrix in spinor space which amounts to
be the invariant metric of USp$\left(  4\right)  .$ This structure guarantees
that the quartet $Z_{c}$ transforms under USp$\left(  4\right)  $ exactly the
same way as $Z.$ Now, similarly to the $\left(  3,4\right)  $ case we
construct a 4$\times2$ matrix $M=\left(  Z,Z_{c}\right)  $ and define an
SU$\left(  2\right)  $ transformation that mixes $\left(  Z,Z_{c}\right)  $
like a doublet. Hence, $M$ is now classified  as $\left(  4,2\right)  $ under
USp$\left(  4\right)  \otimes$ SU$\left(  2\right)  .$ In this way the  8 real
numbers $\mathbf{\bar{r}}^{\alpha}$ become a basis for USp$\left(  4\right)
\otimes$ SU$\left(  2\right)  $ transformations. This is consistent with the
fact that SO$\left(  8\right)  \supset$ SO$\left(  5\right)  \otimes$ SO$\left(
3\right)  ,$ not only in the vector basis but also in the SO$\left(  8\right)
$ spinor basis. Accordingly, our starting point is to re-assign $\mathbf{\bar
{r}}^{\alpha}$ to the spinor basis of SO$\left(  8\right)  $ rather than the
vector basis (recall triality in SO$\left(  8\right)  $). Thus, USp$\left(
4\right)  =$ SO$\left(  5\right)  $ will serve to classify the $D=5$ vector
$\mathbf{r}^{i}$ for the Hatom$_{5}$ and $G_{3}=$ SU$\left(  2\right)  $ will
serve as the gauge symmetry in the relation $\mathbf{r}^{I}\sim\mathbf{\bar
{r}}^{\alpha}\gamma_{\alpha\beta}^{I}\mathbf{\bar{r}}^{\beta}$ in the
canonical transformation (\ref{canon34}), with $I=1,2,3,4,5$ and
$\alpha=1,2,\cdots,8$. This $G_{3}=$ SU$\left(  2\right)  $ gauge group fits
also in the subgroup structure of the hidden symmetry noncompact groups,
Sp$\left(  16,R\right)  \supset$ SO$\left(  6,2\right)  \otimes$ SU$\left(
2\right)  ,$ as it should. With this background we are now able to use the
$4\times4$ five SO$\left(  5\right)  =$ USp$\left(  4\right)  $ gamma matrices
$\Gamma^{I}$ to construct the first half of the canonical transformation
(\ref{canon34}) for the case $\left(  D,\bar{D}\right)  _{G_{D}}=\left(
5,8\right)  _{\text{SU}\left(  2\right)  }$ as follows:
\begin{equation}
\mathbf{r}^{I}\mathbf{=}\frac{1}{8}\left(  Z^{\dagger}\Gamma^{I}%
Z+Z_{c}^{\dagger}\Gamma^{I}Z_{c}\right)  =\frac{1}{4}Z^{\dagger}\Gamma
^{I}Z.\label{rZso5}%
\end{equation}
By taking for example, $z_{1}=\bar{r}_{1}+i\bar{r}_{2},\;z_{2}=\bar{r}%
_{3}+i\bar{r}_{4},\;z_{3}=\bar{r}_{5}+i\bar{r}_{6},\;z_{4}=\bar{r}_{7}%
+i\bar{r}_{8},\;$this relation can be rewritten in the form $\mathbf{r}%
^{I}\sim\mathbf{\bar{r}}^{\alpha}\gamma_{\alpha\beta}^{I}\mathbf{\bar{r}%
}^{\beta}$. The first expression in (\ref{rZso5}) involving both $\left(
Z,Z_{c}\right)  $ displays the SU$\left(  2\right)  $ gauge symmetry, while
the simpler last form is obtained because $Z_{c}^{\dagger}\gamma^{i}%
Z_{c}=Z^{\dagger}\gamma^{i}Z,$ that can be proven by using the properties of
the gamma matrices (namely $C\Gamma^{i}$ are antisymmetric 4$\times4$
matrices$)$. Now, using the Fierz identity for SO$\left(  5\right)  $ gamma
matrices, $\Gamma_{\alpha\beta}^{i}\Gamma_{\gamma\delta}^{i}=2\left(
\delta_{\alpha\delta}\delta_{\gamma\beta}-C_{\alpha\gamma}C_{\beta\delta
}\right)  ,$ and noting $ZCZ=Z^{\dagger}CZ^{\dagger}=0$ due to the
antisymmetry of $C,$ we compute $\mathbf{r}^{2}$,
\begin{equation}
\mathbf{r\cdot r=}\frac{1}{8}\left(  Z^{\dagger}\Gamma^{i}Z\right)  \left(
Z^{\dagger}\Gamma^{i}Z\right)  =\frac{1}{4}\left(  Z^{\dagger}Z\right)
^{2}=\frac{1}{4}\left(  \mathbf{\bar{r}\cdot\bar{r}}\right)  ^{2}%
\mathbf{.}\label{r2so5}%
\end{equation}
This shows agreement with Newton's radial substitution $\left\vert
\mathbf{r}\right\vert =\mathbf{\bar{r}}^{2}/2,$ while we have included all the
angles in both the $D=5$ and $\bar{D}=8$ systems and satisfied the radial
duality condition $\left\vert \bar{l}+\frac{\bar{D}-2}{2}\right\vert
=2\left\vert l+\frac{D-2}{2}\right\vert $ with $\bar{l}-2l.$ The rest of the
canonical transformation involving the momenta is constructed through steps
parallel to the case $\left(  3,4\right)  .$ Further group-theoretical
investigation of the Hatom$_{5}$ and HOsc$_{8}$ spectra, similar to Eq.
(\ref{HHO2}), confirms that the gauge-invariant subset of the HOsc$_{8}$
spectrum exactly matches the full spectrum of the Hatom$_{5}$ spectrum.

In the cases $\left(  D,\bar{D}\right)  _{G_{D}}=\left[  \left(  2,2\right)
_{\text{discrete}};\left(  3,4\right)  _{\text{U}\left(  1\right)  };\left(
5,8\right)  _{\text{SU}\left(  2\right)  }\right]  $ note the perfect match of
the counting of gauge-invariant degrees of freedom for the HOsc$_{\bar{D}}$,
namely $\bar{d}\left(  D\right)  \equiv\left(  \bar{D}-\dim\left(
G_{D}\right)  \right)  =D,$ that is identical to the degrees of freedom of
Hatom$_{D}$. Here $\dim\left(  G_{D}\right)  $ is the number of
\textit{continuous} group parameters in the gauge group $G_{D}.$ In the
$D=2,3,5$ cases we also note that we find $\bar{D}=2\left(  D-1\right)  ,$ so
$\bar{l}=2l$ satisfies the crucial radial duality relationships, $\left\vert
\bar{l}+\frac{\bar{D}-2}{2}\right\vert =2\left\vert l+\frac{D-2}{2}\right\vert
,$ given in (\ref{eqn:lrelation}) or (\ref{j(l)}). How about other dimensions?

Let's start with $D=1,$ with the hidden symmetry of the Hatom$_{1}$ being
SO$\left(  2,2\right)  ,$ and knowing $l=0$ since there is no angular
momentum, as well as $j=0$ according to (\ref{j(l)}). Which value of $\bar{D}$
for HOsc$_{\bar{D}}$ is dual to this system? According to Eq.\ (\ref{j(l)}),
since $j=0$ for the Hatom$_{1}$ tower, we must have the $\bar{l}=0$ tower of
$\bar{j}\left(  0\right)  =\frac{\bar{D}-4}{4}=0.$ So we must have $\bar{D}%
=4$, with the 4 components of $\mathbf{\bar{r}}^{\alpha}$ arranged into a
complex doublet $Z=\left(
%TCIMACRO{\QATOP{z_{1}}{z_{2}}}%
%BeginExpansion
\genfrac{}{}{0pt}{}{z_{1}}{z_{2}}%
%EndExpansion
\right)  $ of the gauge group $G_{1}\equiv$SU$\left(  2\right)  _{L}\subset
$SO$\left(  4\right)  .$ This leads to the SO$\left(  4\right)  $ invariant
canonical transformation (\ref{canon34}), i.e., $r=\frac{1}{2}Z^{\dagger
}Z=\frac{1}{2}\left(  z_{1}^{\ast}z_{1}+z_{2}^{\ast}z_{2}\right)  =\frac{1}%
{2}\mathbf{\bar{r}}^{2},$ which is consistent with Newton's substitution
(\ref{eqn:sub}). Note that angular degrees of freedom have been included in
the canonical transformation, although trivially due to the gauge symmetry
$G_{1}\equiv$SU$\left(  2\right)  _{L}$. The $\left(  D,\bar{D}\right)
=\left(  1,4\right)  $ version of the radial duality condition $\left\vert
\bar{l}+\frac{\bar{D}-2}{2}\right\vert =2\left\vert l+\frac{D-2}{2}\right\vert
$ is also satisfied since $\bar{l}=0$ due to the SO$\left(  4\right)  $ gauge
symmetry while $l=0$ due to $D=1.$ Furthermore, the effective number of gauge
invariant degrees of freedom for HOsc$_{4},$ $\bar{d}\left(  D\right)
=\bar{D}-$dim$\left(  \text{G}_{D}\right)  =4-3=1,$ matches the number of
degrees of freedom $D=1$ for Hatom$_{1}.$

Next consider $D=4$ and analyze all the requirements of the duality we have
discussed up to now to find $\bar{D}$. The radial duality condition
(\ref{eqn:lrelation}) with $D=4$ and assuming $\bar{D}>2,$ becomes $\bar
{l}=2l+\frac{6-\bar{D}}{2}.$ The hidden symmetry of the Hatom$_{4}$ is
SO$\left(  5,2\right)  $ and we must embed this into Sp$\left(  2\bar
{D}\right)  \supset$ SO$\left(  5,2\right)  \otimes G_{4},$ where $G_{4}$ is
the gauge symmetry of the canonical transformation (\ref{canon34}), so $G_{4}$
must also satisfy SO$\left(  \bar{D}\right)  \supset$ SO$\left(  4\right)
\otimes G_{4}.$ These are severe restrictions on $\bar{D}.$ If we consider the
hint, that $\bar{l}=2l$ that worked so far in the cases $D=1,2,3,5$ might also
work for $D=4,$ then we should choose $\bar{D}=6.$ Then $G_{4}$ should satisfy
SO$\left(  6\right)  \supset$ SO$\left(  4\right)  \otimes G_{4},$ where
SO$\left(  4\right)  $ applies to the vector $\mathbf{r}^{i}$ of the
Hatom$_{4},$ while SO$\left(  6\right)  $ applies to the vector $\mathbf{\bar
{r}}^{\alpha}$ of the HOsc$_{6}.$ This pins down $G_{4}=$ U$\left(  1\right)
\otimes$ U$\left(  1\right)  $ that fits the counting of gauge-invariant
degrees of freedom in HOsc$_{6}$, namely $\bar{d}\left(  D\right)  =\bar{D}%
-$dim$\left(  \text{G}_{D}\right)  =6-2=4,$ matching $D=4$ the number of
degrees of freedom in Hatom$_{4}.$ This $G_{4}=$ U$\left(  1\right)  \otimes
$ U$\left(  1\right)  $ is consistent also with the subgroup structure of the
relevant hidden symmetry non-compact groups, Sp$\left(  12,R\right)  \supset$
SO$\left(  5,2\right)  \otimes\left(  \text{U}\left(  1\right)  \otimes
\text{U}\left(  1\right)  \right)  .$ With this information the relation
$\mathbf{r}^{i}\sim\mathbf{\bar{r}}^{\alpha}\gamma_{\alpha\beta}%
^{i}\mathbf{\bar{r}}^{\beta}$ can now be constructed by re-arranging the 6
real numbers of $\mathbf{\bar{r}}^{\alpha}$ into a complex quartet $Z$, but
anticipating that this complex quartet contains only 6 real numbers rather
than the natural 8 real numbers. We begin with the quartet spinor $W$ of
SO$\left(  6\right)  =$ SU$\left(  4\right)  $ that contains four complex
numbers $W_{a}=\left(  w_{1},w_{2},w_{3},w_{4}\right)  $ ~This is also a
spinor of USp$\left(  4\right)  =$ SO$\left(  5\right)  $ and can be used as in
Eq. (\ref{rZso5}) to construct an SO$\left(  5\right)  $ vector out of two
SO$\left(  5\right)  $ spinors, $\mathbf{r}^{I}=\frac{1}{4}W^{\dagger}%
\Gamma^{I}W$ where $I=1,2,3,4,5.$  We now focus on the SO$\left(  4\right)
=$ SU$\left(  2\right)  \otimes$ SU$\left(  2\right)  $ subgroup of SO$\left(
5\right)  =$USp$\left(  4\right)  $ and identity $I\rightarrow i=1,2,3,4$ with
the vector $\mathbf{r}^{i}$ of the Hatom$_{4}.$ Also for $I\rightarrow5$ we
impose the following constraint on the quartet $W$%
\begin{equation}
\mathbf{r}^{5}=\frac{1}{4}W^{\dagger}\Gamma^{5}W=0.
\end{equation}
The solution of this constraint is the quartet is $Z\left(  \mathbf{\bar{r}%
}\right)  \equiv W$(solution) which is parametrized by only 6 real numbers
that can be related to the six $\mathbf{\bar{r}}^{\alpha}$ of the HOsc$_{6}.$
Hence we have
\begin{equation}
\mathbf{r}^{i=1,2,3,4}=\frac{1}{4}Z^{\dagger}\left(  \mathbf{\bar{r}}\right)
\Gamma^{i=1,2,3,4}Z\left(  \mathbf{\bar{r}}\right)  ,\text{ and }Z^{\dagger
}\left(  \mathbf{\bar{r}}\right)  \Gamma^{5}Z\left(  \mathbf{\bar{r}}\right)
=0.\label{rZso4}%
\end{equation}
Clearly, this relation is covariant under the rotation group SO$\left(
4\right)  $ as a subgroup of SO$\left(  5\right)  .$ Moreover, we compute
$\mathbf{r\cdot r}=\frac{1}{16}\left(  Z^{\dagger}\Gamma^{i}Z\right)  \left(
Z^{\dagger}\Gamma^{i}Z\right)  $ by including the vanishing 5$^{th}$
component, $\mathbf{r\cdot r+}\left(  \mathbf{r}_{5}\right)  ^{2}=\frac{1}%
{16}\left[  \left(  Z^{\dagger}\Gamma^{i}Z\right)  \left(  Z^{\dagger}%
\Gamma^{i}Z\right)  +\left(  Z^{\dagger}\Gamma^{5}Z\right)  ^{2}\right]  ,$
because this allows us to use the Fierz identity as in Eq. (\ref{r2so5}) to
find
\begin{equation}
\mathbf{r\cdot r=}\frac{1}{4}\left(  Z^{\dagger}\left(  \mathbf{\bar{r}%
}\right)  Z\left(  \mathbf{\bar{r}}\right)  \right)  ^{2}=\frac{1}{4}\left(
\mathbf{\bar{r}\cdot\bar{r}}\right)  ^{2},
\end{equation}
which agrees with Newton's radial substitution $r=\bar{r}^{2}/2.$ Of course,
our transformation (\ref{rZso4}) includes all the angular variables for both
unit vectors in 4-dimensions $\mathbf{\hat{r}}$ (3 angles) and 6-dimensions
$\widehat{\mathbf{\bar{r}}}$ (5 angles).\textbf{ }To display how $Z\left(
\mathbf{\bar{r}}\right)  $ depends on the 5 angles of $\widehat{\mathbf{\bar
{r}}}$  we work in a basis in which $\Gamma^{5}$ is diagonal, $\Gamma^{5}%
=$diag$\left(  1,1,-1,-1\right)  ,$ and write the quartet $Z\left(
\mathbf{\bar{r}}\right)  $ that satisfies the required properties $Z^{\dagger
}\left(  \mathbf{\bar{r}}\right)  \Gamma^{5}Z\left(  \mathbf{\bar{r}}\right)
=0$ as follows%
\begin{equation}
Z\left(  \mathbf{\bar{r}}\right)  =\left\vert \mathbf{\bar{r}}\right\vert
e^{i\bar{\phi}}\left(
\begin{array}
[c]{l}%
\cos\bar{\theta}_{+}e^{i\bar{\chi}_{+}}\\
\sin\bar{\theta}_{+}e^{-i\bar{\chi}_{+}}\\
\cos\bar{\theta}_{-}e^{i\bar{\chi}_{-}}\\
\sin\bar{\theta}_{-}e^{-i\bar{\chi}_{-}}%
\end{array}
\right)  ,\;Z^{\dagger}\left(  \mathbf{\bar{r}}\right)  \Gamma^{5}Z\left(
\mathbf{\bar{r}}\right)  =0,\;Z^{\dagger}\left(  \mathbf{\bar{r}}\right)
Z\left(  \mathbf{\bar{r}}\right)  =\mathbf{\bar{r}\cdot\bar{r}.}%
\end{equation}
The relation $\mathbf{r}^{i=1,2,3,4}=\frac{1}{4}Z^{\dagger}\Gamma^{1,2,3,4}Z$
has a $G_{4}\equiv$ U$\left(  1\right)  \otimes$ U$\left(  1\right)  $ gauge
symmetry. The first U$\left(  1\right)  $ amounts to an overall phase
transformation on $Z\left(  \mathbf{\bar{r}}\right)  ;$ this can be used to
gauge-fix $\bar{\phi}\rightarrow0.$ The second U$\left(  1\right)  $ amounts
to a translation of $\bar{\theta}_{\pm}$ in opposite directions, $\bar{\theta
}_{\pm}\rightarrow\bar{\theta}_{\pm}\pm\alpha$, so that the sum $\left(
\bar{\theta}_{+}+\bar{\theta}_{-}\right)  $ remains invariant; this can be
used to gauge-fix $\bar{\theta}_{\pm}\rightarrow\bar{\theta}.$ Once gauge
fixed,  $Z\left(  \mathbf{\bar{r}}\right)  $ has only 4 parameters, $\left(
\bar{r},\bar{\theta},\bar{\chi}_{+},\bar{\chi}_{-}\right)  ,$ which is the
expected number of gauge invariants according to $\bar{d}\left(  D\right)
=\bar{D}-$dim$\left(  G_{D}\right)  =6-2=4,$ that matches $D=4$ for
Hatom$_{4}.$ Further study of the spectra of Hatom$_{4}$ versus HOsc$_{6}$
reveals the perfect duality between the respective spectra, once the U$\left(
1\right)  \otimes$ U$\left(  1\right)  $ gauge-invariant subset of states of
HOsc$_{6}$ are identified.

\section{Conclusion and Outlook \label{sec:con}}

We hope the outline given above is sufficient for the cases $D=1,4,5$. We have
not included the group-theoretical details for $D=1,4,5$ in the current paper
because it would take too much space, but if interest persists, we may do it
in a future publication.

We compile a list of the cases we have discussed up to now:
\begin{equation}%
\begin{tabular}
[c]{|l|l|l|l|}\hline
$D$ & $\bar{D}$ & $\underset{\left(  \frac{1}{2}\text{2}^{\frac
{{\scriptsize D}_{\text{{\scriptsize even}}}}{{\scriptsize 2}}}\text{ or
2}^{\frac{{\scriptsize D}_{\text{{\scriptsize odd}}}{\scriptsize -1}%
}{{\scriptsize 2}}}\right)  }{\text{Spinors SO(D)}}$ & $\text{Sp}\left(
2\bar{D},R\right)  \supset\text{SO}\left(  D+1,2\right)  \otimes G$\\\hline
$1$ & $4$ & $1_{\text{real}}\text{ {\scriptsize (four copies)}}$ &
$\text{Sp}\left(  8,R\right)  \supset\text{SO}\left(  2,2\right)
\otimes\text{SU}\left(  2\right)  $\\\hline
$2$ & $2$ & $1_{\text{complex}}$ & $\text{Sp}\left(  4,R\right)
\supset\text{SO}\left(  3,2\right)  \otimes\text{discrete}$\\\hline
$3$ & $4$ & $2_{\text{complex}}$ & $\text{Sp}\left(  8,R\right)
\supset\text{SO}\left(  4,2\right)  \otimes\text{U}\left(  1\right)  $\\\hline
$4$ & $6$ & $4_{\text{complex}}$ $%
%TCIMACRO{\QATOP{\text{plus one}}{\text{constraint}}}%
%BeginExpansion
\genfrac{}{}{0pt}{}{\text{plus one}}{\text{constraint}}%
%EndExpansion
$ & $\text{Sp}\left(  12,R\right)  \supset\text{SO}\left(  5,2\right)
\otimes\text{U}\left(  1\right)  \otimes\text{U}\left(  1\right)  $\\\hline
$5$ & $8$ & $4_{\text{complex}}$ & $\text{Sp}\left(  16,R\right)
\supset\text{SO}\left(  6,2\right)  \otimes\text{SU}\left(  2\right)
$\\\hline
\end{tabular}
\ \ \label{list2}%
\end{equation}
The third column gives information on the re-classification of the SO$\left(
\bar{D}\right)  $ vector \textbf{\={r}} as representations under SO$\left(
D\right)  \times G_{D}.$ In every case a spinor representation of SO$\left(
D\right)  $ is included but in the case $D=4$ we started with a complex
quartet of SO$\left(  \bar{D}=6\right)  $ and imposed constraints to reduce it
effectively to two less real parameters. Furthermore, in the case $D=1$ there
is a repetition of representations (although trivial in this case). These
should be taken as clues for how to proceed for larger dimensions $D\geq6.$

As a result of our experience with $D=1$ to $5$, we witness that $\bar{l}=2l$
applies to all the cases so far. Combined with the radial duality requirement,
$\left\vert \bar{l}+\frac{\bar{D}-2}{2}\right\vert =2\left\vert l+\frac
{D-2}{2}\right\vert ,$ we see that $\bar{D}=2\left(  D-1\right)  ,$ for $2\leq
D\leq5,$ except for the case of $D=1$ for which $\bar{D}=4.$ The pattern
$\bar{l}=2l,$ along with $\bar{D}=2\left(  D-1\right)  ,$ may be taken as a
conjecture for further investigations to determine $\bar{D}$ once $D\geq6$ is
given, but it is not necessary that this conjecture, based on $\bar{l}=2l,$
should hold at larger $D$. In any case, we have enough circumstantial evidence
to expect that there is a duality Hatom$_{D}\overset{\text{G}_{D}%
}{\leftrightarrow}$ HOsc$_{\bar{D}}$ for every dimension $D.$ This is in
harmony with the general prediction from 2T-physics.

We note that so far only the even states of the HOsc$_{\bar{D}}$ participate
in the duality. As outlined in Section \ref{hints}, a subset of the odd states
of the HOsc$_{\bar{D}}$ are supposed to be dual to the anyonic-Hatom according
to 2T-physics. The detailed analysis of this prediction is left to future research.

We generalized Newton's radial duality to a full duality in the form of a
canonical transformation consistent with 2T-physics and its hidden symmetry
SO$\left(  D+1,2\right)  $ that applies to all shadows, beyond the two
shadows, Hatom and HOsc, we discussed here. As emphasized in Section
\ref{general} 2T-physics offers other forms of canonical transformations that
are more complicated and \`{a} priori do not seem compatible with Newton's
radial duality. The basis of the phase spaces in those cases are different
(for example $D=\bar{D}$ for all $D$) but must ultimately be related to the
bases discussed in this paper by some canonical transformations, especially
after solving explicitly all the gauge constraints for the gauge groups
$G_{D}$ introduced in the current paper (since then the HOsc$_{\bar{D}}$
system in the gauge-invariant sector is reduced to a smaller phase space
effectively in $D$ dimensions). We leave the resolution of these and related
questions to future investigations.

\begin{acknowledgments}
Both authors have known Peter Freund for several decades and have both valued
his insights and broad knowledge. For one of us (JLR) he was a major reason
for moving to the University of Chicago. When IB developed his work on super
ternary algebras in late 1970's Freund and Kaplansky's 1975 paper on
superalgebras provided inspiration. We are both grateful to him for directing
us to one another to investigate the subject of this paper. JLR thanks S.
Chandrasekhar, A. Grant, and V. A. Kosteleck\'{y} for helpful discussions.
\end{acknowledgments}

\appendix

\section{Concepts of 2T-physics \label{2Tappendix}}

In this Appendix we briefly outline the essential features of 2T-physics that
are relevant for the reader to better understand the deeper spacetime meaning
of the material in the text. The features of 2T-physics are what prompted
Peter Freund to suggest to J. Rosner that the dualities encountered in the
radial Schr\"{o}dinger equation, that was originally discovered by Newton
\cite{Newton}, and later encountered while analyzing quark-antiquark bound
states \cite{Quigg:1979vr}, could be related to the more general dualities of
the shadows predicted by 2T-physics.

2T-physics is based on having an additional Sp$\left(  2,R\right)  $ gauge
symmetry \textit{acting on phase space} \cite{Bars:1998ph,Bars2Tspinning}
beyond the gauge symmetries that occur in traditional classical or quantum
mechanics, field theory \cite{Bars2TfieldTh,BarsSM,Bars2Tgravity} or string
theory (for reviews, see \cite{BarsReview1998,BarsReview2001,BarsReviewGMann}%
). This gauge symmetry adds gauge degrees of freedom to conventional
1T-physics such that all physical systems in $D$-space and $1$ time dimensions
are elevated to a covariant description in $\left(  D+1\right)  $ space and
$2$ time dimensions with one extra time and one extra space dimension in flat
or curved spacetime.\footnote{The reason for the extra 1+1 dimensions is
closely related to the number and signature of the gauge parameters in
Sp$\left(  2,R\right)  =$ SO$\left(  1,2\right)  ,$ that has 1 spacelike gauge
parameter and two timelike gauge parameters. By contrast, 1T-physics has only
1 timelike gauge parameter as recognized in the familiar $\tau$%
-reparametrization gauge symmetry in the worldline formalism. The difference
then is 1+1 additional gauge degrees of freedom in 2T-physics versus
1T-physics. Even though the number of physical spatial (hence ghost free)
degrees of freedom, $D,$ is the same in both cases, the higher-dimensional
perspective turns out to be much richer in predictions of hidden symmetries
and dualities as compared to 1T-physics. Since we mentioned $\tau
$-reparametrization, it may be interesting to note that $\tau$%
-reparametrization gauge symmetry amounts to general relativity based on local
(i.e., $\tau$-dependent$)$ translation invariance in one temporal dimension.
By contrast Sp$\left(  2,R\right)  =$ SO$\left(  1,2\right)  $ gauge symmetry
amounts to conformal gravity based on local conformal transformations on the
worldline, noting that SO$\left(  1,2\right)  $ is the conformal group in the
space of a single temporal dimension \cite{Bars:1998ph}. This is another way
to understand the extra 1+1 gauge parameters on the worldline and hence the
extra 1+1 gauge degrees of freedom as part of target phase space, and the
associated extra 1+1 constraints.} The higher-dimensional perspective is very
powerful and reveals correlations that actually exist in nature but remain
well hidden and are impossible or very hard to capture in the traditional
1T-physics formalism. However, the predictions can be verified within
1T-physics either through theoretical computations or experiments. The
dualities of the shadows that Peter Freund alluded to are part of general
predictions by 2T-physics.

\textit{Shadows} are 1T-physics systems obtained from 2T-physics by making 1+1
gauge choices (in phase space, not just position space) and solving two
constraints, thus eliminating 1+1 position and momentum degrees of freedom.
Any shadow captures holographically all the gauge-invariant physical phenomena
that occur in $\left(  D+1\right)  +2$ dimensions. For several examples of
shadows, see \cite{Araya:2013bca} and references therein to previous similar
2T gauge choices. The remaining $D$ spatial and 1 temporal dynamical phase
space degrees of freedom, including time and its canonical conjugate
Hamiltonian, are embedded in $\left(  D+1\right)  +2$ dimensional phase space
in an infinite number of geometrical or algebraic configurations that create
very different perspectives for $D+1$ dimensional observers (of
gauge-invariants) to view the phenomena that occur in $\left(  D+1\right)  +2$
spacetime. For this reason, the concepts of time and Hamiltonian are different
for the observers in different shadows. The duality of the shadows is
expressed in the language of 1T-physics via canonical transformations that
include time and Hamiltonian, thus establishing unexpected connections to one
another among many 1T-physics systems with diverse Hamiltonians. All the
connected shadows are actually gauge-fixed forms of the same unique
higher-dimensional system, and these dualities are just Sp$\left(  2,R\right)
$ gauge transformations that take one fixed gauge to another fixed gauge. All
connected shadows in 1T-physics are the same unique system in 2T-physics and
obey the same equations in $\left(  D+1\right)  +2$ dimensions. Thus,
2T-physics in $\left(  4+2\right)  $ dimensions provides an unprecedented
unification that is manifested in our experience in 3+1 dimensions in the form
of dualities and hidden symmetries. The widely recognized conformal symmetry,
SO$\left(  4,2\right)  $ in relativistic 1T-physics in 3+1 dimensions, emerges
from one of the simplest shadows in 2T-physics directly because of the one
extra space and one extra time dimension. This is the shadow for the free
relativistic massless particle; by now we
%have come to
call this case the \textquotedblleft conformal shadow\textquotedblright.
Experimentally observing or theoretically verifying the predicted dualities,
hidden symmetries and their consequences, in particular those related to
conformal symmetry, is one form of experimental evidence for the underlying
$\left(  D+1\right)  +2$ dimensions.

In this paper the simple cases of the shadows for Hydrogen atom, Harmonic
oscillator and a third shadow, closely resembling the conformal shadow (but
with $\mathbf{r}$ and $\mathbf{p}$ interchanged (see Eq.\ (\ref{gaugeH}) and
footnote 1 in \cite{Bars:1998pc})), are discussed in Section
\ref{sec:noncompact} with the purpose of trying to understand if Newton's
radial duality is part of the larger unifying features of 2T-physics.

In the simplest context of 2T-physics, the worldline formalism for a single
spin = 0 particle, the Sp$\left(  2,R\right)  $ gauge symmetry acts on the
phase space degrees of freedom of the particle $\left(  X^{M}\left(
\tau\right)  ,P_{M}\left(  \tau\right)  \right)  ,$ making position and
momentum in $\left(  D+1\right)  +2$ dimensions interchangeable and on equal
footing in the formulation of all classical and quantum physics for each
particle. The signature with two temporal dimensions in target space, no less
and no more, and any number of spatial dimensions$,$ is not an input, but
rather it is an output of the Sp$\left(  2,R\right)  $ gauge symmetry as
explained below. The gauge-invariant subspace of the phase space $\left(
X^{M},P_{M}\right)  $ is unitary, causal, and is physically sensible just like
1T-physics, but with more predictions than the traditional formulation of 1T-physics.

The Sp$\left(  2,R\right)  $ gauge symmetry has three gauge parameters,
$\varepsilon^{a}\left(  \tau\right)  ,a=0,1,2,$ local on the worldline, and
three corresponding generators, $Q_{a}\left(  X,P\right)  ,$ that are
functions of the phase space. Here $Q_{0}$ is the compact generator (as in
SU$\left(  2\right)  $) while $Q_{1},Q_{2}$ are non-compact. The equal-$\tau$
Poisson brackets,\footnote{By definition, the Poisson bracket between any two
functions of phase space is $\left\{  F,G\right\}  \equiv\frac{\partial
F}{\partial X^{M}}\frac{\partial G}{\partial P_{M}}-\frac{\partial F}{\partial
P_{M}}\frac{\partial G}{\partial X^{M}}.$} $\left\{  X^{M},P_{N}\right\}
=\delta_{N}^{M},$ are invariant under the infinitesimal canonical
transformations generated by each $Q_{a},$ namely
\begin{equation}
\delta_{\varepsilon}\left\{  X^{M},P_{N}\right\}  =0,\text{ for }\left\{
\begin{array}
[c]{l}%
\delta_{\varepsilon}X^{M}\equiv\varepsilon^{a}\left(  \tau\right)  \left\{
X^{M},Q_{a}\right\}  =\varepsilon^{a}\left(  \tau\right)  \frac{\partial
Q_{a}}{\partial P_{M}},\\
\delta_{\varepsilon}P_{M}\equiv\varepsilon^{a}\left(  \tau\right)  \left\{
P_{M},Q_{a}\right\}  =-\varepsilon^{a}\left(  \tau\right)  \frac{\partial
Q_{a}}{\partial X^{M}}.
\end{array}
\right.  \label{canon}%
\end{equation}
Note that no spacetime metric $g_{MN}$ is involved in any of these expressions
since $X^{M}$ is defined with a contravariant index and $P_{M}$ is defined
with a covariant index. Therefore, this formalism applies in any curved spacetime.

Generally the $Q_{a}\left(  X,P\right)  $ are non-linear functions of phase
space \cite{Araya:2013bca,BarsReviewGMann,BarsBackgrounds1} when the particle
on the worldline moves in the presence of any set of background fields in
$\left(  D+1\right)  +2$ dimensions, such as gravity, electromagnetism, high
spin fields, etc.. The $Q_{a}$ are required to form the Lie algebra of
Sp$\left(  2,R\right)  $ under classical Poisson brackets,
\begin{equation}
\left\{  Q_{0},Q_{1}\right\}  =Q_{2},\;\left\{  Q_{2},Q_{0}\right\}
=Q_{1},\;\left\{  Q_{1},Q_{2}\right\}  =-Q_{0}, \label{Lie}%
\end{equation}
and similarly under quantum commutators (after quantum ordering the
expressions for the $Q_{a}\left(  X,P\right)  $). The minus sign on the right
hand side of the last commutator is the difference between SU$\left(
2\right)  $ and Sp$\left(  2,R\right)  =$ SL$\left(  2,R\right)  =$ SU$\left(
1,1\right)  =$ SO$\left(  1,2\right)  .$ The requirement of closure as a Lie
algebra turns into a restriction on the background fields such that they must
obey certain subsidiary conditions, covariant in $\left(  D+1\right)  +2$
dimensions, that follow from (\ref{Lie}). Then the physical content of the
background fields amount to the same content of fields (of every spin) in one
less time and one less space dimension, leaving no room for Kaluza-Klein-type
additional degrees of freedom in those background fields
\cite{Araya:2013bca,BarsBackgrounds1}.

The general worldline Lagrangian for the dynamics of a single particle moving
in any background field and subject to the Sp$\left(  2,R\right)  $ gauge
symmetry is given by \cite{Bars:1998ph,Araya:2013bca,BarsReviewGMann}:
\begin{equation}
L\left(  \tau\right)  =\frac{dX^{M}\left(  \tau\right)  }{d\tau}P_{M}\left(
\tau\right)  -A^{a}\left(  \tau\right)  Q_{a}\left(  X\left(  \tau\right)
,P\left(  \tau\right)  \right)  -H\left(  X\left(  \tau\right)  ,P\left(
\tau\right)  \right)  , \label{Lagr}%
\end{equation}
where $A^{a}\left(  \tau\right)  $ is the gauge field for Sp$\left(
2,R\right)  .$ The infinitesimal gauge transformations with local gauge
parameters $\varepsilon^{a}\left(  \tau\right)  $ are based on the canonical
transformations in (\ref{canon}) and the Yang-Mills-type transformation of the
gauge field%
\begin{equation}
\delta_{\varepsilon}X^{M}=\varepsilon^{a}\frac{\partial Q_{a}}{\partial P_{M}%
},\;\delta_{\varepsilon}P_{M}=-\varepsilon^{a}\frac{\partial Q_{a}}{\partial
X^{M}},\;\delta_{\varepsilon}A^{a}=\frac{d\varepsilon^{a}}{d\tau}+\eta
^{ab}\varepsilon_{bcd}A^{c}\varepsilon^{d}. \label{transf}%
\end{equation}
where $\eta^{ad}$ is the Killing metric of Sp$\left(  2,R\right)  ,$
$\varepsilon_{abc}$ is the Levi-Civita symbol and the combinations $\eta
^{ab}\varepsilon_{bcd}$ amount to the structure constants of the Lie algebra
in Eq.\ (\ref{Lie}). Under these gauge transformations the Lagrangian
transforms to a total $\tau$-derivative \cite{BarsReviewGMann}
\begin{equation}
\delta_{\varepsilon}L\left(  \tau\right)  =\frac{d}{d\tau}\left(
\varepsilon^{a}\left(  \tau\right)  \frac{\partial L}{\partial P_{M}\left(
\tau\right)  }P_{M}\left(  \tau\right)  -\varepsilon^{a}\left(  \tau\right)
Q_{a}\left(  X\left(  \tau\right)  ,P\left(  \tau\right)  \right)  \right)  ,
\end{equation}
provided the Hamiltonian $H$ is gauge-invariant, which means it commutes with
the generators under Poisson brackets, $\left\{  H,Q_{a}\right\}  =0.$ Even
when the Hamiltonian is zero the theory based on Eq.\ (\ref{Lagr}) is
extremely rich in physical content. Therefore, in almost all discussions of
2T-physics in the literature so far, including in the present paper, the
Hamiltonian is chosen to be zero, $H=0.$

When the $\left(  D+1\right)  +2$ dimensional spacetime is flat and there are
no background fields, the three Sp$\left(  2,R\right)  $ real (or Hermitian)
generators are rearranged to simple expressions,
\begin{equation}
Q_{a}\rightarrow\left(  \left(  Q_{0}-Q_{1}\right)  ,\left(  Q_{0}%
+Q_{1}\right)  ,Q_{2}\right)  =\left(  \frac{X\cdot X}{2},\frac{P\cdot P}%
{2},\frac{X\cdot P}{2}\right)  , \label{flatQ}%
\end{equation}
where a flat spacetime metric $\eta_{MN}$ with signature $\left(  \left(
D+1\right)  ,2\right)  $ is used for the dot products$.$ This is how
SO$\left(  D+1,2\right)  $ becomes relevant as a global symmetry of the
2T-physics action (\ref{Lagr}). Note that under Poisson brackets these
quadratic expressions of phase space close to form the Lie algebra of
Sp$\left(  2,R\right)  $ as required in (\ref{Lie}). The canonical
transformations in Eqs. (\ref{canon},\ref{transf}) reduce to a linear
transformation on the phase space that treats $\left(  X^{M},P^{M}\right)  $
as a collection of Sp$\left(  2,R\right)  $ doublets, one for every spacetime
direction $M.$ The finite (as opposed to infinitesimal) Sp$\left(  2,R\right)
$ gauge transformation is then linear
\begin{equation}
\left(
\begin{array}
[c]{l}%
X^{M}\left(  \tau\right) \\
P^{M}\left(  \tau\right)
\end{array}
\right)  \rightarrow\exp\left(
\begin{array}
[c]{cc}%
\varepsilon^{2}\left(  \tau\right)  & \varepsilon^{1}\left(  \tau\right)
+\varepsilon^{0}\left(  \tau\right) \\
\varepsilon^{1}\left(  \tau\right)  -\varepsilon^{0}\left(  \tau\right)  &
-\varepsilon^{2}\left(  \tau\right)
\end{array}
\right)  \left(
\begin{array}
[c]{l}%
X^{M}\left(  \tau\right) \\
P^{M}\left(  \tau\right)
\end{array}
\right)  . \label{linear}%
\end{equation}
So in this special case of linear Sp$\left(  2,R\right)  $ transformations,
Sp$\left(  2,R\right)  $ is the same as a 2$\times2$ SL$\left(  2,R\right)  $
matrix $\left(
%TCIMACRO{\QATOP{a}{c}}%
%BeginExpansion
\genfrac{}{}{0pt}{}{a}{c}%
%EndExpansion%
%TCIMACRO{\QATOP{b}{d}}%
%BeginExpansion
\genfrac{}{}{0pt}{}{b}{d}%
%EndExpansion
\right)  $ with unit determinant$.$ Note that the $\varepsilon_{0}$
transformation is compact and as a subgroup it is a local SO(2)
transformation. To treat $\left(  X^{M},P^{M}\right)  $ as a Sp$\left(
2,R\right)  $ doublet as well as a SO$\left(  D+1,2\right)  $ vector, the $M$
index is raised for the momentum, $P^{M}=\eta^{MN}P_{N},$ by using the inverse
metric $\eta^{MN}$ that is introduced in this flat background. In this form it
is evident that the Sp$\left(  2,R\right)  $ gauge transformation commutes
with the global SO$\left(  D+1,2\right)  $ Lorentz-type target spacetime transformations.

In the flat background, the Sp$\left(  2,R\right)  $ generators in
(\ref{flatQ}), as well as the corresponding Lagrangian (\ref{Lagr}), are
symmetric under linear SO$\left(  D+1,2\right)  $ Lorentz transformations of
the elementary degrees of freedom $\left(  X^{M},P^{M}\right)  .$ Using
Noether's theorem, the generators of the conserved SO$\left(  D+1,2\right)  $
\textit{global symmetry} are constructed and verified that they commute with
the $Q_{a}$:
\begin{equation}
Q_{a}\equiv\left(  X^{2},P^{2},X\cdot P\right)  ,\;L^{MN}=\left(  X^{M}%
P^{N}-X^{N}P^{M}\right)  ,\;\left[  Q_{a},L^{MN}\right]  =0. \label{invariant}%
\end{equation}
Since the Sp$\left(  2,R\right)  $ generators are SO$\left(  D+1,2\right)
$-invariant dot products, they had to commute with the $L^{MN}$ (classically
using Poisson brackets, and quantum mechanically using quantum commutators
based on the fundamental commutator, $\left[  X^{M},P^{N}\right]  =i\eta^{MN}%
$). This statement also means that the $L^{MN}$ are Sp$\left(  2,R\right)  $
gauge-invariants since they commute with the $Q_{a}$. The gauge-invariant
sector of phase space is identified as the observables $F$ (functions of phase
space) that commute with the Sp$\left(  2,R\right)  $ generators, $\left[
Q_{a},F\left(  X,P\right)  \right]  =0.$ All gauge invariants are all possible
functions of the $L^{MN}.$ Hence, for the flat background case, these are all
the Sp$\left(  2,R\right)  $ gauge-invariant physical observables:
\begin{equation}
\text{ All physical observables in flat background:\ }F\left(  L^{MN}\right)
. \label{observables}%
\end{equation}
These functions $F\left(  L^{MN}\right)  $ need not be SO$\left(
D+1,2\right)  $-invariant. Since we have identified the gauge invariants, if
one wishes, one may add to the Lagrangian (\ref{Lagr}) any Hamiltonian,
$H\left(  X,P\right)  =H\left(  L^{MN}\right)  ,$ as long as it is any
function of the $L^{MN}.$ Such a Hamiltoian may break the global SO$\left(
D+1,2\right)  $ symmetry without destroying the essential Sp$\left(
2,R\right)  $ gauge symmetry. Even with a broken global SO$\left(
D+1,2\right)  $ there still remains an underlying SO$\left(  D+1,2\right)  $
group structure that can be used to keep track of the hidden
higher-dimensional nature of all related physics. In most of the 2T-physics
literature the discussion has concentrated on the case of a zero Hamiltonian
and unbroken SO$\left(  D+1,2\right)  $.

The equation of motion derived from the Lagrangian, by minimizing with respect
to the gauge field, $\partial L/\partial A^{a}=0,$ demands the constraints
$Q_{a}=0.$ This defines the physical space as being the gauge invariants for
which the gauge generators must vanish. In the flat background (\ref{flatQ})
this restricts the classical phase space to only the solutions of the simple
constraints
\begin{equation}
\text{If flat background: }X^{2}=0,\;P^{2}=0,\;X\cdot P=0. \label{consFlat}%
\end{equation}
The vanishing of the generators on shell is the simple statement that the
physical subspace of the phase space that obeys these equations is gauge
invariant. Here is where the reader can see why two times are required for
non-trivial physical content in the solution of these constraints. If the flat
background metric $\eta_{MN}$ were purely Euclidean (no timelike dimensions in
target space, but there still is the evolution parameter $\tau$), the only
solution is $X^{M}\left(  \tau\right)  =P^{M}\left(  \tau\right)  =0,$ which
is zero physics content. If the flat background metric $\eta_{MN}$ were
Minkowski with only one timelike direction in target space, then the only
solution would be that $X^{M}\left(  \tau\right)  $ should be parallel to
$P^{M}\left(  \tau\right)  $ and both lightlike; but this has zero angular
momentum $L^{MN}=0,$ which implies there are no nontrivial gauge-invariant
$F\left(  L^{MN}\right)  $; so again no physical content. If the flat
background metric $\eta_{MN}$ contained three or more timelike dimensions,
then the gauge-invariant sector of the theory would violate causality and also
have ghosts (negative norm states in the quantum treatment) because Sp$\left(
2,R\right)  $ is insufficient gauge symmetry to remove them. So less than two
times and more than two times are eliminated as unphysical theories. Therefore
the Sp$\left(  2,R\right)  $ gauge symmetry is the fundamental underlying
principle that requires two timelike directions in target
spacetime,\footnote{This is a conceptually fundamental viewpoint because it
attributes the signature and dimension of spacetime to emanate from the
properties of the Sp$\left(  2,R\right)  $ gauge symmetry. It may be compared
to the statement that the SU$\left(  3\right)  \times$SU$\left(  2\right)
\times$U$\left(  1\right)  $ gauge symmetry is the underlying fundamental
principle for the existence and nature of the electroweak and QCD forces,
because it is the gauge symmetry that requires the introduction of the
Yang-Mills gauge fields as the carriers of forces with the patterns of
interactions in the Standard Model of particles and fields. Similar statements
also apply to the gauge symmetries in general relativity and string theory.
Extending the Sp$\left(  2,R\right)  $ gauge symmetry concept to field theory
has in fact demanded that all known physics be (and in fact is) formulated
with an additional space and an additional time dimension.} no less and no
more, in order to have unitarily and causally sensible non-trivial physical
content.\footnote{The question arises of whether a higher gauge symmetry could
allow more timelike dimensions. This has been tried many times over the past
20 years but it has never worked out in the sense that either the physical
content is empty (too much gauge symmetry, too many constraints) or the
structure of the noncompact gauge group (signature of the gauge parameters
$\varepsilon^{a}$) is incompatible with the signature structure of spacetime
to remove all the ghosts. It seems very likely that there is a no-go theorem,
but such a theorem has not been conclusively proven.}

When the metric has just two timelike dimensions, no less and no more, the
gauge-invariant sector of the theory is causal and has no ghosts. When the
background is not flat one can similarly construct a parallel argument. A more
general reasoning confirms this conclusion: the signature of the gauge
parameters $\varepsilon^{a}$ is such that the non-compact parameters
$\varepsilon^{1,2}$ are timelike and the compact parameter $\varepsilon^{0}$
is spacelike. This means that $\varepsilon^{a}\left(  \tau\right)  ,$ together
with the constraints $Q_{a}=0,$ can remove from the $\left(  \left(
D+1\right)  +2\right)  $ dimensional phase space $\left(  X^{M},P_{M}\right)
$ precisely one space and two timelike degrees of freedom, leaving behind only
$D$ independent \textit{spacelike physical degrees of freedom}. This fully
gauge-fixed \textit{spacelike} phase space configuration has no ghosts and is
causal (evolving with $\tau$) , just as in ordinary non-relativistic or
relativistic 1T-physics. This is why two times in target space, no less and no
more, amounting to one extra space and one extra time dimension (compared to a
1T-worldline theory with $\tau$-reparametrization gauge symmetry) is predicted
by the larger gauge symmetry Sp$\left(  2,R\right)  .$

In a covariant quantization formalism (without making gauge choices to solve
the constraints (\ref{consFlat})) the gauge transformations already vanish for
the observables identified above as $F\left(  L^{MN}\right)  $ since $\left[
Q_{a},L^{MN}\right]  =0,$ hence these observables are gauge-invariant. To
implement the vanishing of the $Q_{a}$ in the covariantly quantized theory,
one requires gauge-invariant states, namely $Q_{a}|$gauge invariant$\rangle
=0.$ To identify these physical states, one begins by classifying all the
quantum states (gauge-invariant and non-invariant) by the commuting symmetries
Sp$\left(  2,R\right)  \otimes$SO$\left(  d,2\right)  $ of the action
(\ref{Lagr}), where $d=D+1$ refers to the spatial dimensions,%
\begin{equation}
|\text{all states}\rangle=|\text{Sp}\left(  2,R\right)  ,\text{SO}\left(
d,2\right)  \rangle.
\end{equation}
All possible unitary representations of both noncompact symmetries may appear.
The gauge-invariant subset of quantum states, $Q_{a}|$Sp$\left(  2,R\right)
,$SO$\left(  d,2\right)  \rangle=0$, can only be the unique singlet of
Sp$\left(  2,R\right)  ,$ so this nails down the physical states as being
singlets under Sp$\left(  2,R\right)  $ and unitary representations under
SO$\left(  d,2\right)  .$ The question still remains: Which unitary
representations of SO$\left(  D+1,2\right)  $? This depends on the background
fields from which the $Q_{a}\left(  X,P\right)  $ are constructed. In the flat
background there is a definite prediction.

In the flat background the physical state condition takes the form
\begin{equation}
\text{If flat background: }X\cdot X|\psi_{phys}\rangle=0,\;P\cdot
P|\psi_{phys}\rangle=0,\;\left(  X\cdot P+P\cdot X\right)  |\psi_{phys}%
\rangle=0. \label{quantumConstr}%
\end{equation}
These quantum states must now automatically fall into irreducible
representations of the global symmetry SO$\left(  D+1,2\right)  .$ Therefore
all gauge-invariant quantum physics derived from the Lagrangian (\ref{Lagr}),
both quantum observables $F\left(  L^{MN}\right)  $ as well quantum states
$|\psi_{phys}\rangle$, are automatically predicted to have an underlying
SO$\left(  D+1,2\right)  $ structure that reveals the underlying $\left(
D+1\right)  +2$ dimensions, even if the SO$\left(  D+1,2\right)  $ global
symmetry may be broken by adding some SO$\left(  D+1,2\right)  $ non-invariant
Hamiltonian in (\ref{Lagr}). This provides the inescapable prediction of
2T-physics regarding the presence of the underlying $\left(  D+1\right)  +2$
dimensions. This aspect may remain hidden in the usual formalism of
1T-physics, but specific predictions made by 2T-physics become the practical
tool for uncovering the hidden $\left(  D+1\right)  +2$ structure in 1T-physics.

What are the physical unitary representations of SO$\left(  D+1,2\right)  $
that are predicted for the system (\ref{Lagr})? For this we should compute the
predicted Casimir eigenvalues. At the classical level (ignoring quantum
ordering) the quadratic Casimir is $C_{2}=\frac{1}{2}L^{MN}L_{MN}=\left(
X^{2}P^{2}-\left(  X\cdot P\right)  ^{2}\right)  =0,$ where the vanishing
occurs only in the physical sector that satisfies the constraints
(\ref{consFlat}). Similarly, all Casimir operators,
\begin{equation}
C_{k}=\frac{\left(  i\right)  ^{k}}{k!}\left(  L^{M_{1}M_{2}}L_{M_{2}M_{3}%
}\cdots L^{M_{n-1}M_{k}}L_{M_{k}M_{1}}\right)  , \label{CasimirDefSO}%
\end{equation}
vanish in the physical sector of the phase space, in the classical theory.
However, in the quantum theory, by implementing the quantum constraints
(\ref{quantumConstr}) while respecting the ordering of the quantum operators
$\left(  X,P\right)  $ as they appear in $C_{k},$ one finds that the Casimir
operators are diagonal on the physical states, $C_{k}|\psi_{phys}%
\rangle=\lambda_{k}|\psi_{phys}\rangle,$ where the $\lambda_{k}$ are definite
non-trivial eigenvalues given by\footnote{The reader is alerted that the
definition of $C_{n}$ given here may differ from previous 2T-physics papers by
inessential overall normalization factors for the cases $n\geq3$. Moreover, in
the broader literature, the Casimirs for $n\geq3$ may in some definitions
amount to a linear combination of our $C_{n}$.} (see Eq.\ (9) in
\cite{Bars:1998pc} and Eq.\ (2.9) in \cite{BarsPicon})
\begin{equation}%
\begin{array}
[c]{l}%
C_{2}=\left(  1-\frac{\left(  D+1\right)  ^{2}}{4}\right)  ,\;C_{3}=\frac
{D+1}{3!}\left(  1-\frac{\left(  D+1\right)  ^{2}}{4}\right)  ,\;\\
C_{4}=\frac{1}{4!~2}\left(  1-\frac{\left(  D+1\right)  ^{2}}{4}\right)
\left(  1+\frac{3\left(  D+1\right)  ^{2}}{4}\right)  ,\cdots
\end{array}
\label{CasimirsSOd2}%
\end{equation}
This is just a single infinite-dimensional unitary representation which is
identified as the \textquotedblleft singleton\textquotedblright%
\ representation of SO$\left(  D+1,2\right)  $. So all gauge-invariant
physical quantum states of the system must be assembled into the unique
singleton representation with the specific $C_{k}$ quantum numbers given above.

This is the result of \textquotedblleft covariant
quantization\textquotedblright\ (without choosing any gauges) of the
2T-physics particle given by the Lagrangian (\ref{Lagr}) with $H=0$ and a
\textit{flat background}. The full set of quantum states corresponds to the
states of the infinite-dimensional unitary \textit{singleton} representation
of SO$\left(  D+1,2\right)  $ taken in any basis. That is, along with the
Casimir operators, a simultaneously diagonalizable subset of operators
constructed from $L^{MN}$ that defines the basis is not specified, so any such
basis will do. There clearly are an infinite set of combinations $F\left(
L^{MN}\right)  $ of simultaneously diagonalizable observables, so there are an
infinite set of quantum bases. Adding in (\ref{Lagr}) a nontrivial
gauge-invariant Hamiltonian $H\left(  L^{MN}\right)  \neq0,$ or choosing a
gauge for the $\left(  X^{M},P^{M}\right)  $ that favors some orientation
within SO$\left(  D+1,2\right)  ,$ would influence the choice of basis, but
this would not change the singleton representation that is already fixed by
the $C_{k}.$

The same gauge theory may also be treated by working in specific gauge choices
and solving the constraints $Q_{a}=0$ explicitly. Then one finds an infinite
number of solutions of the constraints in Eq.\ (\ref{consFlat}) some of which
are discussed explicitly in several papers \cite{Bars:1998pc, Araya:2013bca,
BarsReview2001}. These solutions are called \textquotedblleft\textit{shadows}%
\textquotedblright. The shadows are gauge-fixed versions of the phase space
$\left(  X^{M},P^{M}\right)  _{\text{fixed}}$ that solve the constraints,
$X_{\text{fixed}}^{2}=P_{\text{fixed}}^{2}=X_{\text{fixed}}\cdot
P_{\text{fixed}}=0.$ When only two out of three gauge choices are made, and
two out of three constraints are solved explicitly, each solution, $\left(
X_{\left(  k\right)  }^{M},P_{\left(  k\right)  }^{M}\right)  $ labelled by
$k=1,2,3,\cdots,$ is parametrized in terms of a 1T-sub-phase-space, $\left(
\mathbf{r}_{\left(  k\right)  }\mathbf{,p}_{\left(  k\right)  },t_{\left(
k\right)  },h_{\left(  k\right)  }\right)  ,$ in \textit{one less space and
one less time dimension},
\begin{equation}
\text{Shadows:\ }X_{\left(  k\right)  }^{M}\left(  \mathbf{r}_{\left(
k\right)  }\mathbf{,p}_{\left(  k\right)  },t_{\left(  k\right)  },h_{\left(
k\right)  }\right)  ,\;P_{\left(  k\right)  }^{M}\left(  \mathbf{r}_{\left(
k\right)  }\mathbf{,p}_{\left(  k\right)  },t_{\left(  k\right)  },h_{\left(
k\right)  }\right)  . \label{shadows}%
\end{equation}
Here $\left(  t_{\left(  k\right)  }\left(  \tau\right)  ,h_{\left(  k\right)
}\left(  \tau\right)  \right)  $ is a temporal canonical pair at the same
level as the $D$ spatial canonical pairs $\left(  \mathbf{r}_{\left(
k\right)  }\left(  \tau\right)  \mathbf{,p}_{\left(  k\right)  }\left(
\tau\right)  \right)  .$ An infinite set of shadows exist due to the fact that
$D$ spatial plus $1$ temporal phase space can be embedded in $\left(
D+1\right)  +2$ dimensional phase space in an infinite number of non-linear
geometric or algebraic ways. At this stage the gauge-fixed Lagrangian
(\ref{Lagr}) takes the form of a particle on the worldline in 1T-physics with
a remaining gauge symmetry and a remaining constraint%
\begin{equation}
L_{\text{2fixed}}^{\left(  k\right)  }=\left[  \dot{x}_{\left(  k\right)
}^{\mu}p_{\mu\left(  k\right)  }-A\left(  \tau\right)  Q\left(  x_{\left(
k\right)  },p_{\left(  k\right)  }\right)  \right]  ,\;x_{\left(  k\right)
}^{\mu}\equiv\left(  t_{\left(  k\right)  },\mathbf{r}_{\left(  k\right)
}\right)  ,\;p_{\mu\left(  k\right)  }\equiv\left(  h_{\left(  k\right)
},\mathbf{p}_{\left(  k\right)  }\right)  . \label{L2fixed}%
\end{equation}
A total $\tau$-derivative $\frac{d\Lambda_{k}}{d\tau}\left(  x^{\mu}%
\mathbf{,}p_{\mu},\tau\right)  $ is dropped from $L_{\text{2fixed}}^{\left(
k\right)  }$ since it does not affect the physics (see \cite{Araya:2013bca}
for a nontrivial role of this total derivative for building canonical
transformations among shadows). When the remaining third gauge choice is made
by taking $t_{\left(  k\right)  }\left(  \tau\right)  =\tau,$ the solution of
the third constraint, $Q\left(  x,p\right)  =0,$ yields an expression for an
$h_{\left(  k\right)  }$ that depends on the remaining dynamical spatial
degrees of freedom, $h_{\left(  k\right)  }=H_{k}\left(  \mathbf{r}_{\left(
k\right)  }\mathbf{,p}_{\left(  k\right)  },t_{\left(  k\right)  }\right)  .$
Then the gauge-fixed form of the original action (\ref{Lagr}) or of
(\ref{L2fixed}) for the $k^{th}$ shadow takes the standard form in
1T-physics:
\begin{equation}
L_{3\text{fixed}}^{\left(  k\right)  }=\mathbf{\dot{r}}_{\left(  k\right)
}\mathbf{\cdot p}_{\left(  k\right)  }-H_{k}\left(  \mathbf{r}_{\left(
k\right)  }\mathbf{,p}_{\left(  k\right)  },t_{\left(  k\right)  }\right)  .
\label{L3Fixed}%
\end{equation}
Here the emerging Hamiltonians $H_{k}\left(  \mathbf{r}_{\left(  k\right)
}\mathbf{,p}_{\left(  k\right)  },t_{\left(  k\right)  }\right)  $ for the
shadows are different for each distinct solution labelled by $k.$ Examples of
shadows that emerge from the Lagrangian (\ref{Lagr}) with a flat background
and $H=0$ include: the free massless relativistic particle, free massive
relativistic particle, free massive non-relativistic particle, Hatom, HOsc,
particle moving in various curved backgrounds including the expanding
universe, twistor equivalent of all these, and many others. All of these
systems (an infinite set, but only a few studied) are united by the fact that
they obey the same equations in the higher dimensions, namely $X^{2}%
=P^{2}=X\cdot P=0,$ that's all! In the lower $D$ dimensions these shadows are
all duals to each other; since they are gauge equivalent, each shadow in
$D$-space and $1$-time dimension holographically captures all the
gauge-invariant physics content available in the $\left(  D+1\right)  $-space
and $2$-time dimensions, as described below. The physics interpretation for 1T
observers, like us humans, is different for each shadow, because the gauge
choice of time and Hamiltonian, $\left(  t_{\left(  k\right)  }=\tau
,h_{\left(  k\right)  }=H_{\left(  k\right)  }=\left(  \mathbf{r}_{\left(
k\right)  }\mathbf{,p}_{\left(  k\right)  },\tau\right)  \right)  ,$ as
embedded in $\left(  D+1\right)  +2$ dimensions creates \textit{many different
1T observational perspectives of the same phenomena that occur in }$\left(
D+1\right)  +2$\textit{ dimensions.}

It is now evident that the shadow Lagrangian $L_{\left(  k\right)  }$ in
(\ref{L3Fixed}) has a hidden SO$\left(  D+1,2\right)  $ symmetry since it is
merely a gauge-fixed form of the original action (\ref{Lagr}) that has the
global SO$\left(  D+1,2\right)  $ symmetry that commutes with the gauge
symmetry. The transformation laws for the hidden symmetry are given by
computing equal-$\tau$ Poisson brackets of $\left(  \mathbf{r}_{\left(
k\right)  }\mathbf{,p}_{\left(  k\right)  }\right)  $ with the $L^{MN}$
evaluated for that shadow,
\begin{equation}%
\begin{array}
[c]{l}%
L_{\left(  k\right)  }^{MN}\equiv X_{\left(  k\right)  }^{M}\left(
\mathbf{r}_{\left(  k\right)  }\mathbf{,p}_{\left(  k\right)  },\tau\right)
P_{\left(  k\right)  }^{N}\left(  \mathbf{r}_{\left(  k\right)  }%
\mathbf{,p}_{\left(  k\right)  },\tau\right)  -P_{\left(  k\right)  }%
^{N}\left(  \mathbf{r}_{\left(  k\right)  }\mathbf{,p}_{\left(  k\right)
},\tau\right)  X_{\left(  k\right)  }^{M}\left(  \mathbf{r}_{\left(  k\right)
}\mathbf{,p}_{\left(  k\right)  },\tau\right)  ,\\
\delta_{\omega}\mathbf{r}_{\left(  k\right)  }=\frac{\omega_{MN}}{2}%
\frac{\partial L_{\left(  k\right)  }^{MN}}{\partial\mathbf{p}_{\left(
k\right)  }},\;\delta_{\omega}\mathbf{p}_{\left(  k\right)  }=-\frac
{\omega_{MN}}{2}\frac{\partial L_{\left(  k\right)  }^{MN}}{\partial
\mathbf{r}_{\left(  k\right)  }}\;\Rightarrow\;\delta_{\omega}L_{\left(
k\right)  }=\text{total }\tau\text{-derivative.}%
\end{array}
\label{ShadowInvariance}%
\end{equation}
Then the shadow Lagrangian transforms to a total derivative, so the action,
$\int d\tau L_{\text{3fixed}}\left(  \tau\right)  ,$ is SO$\left(
D+1,2\right)  $ invariant. For examples, see \cite{Araya:2013bca,BarsHidden2}.
Applying Noether's theorem by starting from the transformation rules above
(without being informed that they are assembled into $L^{MN}$) leads to the
derivation of the conserved charges $\frac{d}{d\tau}L_{\left(  k\right)
}^{MN}\left(  \mathbf{r}_{\left(  k\right)  }\mathbf{,p}_{\left(  k\right)
},\tau\right)  =0.$

After quantization, quantum ordering for each shadow must be performed such
that the $L_{\left(  k\right)  }^{MN}\left(  \mathbf{r}_{\left(  k\right)
}\mathbf{,p}_{\left(  k\right)  },\tau\right)  $ close correctly, under equal
$\tau$ quantum commutators based on $\left[  \mathbf{r}_{\left(  k\right)
}^{i}\mathbf{,p}_{\left(  k\right)  }^{j}\right]  =i\delta^{ij},$ to form the
SO$\left(  D+1,2\right)  $ Lie algebra, and yield the same Casimir eigenvalues
given in Eq.\ (\ref{CasimirsSOd2}) (for examples of such properties of the
shadows see \cite{Bars:1998pc,BarsHidden2}).

The $L^{MN}$ are gauge-invariant because they commute with the SL$\left(
2,R\right)  $ generators as seen in (\ref{invariant}). So, each $L^{MN}$ is
independent of the shadow, even when it is evaluated in terms of a given
shadow $\left(  X_{\left(  k\right)  }^{M},P_{\left(  k\right)  }^{M}\right)
$. The physical gauge-invariant observables $F\left(  L^{MN}\right)  $ in
(\ref{observables}) are then identified as functions of a smaller
$D$-dimensional Euclidean phase space for each shadow, $F\left(  L_{\left(
k\right)  }^{MN}\right)  .$ An infinite set of duality relations between
gauge-invariant observables of shadow $k_{1}$ and shadow $k_{2}$ are predicted
by evaluating any given function of the $L^{MN}$ in those two different
shadows%
\begin{equation}
\text{Dualities for every function}\;\text{: }F\left(  L_{\left(
k_{1}\right)  }^{MN}\right)  =F\left(  L_{\left(  k_{2}\right)  }^{MN}\right)
=F\left(  L^{MN}\right)  . \label{dualitiesF}%
\end{equation}
These are an infinite set of measurable predictions from 2T-physics for the
dynamics of 1T-physics. From these gauge-invariant predictions we can extract
the canonical transformations for the phase spaces $\left(  \mathbf{r}%
_{\left(  k_{1}\right)  }\mathbf{,p}_{\left(  k_{1}\right)  },t_{\left(
k_{1}\right)  },H_{\left(  k_{1}\right)  }\right)  \leftrightarrow~\left(
\mathbf{r}_{\left(  k_{2}\right)  }\mathbf{,p}_{\left(  k_{2}\right)
},t_{\left(  k_{2}\right)  },H_{\left(  k_{2}\right)  }\right)  .$ For
examples, see \cite{Araya:2013bca}.

For spinning particles of spin $s$, fermions $\psi_{i}^{M}$ with
$i=1,2,\ldots,2s,$ are added to the bosonic phase space $\left(  \psi_{i}%
^{M},X^{M},P_{M}\right)  ,$ and then the gauge group is OSp$\left(
2s|2\right)  $ that contains the previous Sp$\left(  2,R\right)  $
\cite{Bars2Tspinning}. Fermions can also be added via spacetime supersymmetry
(\cite{BarsSUSY} and Section 3.1 in \cite{BarsTwistor}). Repeating the same
reasoning that leads to the shadows, now we obtain 1T-physics that includes
spin, and the associated predictions for the corresponding shadows. In
particular one bit of information relevant for the current paper is the
SO$\left(  D+1,2\right)  $ Casimir eigenvalues analogous to
(\ref{CasimirsSOd2}) that unite all the shadows in the same unitary
representation. For spin $s$ it is given by \cite{Bars2TfieldTh}
\begin{equation}
C_{2,s}=\frac{s-1}{4}(\left(  D+1\right)  +2)(\left(  D+1\right)
+2s-2),~C_{3,s},\cdots\label{casimirs-S}%
\end{equation}
The resulting dualities are far richer than just Hatom$\leftrightarrow$ HOsc
and includes many shadows with spin that fit in the same spinning
representation of SO$\left(  D+1,2\right)  $ with the $C_{k,s}$ given above.
The spinning 1T-physics shadows include the massless spin 1/2 particle (Dirac
equation when quantized \cite{Bars2Tspinning}) and the dyonic-Hatom whose
properties are outlined in Section \ref{hints}.

Moreover, instead of flat backgrounds, the generalized version of 2T-physics
on the worldline includes background fields on which the phase space $\left(
\psi_{i}^{M},X^{M},P^{M}\right)  $ propagates \cite{BarsBackgrounds1}.
Although there are an infinite set of shadows, only some of the shadows,
including backgrounds that provide interactions, have been explicitly explored
in the language of canonical transformations in the 1T-physics framework
\cite{Araya:2013bca}. Little use has been made of the physics predictions
provided by these classical or quantum dualities. These can be useful both for
experimental predictions to verify aspects of the hidden dimensions as well as
for trying to solve difficult problems that may be more tractable in some dual
shadow version \cite{Araya:2013bca}.

In addition, these concepts have been generalized to field theory
\cite{Bars2TfieldTh}\cite{BarsReview2001}\cite{BarsReviewGMann} including the
Standard Model in $\left(  4+2\right)  $ dimensions \cite{BarsSM}, gravity
\cite{Bars2Tgravity}, their coupling to each other, and their supersymmetric
and higher-dimensional generalizations. So, 2T-physics in $\left(  4+2\right)
$ dimensions presents all the physics we know that actually works as a
shadow\ in $\left(  3+1\right)  $ dimensions. The emerging theory in the
\textquotedblleft conformal shadow\textquotedblright\ is closely related to
the familiar $\left(  3+1\right)  $ dimensional Standard Model coupled to
gravity but yields an improved standard theory \cite{BarsSteinTurok} with a
predicted local scale invariance (Weyl symmetry). The familiar form of Weyl
transformation in $3+1$ is shown to come from local general coordinate
transformations in the extra 1+1 dimensions and is present because of the
local Sp$\left(  2,R\right)  $ that acts on the phase space $\left(
X^{M},P^{M}\right)  $. These are manifestations of the underlying 4+2
dimensions in forms that are recognizable in 3+1 dimensional theories of
fundamental nature. Due to the Weyl symmetry the improved theory
\cite{BarsSteinTurok} is geodesically complete at cosmological and black
hole-type singularities. It also shows how all dimensionful parameters in the
Standard Model of particle physics (Newton constant, Higgs vacuum,
cosmological constant) come from a single source that spontaneously breaks the
Weyl symmetry \cite{BarsSteinTurok,BarsEFTC}, thus opening a window to the
extra dimensions. These consequences of the improved standard theory are
consistent with all we know while providing new insight not available before
for the source of mass and its relation to the symmetries of the underlying
4+2 dimensions.

2T-physics also potentially predicts the existence of many field theories that
are dual to the Standard Model and general relativity which may be put to use
as computational tools that take advantage of dualities in field theory
\cite{BarsFieldThDuals}. A small part of this duality was used to explore the
geodesic completeness of the improved standard theory \cite{BarsSteinTurok}
based on its Weyl symmetry and explore the new antigravity region behind the
cosmological singularity \cite{BCST,BJantigravity,BarsEFTC} as well as beyond
the black hole singularity \cite{ABJblackhole}. Little effort has been made to
expand on such predictions of 2T-physics and the more general dualities in
field theory. Hence a lot more research needs to be dedicated to further
exploration in these directions.

\end{document}